\documentclass[aps,prx,longbibliography,twocolumn,notitlepage,citeautoscript,superscriptaddress]{revtex4-2}
\usepackage{amsmath,amssymb,bm,graphicx,color,gensymb,hyperref}
\usepackage[utf8]{inputenc}
\usepackage{xcolor}

\usepackage{cancel}

\usepackage[normalem]{ulem} 
\begin{document}
\title{Easy-plane anisotropic-exchange magnets on a honeycomb lattice:\\quantum effects and dealing with them}
\author{P. A. Maksimov}
\affiliation{Bogolyubov Laboratory of Theoretical Physics, Joint Institute for Nuclear Research, 
Dubna, Moscow region 141980, Russia}
\author{A. L. Chernyshev}
\affiliation{Department of Physics and Astronomy, University of California, Irvine, California 92697, USA}
\date{\today}
\begin{abstract}
We provide analytical and numerical insights into the phase diagram and other properties of the extended Kitaev-Heisenberg model on the honeycomb lattice in the {\it easy-plane} limit,  in which  interactions are only between  spin components that belong to the plane of magnetic ions. This parameter subspace allows for a much-needed systematic {\it quantitative}  investigation of spin excitations in the ordered phases and of their generic features. Specifically,
we demonstrate that  in this limit one can consistently take into account  magnon interactions in both zero-field zigzag and field-polarized  phases. For  the nominally polarized phase, we propose a regularization of the unphysical divergences that occur at the critical field and are plaguing the $1/S$-approximation in this class of models. For the explored parameter subspace, all symmetry-allowed terms of the standard parametrization of the extended Kitaev-Heisenberg model, such as $K$, $J$, and $\Gamma$, are significant, making the offered consideration relevant to a much wider parameter space.
The dynamical  structure factor near paramagnetic critical point illustrates this relevance by showing features that are reminiscent of the ones observed in $\alpha$-RuCl$_3$, underscoring that they are not unique and should be common to a wide range of parameters of the model and, by an extension, to other materials.
\end{abstract}
\maketitle
\section{Introduction}
Recently, strongly correlated materials with a sizable spin-orbit coupling (SOC) have been the subject of a significant research interest \cite{Witczak_Krempa14}. An interplay of SOC with the crystal electric fields yields anisotropic exchange interactions between low-energy spin degrees of freedom \cite{rau_spin-orbit_2016}, raising the prospect of realizing  a variant of the compass model \cite{nussinov_compass_2015} with a spin liquid ground state and fractionalized Majorana excitations, known as the Kitaev model \cite{KITAEV2006}, in some transition-metal insulators \cite{Jackeli}. A considerable theoretical and experimental effort has been dedicated to the honeycomb-lattice materials such as  $\alpha$-RuCl$_3$ \cite{plumb2014,chaloupka2015,yadav2016,kim2016,hou_2017,Moore_swt_2018,Okamoto_dft_2020,%
kaib_me_2021,kubota2015,banerjee2016,Leahy2017,Buchner_NMR_2017,Aguilar_Raman_2019,Lebert_2020,%
Hess2020,Suzuki2021,lefranccois_phonon_2021}, with many other transition-metal and rare-earth compounds of various lattice geometries also investigated in this context \cite{Gegenwart_2012,Winter_review,Takagi_review,%
halides_review,Ni_Yb2020,YbBr3_2020,Pr_Doi_2006,Motome_Pr_2020,Khaliullin_3d_2020,%
Park_Co_2020,Wildes_Co_2017,Zvereva_Ni_2017,MM,SciRep,gardner_magnetic_2010,Ross_PRX11,%
hyper_Takagi_2015,hyper_Perkins_2018}.

While it has been well-understood that the aforementioned interplay of SOC and crystal-field effects generally leads to the other anisotropic terms beyond the coveted Kitaev interaction \cite{valenti16,Berlijn19,Suga2018}, it is only recently that the richness provided by these terms has become a central focus of the wider studies.
The so-called extended Kitaev-Heisenberg (KH) model that includes all  terms  allowed by the lattice symmetries  has shown  exceptionally fertile  phase diagram \cite{rau2014trigonal,Katukuri,Perkins_Gamma_17,Catuneanu2018,Pollet_2_21,Xin_KG_20,Xin_KG_SL_20,%
Kee_chiral_20,Liu_KG_DM_21}.  Moreover, many materials that are expected to be closely described by this model 
have also demonstrated a remarkable variety of unusual phases that  include uncommon collinear states
originating from competing interactions \cite{Coldea12,sears2015,NCSO_zigzag16} and exotic non-collinear counter-rotating spirals \cite{Coldea_spirals,Cava_2020_BaCo,Chun_Li_21}, some of which are poorly understood. Notably, the KH model on a triangular lattice also offers a wide array of exotic states, such as $Z_2$ vortex crystal and multi-$\mathbf{Q}$ spirals \cite{Ioannis,Z2_spectrum_Perkins, multiQ}. 

It is impossible for us to express sufficient amazement at the complexity of the overall problem, 
as even an approximate, quasiclassical exploration of these  states and their excitations is often 
not a trivial issue. To challenge the theoretical progress further, the anisotropic-exchange terms can result in pronounced quantum fluctuations, affecting ground state and excitation spectra in both zero-field ordered states \cite{Perkins_16_zigzag,YBK_KG18} and in the fluctuating nominally polarized phases \cite{Trebst11_hk_field,Valenti2018,Mandrus18,Gedik19,Sears2020,Balz2019,McClarty2018,%
Pollman2018,Vojta16_hk_field,Vojta17_hkg_field,Perkins17_hk_field,YBK2020,%
YBK2020_spectrum}, making the analysis of them less than straightforward.

It has been suggested that the excitation spectra in all  ordered phases should  demonstrate universal features in the form of the broad, continuum-like modes, owing to strong magnon interaction \cite{us_rucl3} that comes from the strong coupling of all spin components due to SOC. While fascinating on their own, these features may also  give a potential ``false-positive'' signal of a spin liquid if the broad spectrum on itself is naively taken as a sign of fractionalized excitations instead. Therefore, a constructive discussion of  the quantum effects in the ordered phases is crucial for an understanding of the properties of the model and of experiments.

Unfortunately, a consistent consideration of excitations in the extended Kitaev-Heisenberg model is, generally, a cumbersome procedure, as even the linear approximation requires a numerical diagonalization of large matrices due to complexity of the Hamiltonian and the states themselves \cite{Colpa,Smit20}. However, as  was first pointed out in Ref.~\cite{Smit20}, one can search for a subset of the parameter space that  allows for a more straightforward calculations of the non-linear effects, but is still representative of the wider  phase diagram.

In this work, we provide further insights into the nature of  quantum effects in the spectra of  generalized
Kitaev-Heisenberg model by exploring a different part of the phase diagram that is complementary to the region discussed previously \cite{Smit20}. We refer to this parameter subspace as to the {\it easy-plane},
because if written in crystallographic axes related to the planes of magnetic ions  \cite{chaloupka2015},  bond-dependent and bond-independent terms of the 
model  contain   interactions only between the in-plane spin components. As we demonstrate,
diagonalization of the bosonic spin-wave Hamiltonian in this subspace can be done via a simple combination of the unitary and  paraunitary transformations,  making the subsequent  calculation of the non-linear terms rather unburdensome. Our approach also demonstrates, once again, that a parametrization of extended Kitaev-Heisenberg model in crystallographic axes can be extremely beneficial for an exploration and better understanding of its phase diagram \cite{us_PRR}.

In the present study, we focus on quantum effects in the zigzag and field-induced paramagnetic phases  for a representative set of parameters from the easy-plane subspace. Remarkably, the region that we study,  corresponds to an extension of the honeycomb 120${\degree}$ compass model, which has drawn attention in the past in the context of models and compounds with orbital degeneracy \cite{nussinov_compass_2015}. We point out that our easy-plane choice of parameters, rewritten using cubic axes parametrization in which generalized KH model is typically written, corresponds to all principal symmetry-allowed terms of the model, $K$, $J$, and $\Gamma$, being of the same order. This makes a convincing case that although we do not attempt to study the entire phase diagram of an extended KH model, our consideration is relevant to a much wider parameter space, which should retain all the thought-provoking spectral features that we demonstrate.

Utilizing the benefits of analytical approach to the  spin-wave theory, we can calculate   magnon self-energy explicitly. We  show that quantum effects, such as spontaneous magnon decays and spectrum renormalization \cite{decay_review}, are sizable in both zero-field zigzag and  high-field polarized phases.  In the polarized phase, a strong divergence
in the spectrum renormalization at the critical field is also observed. Previously, it has been suggested that this singularity signifies a renormalization of the critical field, see Ref.~\cite{Vojta2020_NLSWT}, but the problem with the unphysical spectrum was left unresolved. In this study, we  present a method  to regularize such a divergence that should be relevant to a large class of anisotropic-exchange models with complex ground states.

We have also analyzed quantum effects in the phase diagram of the considered model using numerical density matrix renormalization group (DMRG) method \cite{white_density_1992,itensor}. We have confirmed that the strong spectrum renormalization effects that we find in the zigzag and polarized phase correspond to a shift of the phase boundaries. According to the DMRG results, the area occupied by the zigzag phase contracts in the regions where the spin-wave theory predicts strong renormalization that can yield spectrum instabilities. The numerical calculations in the polarized phase also confirm the downward renormalization of the critical fields relative to their classical values.

Recent spectroscopic experiments in $\alpha$-RuCl$_3$, such as electron spin resonance (ESR), time-domain terahertz spectroscopy, and Raman scattering  \cite{Loidl17,Zvyagin17,Sahasrabudhe20,Wulferding20,Zvyagin20}, have demonstrated unusual features in the polarized phase.  These include  nearly gapless $\mathbf{q}=0$ excitations and a significant spectral weight of the continuum above the one-magnon mode at the critical field, both in contrast with the linear spin-wave theory prediction. In the present work, by using the non-linear  $1/S$-approach, we reproduce these features of the dynamical spin structure factor in the polarized phase for representative sets of parameters of our model . Although our model is not directly applicable to $\alpha$-RuCl$_3$ \cite{us_PRR}, it clearly points to the fact that this phenomenology is not unique and should be general for the extended Kitaev-Heisenberg model and related compounds.

Our paper is structured as follows. We introduce the {\it easy-plane} anisotropic-exchange Hamiltonian and map out its classical phase diagram in Sec.~\ref{sec_model}. The linear  and non-linear spin-wave spectra  in the zigzag state are discussed in Sec.~\ref{sec_zz}. Section~\ref{sec_xfield} presents results for the spectrum in the field-polarized phase in the on-shell and off-shell approximations. The approach to a regularization of the singularity at the critical field is presented in Sec.~\ref{sec_onshell}. The dynamical structure factor and its features that are relevant to the experiments are discussed in Sec.~\ref{sec_rucl3}. We conclude by Sec.~\ref{sec_discussion} and  additional details are provided in  Appendixes.

\section{Easy-plane anisotropic-exchange model and classical phase diagram}
\label{sec_model}

\begin{figure*}
\centering
\includegraphics[width=2.0\columnwidth]{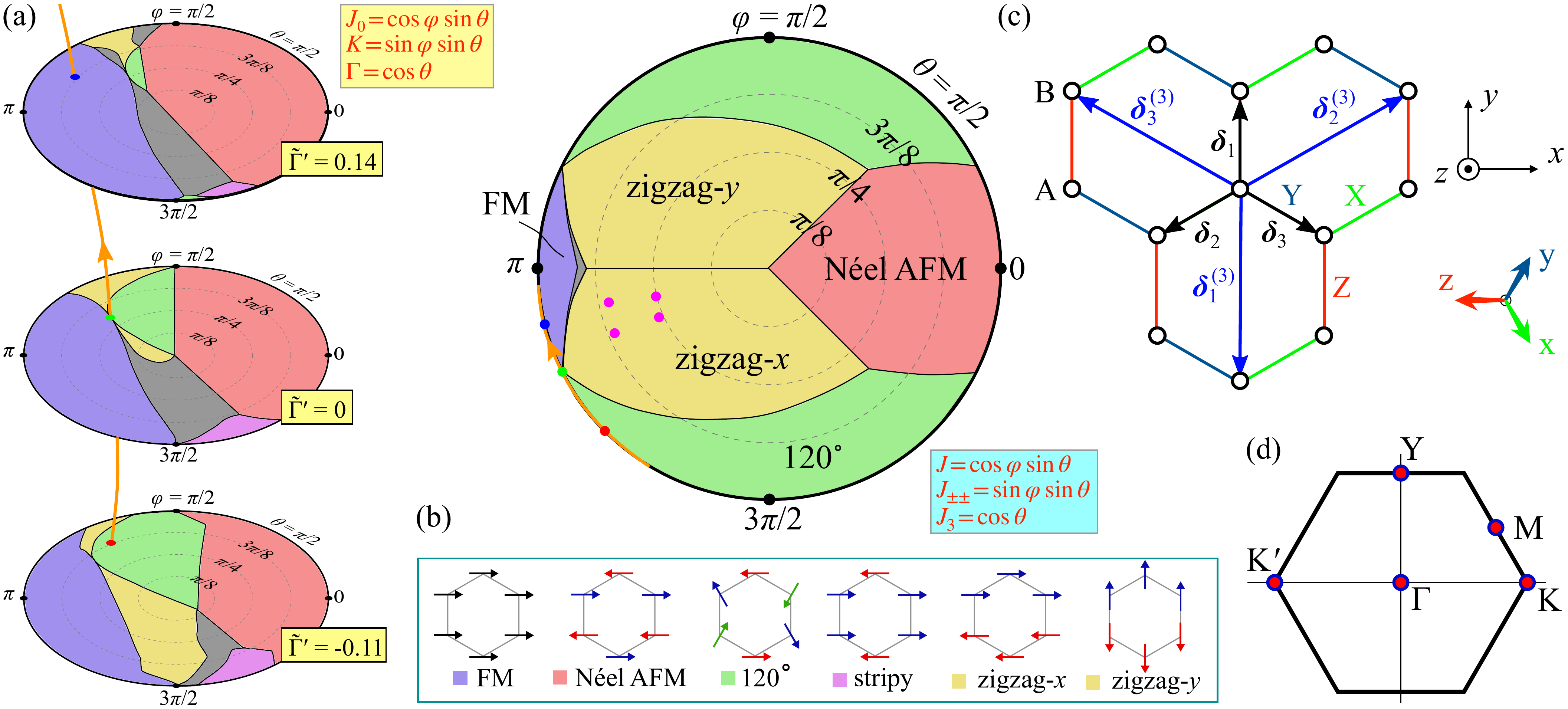}
\caption{(a) The classical phase diagrams of the model (\ref{eq_H_JKGGp}) in the $J_0$-$K$-$\Gamma$ parameter space for $\Gamma>0$ and for three representative values of $\Gamma'$ in units of $\sqrt{J_0^2+K^2+\Gamma^2}$.  The legend shows sketches of the  single-$\mathbf{Q}$ phases, incommensurate phases are shown in gray. The blue, green, and red points in (a) and (b) are the same and are related via Eq.~\eqref{eq_jkg_xy_transform}, see the text. (b) The classical phase diagram of the $J$-$J_{\pm\pm}$-$J_3$ model \eqref{HJpm_xy} for $J_3>0$. The  magenta points in the zigzag phase are representative parameter sets used later in this work. (c) The honeycomb lattice of magnetic ions with A and B sublattices, the nearest-  and third-nearest-neighbor vectors, $\bm{\delta}_\alpha$ and $\bm{\delta}^{(3)}_\alpha$,  ${\rm \{X,Y,Z\}}$ types of bonds for the KH model \eqref{eq_H_JKGGp},  and the cubic, $\{\text{x,y,z}\}$, and crystallographic, $\{x,y,z\}$, axes. (d) Brillouin zone (BZ) of the honeycomb lattice with the  ordering vectors of the single-$\mathbf{Q}$ phases.}
\label{fig_lattice}
\end{figure*}

Generally, the anisotropic-exchange models of insulators with strong SOC  do not retain the SU(2) spin symmetry. Due to crystal-field effects, all anisotropic terms beyond Heisenberg interaction are  allowed, provided they respect  discrete symmetry of the lattice \cite{rau_spin-orbit_2016,Rau18,rau_jkg,valenti16}. Therefore, the general form of a bilinear interaction of the low-energy spin Hamiltonian can be written as
\begin{align}
\hat{\cal H}=\sum_{\langle ij\rangle_n} \mathbf{S}^{\rm T}_i \hat{\bm J}^{(n)}_{ij} \mathbf{S}_j
\label{eq_Hij}
\end{align}
where ${\bf S}_i^{\rm T}\!=\!\left(S_i^{x},S_i^{y},S_i^{z}\right)$, $\langle ij\rangle_n$ denotes $n$-th nearest-neighbor sites, and  $\hat{\bm J}^{(n)}_{ij}$ is a $3\!\times\! 3$ exchange matrix. The elements of $\hat{\bm J}^{(n)}_{ij}$ matrices are constrained by the symmetry of the lattice and typically depend on the orientation of the bond. 

In this work, we study  extended Kitaev-Heisenberg spin-exchange model on the honeycomb lattice of magnetic ions with the lattice symmetry given by the edge-sharing octahedral environment of ligands \cite{Rau18}. In addition to the nearest-neighbor interaction $\hat{\bm J}^{(1)}_{ij}$, we also include third-nearest neighbor term $\hat{\bm J}^{(3)}_{ij}$ as a proxy of all further-neighbor interactions and also because  it is often found to be significant in various transition-metal insulators with the honeycomb-lattice structure   \cite{Winter_review,dejongh1990,valenti16,Regnault_BaCoAsO_1977,Regnault_BaNiPO_1980,%
Ross_BaCoPO_2018,Songvilay_Co_2020,Dai_BaNiAsO_2021,Paramekanti_Co_2021}. 

\subsection{Nearest-neighbor model}

In the presence of anisotropy, the form of $\hat{\bm J}^{(n)}_{ij}$ is not invariant under the rotation of the spin quantization axes. Because of the perceived prevalence of anisotropic terms, the conventional choice for such axes are the ones associated with the metal-ligand bonds in the idealized octahedra, referred to as the cubic axes $\{\text{x,y,z}\}$, illustrated in Fig.~\ref{fig_lattice}(c), not the ones affiliated with the planes of magnetic ions. In these axes, the nearest-neighbor part of the model \eqref{eq_Hij}, with four terms that are allowed by the lattice symmetry, is the generalized Kitaev-Heisenberg model
\begin{align}
\hat{\cal H}_1=\sum_{\langle ij \rangle_1}&\Big\{
J_0 \mathbf{S}_i \cdot \mathbf{S}_j +K S^\gamma_i S^\gamma_j 
+\Gamma \left( S^\alpha_i S^\beta_j +S^\beta_i S^\alpha_j\right)\nonumber\\
+&\Gamma' \left( S^\gamma_i S^\alpha_j+S^\gamma_i S^\beta_j+S^\alpha_i S^\gamma_j+S^\beta_i S^\gamma_j\right)\Big\}.
\label{eq_H_JKGGp}
\end{align}
Here the sum is taken over  three types of bonds of the honeycomb lattice that are denoted as {X,Y, and Z}, with corresponding nearest-neighbor vectors $\bm{\delta}_{2}$, $\bm{\delta}_{3}$, and $\bm{\delta}_{1}$, shown in Fig.~\ref{fig_lattice}(c). Thus, $\{\alpha,\beta,\gamma\}=\{\text{x,y,z}\}$ for the Z-type bond and interactions on the other bonds are obtained via a cyclic permutation of  indices.

This model \eqref{eq_H_JKGGp} has been thoroughly explored in the quantum $S\!=\!1/2$ and classical limits, with the single-$\mathbf{Q}$ ordered phases that include ferromagnetic (FM), N{\'e}el antiferromagnetic, 120${\degree}$, stripy, and zigzag phases, as well as various incommensurate ordered and quantum disordered states identified in its phase diagram \cite{chaloupka2010,Kimchi11,chaloupka_zigzag_2013,Oles_KH_16,Chaloupka_jkg_2019,%
Liu_KGGp_20,Pollet21,buessen2021functional}. The middle panel of Fig.~\ref{fig_lattice}(a) shows the classical phase diagram for $\Gamma'=0$ and $\Gamma>0$, with the bottom and top panels showing its evolution with finite $\Gamma'$. The legend of Figure 1 sketches the single-$\mathbf{Q}$ ordered states and Fig.~\ref{fig_lattice}(d) explicates the ordering $\mathbf{Q}$-vectors associated with them: $\Gamma$-point for the FM and N{\'e}el, K or K$^\prime$ for the 120${\degree}$, and M or Y for the stripy and zigzag states. These results are obtained using Luttinger-Tisza approach \cite{lt_original,Ioannis,us_PRR,Lyons_Kaplan_1960,Friedman_1974,Litvin_1974,Niggemann_2019} to the model \eqref{eq_H_JKGGp}, see details in  Appendix \ref{app_model}. This phase diagram also agrees with the prior work on the same model \cite{rau_jkg,Smit20,us_PRR}.

We would like to focus on a particular parameter set
\begin{align}
\label{eq_point_jkg}
\Gamma'=0, \ \ K=\Gamma=-J_0,
\end{align}
for $J_0<0$, which is marked in the phase diagram in the middle panel of Fig.~\ref{fig_lattice}(a) by the green dot. This point is tricritical between the ferromagnetic, zigzag, and $120{\degree}$ phases. As was pointed out in the earlier studies \cite{rau_jkg}, the macroscopic degeneracy can be easier seen if the model is rewritten as a single-parameter honeycomb $120{\degree}$ compass model \cite{nussinov_compass_2015}
\begin{align}
\hat{\cal H}_c = J_c\sum_{\langle ij\rangle_1} \big(\hat{{\bm \delta}}_{ij}\cdot\mathbf{S}_i\big)\big(\hat{{\bm \delta}}_{ij}\cdot\mathbf{S}_j\big),
\label{eq_120compass}
\end{align}
where $J_c\!=\!-2K$ and $\hat{{\bm \delta}}_{ij}$ are the unit vectors of the nearest-neighbor bond direction of the  lattice. This model has attracted interest in the context of the $p$-band Mott insulators \cite{120_ObD1} and of the $e_g$ Kugel-Khomskii model on the honeycomb lattice \cite{120_dimers,ishihara_eg,mila_su4}. Lifting the macroscopic degeneracy in the model \eqref{eq_120compass} has been considered in both classical and quantum $S\!=\!1/2$ limits \cite{120_ObD2,120_continuum,120_VBS,120_SL}.

One can lift the degeneracy of this special point by introducing further terms to the compass model \eqref{eq_120compass}. A particularly straightforward way that retains the co-planar character of the coupling in \eqref{eq_120compass} is to enrich it by the easy-plane ($XY$) exchange interactions
\begin{align}
\hat{\cal H} = \hat{\cal H}_c +\sum_{\langle ij\rangle_1} J^{(1)}_{xy}\left( S^x_i S^x_j +S^y_i S^y_j\right).
\label{eq_120compass_xy}
\end{align}
Note that the model \eqref{eq_120compass_xy} is already written in a basis that is different from the cubic axes in \eqref{eq_H_JKGGp} and is naturally tied instead to the plane of magnetic ions, which we will refer to as the crystallographic axes $\{x,y,z\}$, see Fig.~\ref{fig_lattice}(c) for a specific choice of them, with $y$ axis being the $C_2$-symmetry axis.

It is particularly enlightening now to rewrite the resultant model   \eqref{eq_120compass_xy} one more time 
using a more common ``ice-like''  convention for the parametrization of the nearest-neighbor exchange matrix
$\hat{\bm J}^{(1)}_{ij}$ \cite{Curnoe08,Onoda_2011,Ross_PRX11,Rau18} to make its {\it easy-plane} character explicit
\begin{align}
\label{HJpm_xy1}
\hat{\cal H}=&\sum_{\langle ij\rangle_1}\Big\{
J \Big(S^{x}_i S^{x}_j+S^{y}_i S^{y}_j\Big)\\
-&2 J_{\pm \pm} \Big( \Big( S^x_i S^x_j - S^y_i S^y_j \Big) c_\alpha 
-\Big( S^x_i S^y_j+S^y_i S^x_j\Big) s_\alpha \Big)\Big\}.\nonumber
\end{align}
Here the shorthand notations are $c_\alpha\equiv\cos\tilde{\varphi}_\alpha$ and $s_\alpha\equiv\sin\tilde{\varphi}_\alpha$,  bond-dependent phases $\tilde{\varphi}_\alpha\!=\!\{0,2\pi/3,-2\pi/3\}$ are the angles of the nearest-neighbor vectors ${\bm \delta}_\alpha$ shown in Fig.~\ref{fig_lattice}(c) with the $y$ axis, and spin components are in the $\{x,y,z\}$ crystallographic axes. The ``ice-like''  \eqref{HJpm_xy1} and the ``compass-like''  \eqref{eq_120compass_xy} model parameters are related via $J_c=2J_{\pm\pm}$ and $J^{(1)}_{xy}=J-2J_{\pm\pm}$, with the pure compass $J_c$-only model \eqref{eq_120compass} corresponding to $J_{\pm\pm}=J/2$. 

It is now essential to reflect on the fact that the lattice symmetry of the honeycomb lattice with an edge-sharing octahedral environment of ligands allows four independent parameters in the  nearest-neighbor anisotropic-exchange model, regardless of the parametrization of the latter. It is clear that the model \eqref{HJpm_xy1} is restricted compared to the extended KH model \eqref{eq_H_JKGGp}, as it only contains two in-plane terms, $J$ and $J_{\pm\pm}$, and is missing the couplings involving the out-of-plane spin components, the $XXZ$-term $\Delta S^z_i S_j^z$ and the $J_{z\pm}$-term that couples $S^z$ to the in-plane $S^{x(y)}$ components, see  Appendix~\ref{app_model} for the full model in the ``ice-like'' parametrization. 

It is important to connect the easy-plane anisotropic-exchange model \eqref{HJpm_xy1} back to the cubic-axis parametrization of the extended KH model \eqref{eq_H_JKGGp}. A straightforward linear transformation, corresponding to a rotation from the cubic to the crystallographic axes, see Appendix~\ref{app_model}, yields a simple translation table 
\begin{align}
J_0&=\frac{2}{3}\big( J+J_{\pm\pm}\big),~K=-2J_{\pm\pm},\nonumber\\
\Gamma&=-\frac{1}{3} \big( J+4J_{\pm\pm}\big),~\Gamma'=-\frac{1}{3} \big(J-2J_{\pm\pm}\big),
\label{eq_jkg_xy_transform}
\end{align}
which implies that, generally, all four interactions of the extended Kitaev-Heisenberg model are non-zero in the parameter subspace spanned by the easy-plane model. 

Perhaps a  more informative result   is  the relations between the parameters of the   extended KH model that are imposed  by the easy-plane  parameter subspace. Such relations can be easily  inferred from Eq.~(\ref{eq_jkg_xy_transform})
\begin{align}
J_0 =-\big( \Gamma+\Gamma^\prime\big),~K=\Gamma-\Gamma^\prime,
\label{eq_KH_constrained}
\end{align}
where $\Gamma$ and $\Gamma^\prime$ are chosen as the two independent parameters of such a restricted KH model. Clearly, $\Gamma^\prime\!=\!0$ brings us back to the special tricritical $K\!=\!\Gamma\!=\!-J_0$ point of Eq.~(\ref{eq_point_jkg}) that has inspired this consideration.

The one-dimensional (1D) path illustrated in Fig.~\ref{fig_lattice}(a) passes through the tricritical point (\ref{eq_point_jkg}) in the $\Gamma'=0$ panel and is going through different phases for  different choices of $\Gamma'$ according to (\ref{eq_KH_constrained}). The relation of this path to the easy-plane model (\ref{HJpm_xy1}) is demonstrated in Fig.~\ref{fig_lattice}(b).  The  tricritical point in the ice-like parameters is given by $J_{\pm\pm}=J/2<0$ and is shown as a green dot on the outer rim ($J_3=0$) of the $J$-$J_{\pm\pm}$-$J_3$ phase diagram in Fig.~\ref{fig_lattice}(b).
The two other representative points along this rim, which are shown by the red and blue dots in Fig.~\ref{fig_lattice}(b), correspond to the choice of  $J_{\pm\pm}\!=\!J$ and $J_{\pm\pm}\!=\!0.25J$, respectively, and to  the red and blue dots in the upper and lower panels of Fig.~\ref{fig_lattice}(a). 
The choice of the constant $\Gamma'$ (in units of $\sqrt{J_0^2+K^2+\Gamma^2}$) in these panels corresponds to its value at these representative points that can be deducted from Eq.~\eqref{eq_jkg_xy_transform}.
Thus, each point on the outer rim of the $J$-$J_{\pm\pm}$-$J_3$ phase diagram in Fig.~\ref{fig_lattice}(b) corresponds to a single point in a $J_0$-$K$-$\Gamma$ slice of the  three-dimensional  (3D) phase diagram of the  model (\ref{eq_H_JKGGp}) for a fixed  $\Gamma'$.  

Altogether, the two-parameter easy-plane  anisotropic-exchange model (\ref{HJpm_xy1}) offers a 1D exploratory path through the  3D space of the  four-parameter extended KH model \eqref{eq_H_JKGGp}. This path  explores all of the commensurate ordered phases of the full model, with all parameters of the extended KH model present but without imposing a fine-tuning on any of them.  
Yet, this model (\ref{HJpm_xy1}) allows for significantly more straightforward calculations of the non-linear effects that we pursue in this work, remaining representative of the wider phase diagram.
The possibility of such a simplified consideration,  suggested by the model (\ref{HJpm_xy1}), highlights  the benefit of using different parametrizations of the anisotropic-exchange models~\cite{us_anisotropic}.

\subsection{Role of $J_3$}

As was mentioned above, in the following  we will include the third-nearest neighbor exchange term  in addition to the nearest-neighbor terms of the model (\ref{HJpm_xy1}). This extension abbreviates all further-neighbor interactions into one and takes into account a significance of the third-neighbor terms in real materials~\cite{Winter_review,dejongh1990,valenti16,Regnault_BaCoAsO_1977,Regnault_BaNiPO_1980,%
Ross_BaCoPO_2018,Songvilay_Co_2020,Dai_BaNiAsO_2021,Paramekanti_Co_2021}.

To maintain the easy-plane character of the model (\ref{HJpm_xy1}), the extra $J_3$-terms are taken in the same form as the $J^{(1)}_{xy}$ extension of the pure compass model in (\ref{eq_120compass_xy}), yielding the three-parameter $J$-$J_{\pm\pm}$-$J_3$ model 
\begin{align}
\hat{\cal H}=&\sum_{\langle ij\rangle_1}\Big\{
J \Big(S^{x}_i S^{x}_j+S^{y}_i S^{y}_j\Big)\nonumber\\
-&2 J_{\pm \pm} \Big( \Big( S^x_i S^x_j - S^y_i S^y_j \Big) c_\alpha 
-\Big( S^x_i S^y_j+S^y_i S^x_j\Big) s_\alpha \Big)\Big\}\nonumber \\
\label{HJpm_xy}
+&J_3 \sum_{\langle ij\rangle_3}\Big(S^{x}_i S^{x}_j+S^{y}_i S^{y}_j\Big). 
\end{align}
It is written in the same ``ice-like'' notations as the nearest-neighbor model (\ref{HJpm_xy1}) and this is the parametrization that will be used exclusively from now on.

Fig.~\ref{fig_lattice}(b) demonstrates the role played by the third-neighbor term in the classical phase diagram of the easy-plane anisotropic-exchange model \eqref{HJpm_xy} for  $J_3>0$. The phase diagram contains the same commensurate single-$\mathbf{Q}$ states as the $J_0$-$K$-$\Gamma$ diagrams in Fig.~\ref{fig_lattice}(a) except for the stripy phase, which is stabilized by $J_3<0$. The $J_3<0$ hemisphere of the phase diagram and classical energies of all  commensurate phases are given in Appendix \ref{app_model}.

In the rest of the paper we will be focusing on the zigzag portion of the phase diagram in Fig.~\ref{fig_lattice}(b) because of its relative simplicity and also motivated by a large number of materials with strong anisotropic exchange interactions that realize such a zigzag order in their ground states  \cite{Hill11,Coldea12,sears2015,johnson2015,NCTO_zigzag16,NCSO_zigzag16}.
As one can see in Fig.~\ref{fig_lattice}(b),  the zigzag state occupies a large portion of the phase diagram of the  model \eqref{HJpm_xy}.  However, we emphasize that the third-nearest neighbor interaction is crucial for its stability, because  in the nearest-neighbor model \eqref{HJpm_xy1} the zigzag state is stable only at the two tricritical points, $J_{\pm\pm}=\pm J/2$. The scenario that the zigzag order is stabilized by a $J_3$-term has been  discussed and significantly substantiated in a number of studies of closely related models and materials
\cite{Winter_review,us_rucl3,dejongh1990,Paramekanti_Co_2021,us_PRR}

As is shown in  Fig.~\ref{fig_lattice}, there are two types of the zigzag state that differ by the mutual orientation of the zigzag and spin patterns, which we refer to as to the zigzag-$x$ and zigzag-$y$ states. In the former, favored by $J_{\pm\pm}<0$, spins align with the zigzag direction, while for the latter, $J_{\pm\pm}>0$, spins are perpendicular to it.

The main focus of this work is on the quantum corrections to the  spin-wave theory in the zigzag and high-field polarized phases of the model \eqref{HJpm_xy}. Generally, for the complex multiple-sublattice  ordered states in the non-Bravais lattices shown in Fig.~\ref{fig_lattice}, even the linear spin-wave theory (LSWT)  consideration requires numerical diagonalization, making the problem of the non-linear effects virtually intractable. The fact that such kinds of calculations  can be done analytically for the easy-plane anisotropic-exchange model \eqref{HJpm_xy} is a strong motivation for its exploration. The next two Sections provide a thorough investigation of the quantum effects in the zero-field zigzag and the affiliated field-induced polarized states of the model \eqref{HJpm_xy}. 
We argue that the obtained results are general and the studied model is able to provide insights into experimentally observed features in real materials. 

\section{Zero-field zigzag state}
\label{sec_zz}

Similarly to the same model on the triangular lattice \cite{us_anisotropic}, the structure of the zigzag state in the model \eqref{HJpm_xy} depends on the sign of $J_{\pm\pm}$. The classical energy of the zigzag state  is given by
\begin{align}
e_{\rm cl}=\frac{E_{\rm cl}}{NS^2}=J+4J_{\pm\pm}\cos 2\varphi-3J_3,
\label{Ecl}
\end{align}
where $N$ is the number of atomic unit cells and $\varphi$ is the angle of spins with the $x$-axis. Minimization yields $\varphi=\pi/2$ for $J_{\pm\pm}>0$ and $\varphi=0$ for $J_{\pm\pm}<0$, zigzag-$y$ and zigzag-$x$ states in Fig.~\ref{fig_lattice}, respectively, making $e_{\rm cl}\!=\!J-4|J_{\pm\pm}|-3J_3$ for both cases.

\subsection{Two-sublattice approach}
\label{Sec_lswt}

The spin-wave theory requires a diagonalization of the $2N_s \times 2N_s$ matrix after bosonization of the exchange Hamiltonian \cite{Colpa}, where $N_s$ is the number of magnetic sublattices. Generally, for $N_s\geq 2$ this procedure can only be done numerically. However, it is known that there are cases in which diagonalization can be performed analytically even for the complex ordered states and lattices \cite{Smit20,Kopietz,us_kagome,ushoney16}.
The procedure that we demonstrate here is based on a combination of the unitary and para-unitary transformations, which also utilizes  higher symmetry of the zigzag state in model (\ref{HJpm_xy}).

Within the spin-wave theory, the key transformation is the rotation of the spin quantization axis from the laboratory reference frame to the local one, with the local $z$-direction   on each site given by the spin 
configuration obtained from  minimization of the classical energy.

In the zigzag-$x$ state, using its sketch in Fig.~\ref{fig_lattice} as a guidance,  spin transformation from the laboratory  to the local reference frame  for the A(B) sublattices  is given by
\begin{align}
\label{rotation}
\big(S_{i}^{x}, S^y_i, S_{i}^{z}\big)_{\rm lab} = 
\big(\pm e^{i{\bf Q}{\bf r}_{\ell}}S_{i}^{z}, S^x_i, \pm e^{i{\bf Q}{\bf r}_{\ell}}S_{i}^{y}\big)_{\rm loc},
\end{align}
where we choose ${\bf Q}\!=\!\left(0,2\pi/3a\right)$
as the ordering vector of the zigzag structure with the horizontal direction of the zigzag pattern, $a$ is the interatomic distance, coordinates ${\bf r}_{\ell}\!=\!n{\bf a}_1\!+\!m{\bf a}_2$ correspond to the A-sublattice with the primitive vectors of the honeycomb lattice ${\bf a}_{1}\!=\! \bm{\delta}_2-\bm{\delta}_1$ and ${\bf a}_{2}\!=\! \bm{\delta}_3-\bm{\delta}_1$, and 
${\bf r}_i={\bf r}_\ell+{\bm\rho}_\alpha$ are the coordinates of the atoms, with
${\bm\rho}_1\!=\! (0,0)$ and ${\bm\rho}_2\!=\! \bm{\delta}_1\!
=\!\left(0,a\right)$. This transformation introduces axes that follow the 
staggered pattern of the zigzag state with the phase factor
$e^{i{\bf Q}{\bf r}_{\ell}}\!=\!(-1)^{n+m}$,  retaining the two-sublattice  structure of the honeycomb lattice. 

Note that the Hamiltonian \eqref{HJpm_xy} is invariant under  simultaneous in-plane $\pi/2$-rotation of spins and  change of the sign of the $J_{\pm\pm}$-term. This is identical to the anisotropic model on the triangular lattice for the stripe-$x$ and stripe-$yz$ states \cite{us_anisotropic}. Due to this symmetry, the Hamiltonians for the 
zigzag-$x$ and zigzag-$y$ states reduce to the same form when written in the local spin axes
\begin{align}
\hat{{\cal H}}_\text{loc}=&\sum_{i,{\bm \delta}_1}\Big\{
\big(J- 2 \left| J_{\pm \pm}\right|\big)S^{x}_i S^{x}_j-\big(J+2 \left| J_{\pm \pm}\right|\big)S^{z}_i S^{z}_j\Big\}\nonumber\\
+&\sum_{i,{\bm \delta}_{2,3}}\Big\{
\big(J+  \left| J_{\pm \pm}\right|\big)S^{x}_i S^{x}_j+\big(J- \left| J_{\pm \pm}\right|\big)S^{z}_i S^{z}_j\Big) 
\nonumber\\
&\quad \quad \quad \quad \ \ \ \ -2 \left| J_{\pm \pm}\right|e^{i{\bf Q}{\bf r}_{i}}\Big( S^x_i S^z_j+S^z_i S^x_j\Big) s_\alpha \Big\}\nonumber\\
  +&J_3\sum_{i,{\bm \delta}_\alpha^{(3)}} \Big(S^{x}_i S^{x}_j-S^{z}_i S^{z}_j\Big),
\label{HJpm_xy2}
\end{align}
where $i\!\in$~A, ${\bf r}_j\!=\!{\bf r}_i+{\bm \delta}({\bm \delta}^{(3)})_\alpha$,   and we  used explicit values of 
$\cos\tilde{\varphi}_\alpha\!=\!\{1,-1/2,-1/2\}$ and $\sin\tilde{\varphi}_1\!=\!0$ for the bond-dependent 
phases $\tilde{\varphi}_\alpha$ in (\ref{HJpm_xy1}), leaving the shorthand notation of $s_\alpha\!=\!\sin\tilde{\varphi}_\alpha\!=\!\pm\sqrt{3}/2$ for the ${\bm \delta}_{2(3)}$ bonds.

We note that already at this stage, the simplified nature  of the easy-plane model (\ref{HJpm_xy}) and of the coplanar ground state manifest themselves in an explicitly two-sublattice form of the ``diagonal'' part of the Hamiltonian  (\ref{HJpm_xy2}), which yields the LSWT below, as they do not contain phase factors associated with the zigzag ordering vector ${\bf Q}$. The  terms that do are ``off-diagonal'' and  contribute only to the  higher $1/S$ order. This is different in a more general model because extra terms and the out-of-plane tilt of the zigzag structure generally do not permit the four-to-two-sublattice reduction, see Ref.~\cite{Smit20}.

\subsection{Linear spin-wave formalism}
\label{sec_lswt_formalism}

Here we present  basic steps of the linear spin-wave theory formalism and outline the key 
features of our model that allow for a simplified analytical treatment.
 
The standard bosonization of spin operators is the Holstein-Primakoff transformation, 
\begin{equation}
S^z_{\ell,\alpha} = S-a_{\ell,\alpha}^\dag a^{\phantom{\dag}}_{\ell,\alpha},\ \ \ S^+_{\ell,\alpha}\approx \sqrt{2S}\, a_{\ell,\alpha},
\label{eq_HP}
\end{equation} 
where $z$ is the local quantization axis, ${\ell}$ and $\alpha\!=\!1,2$ numerate  unit cells and bosonic species on the sublattices A and B, respectively,  and  below we  use $a_{1(2)}\!\equiv\!a(b)$  interchangeably.
Next is the Fourier transform
\begin{eqnarray}
a^{\phantom{\dag}}_{\ell,\alpha}=
\frac{1}{\sqrt{N}} \sum_{{\bf k}} e^{i{\bf k}({\bf r}_\ell+{\bm \rho}_\alpha)}
a^{\phantom{\dag}}_{\alpha {\bf k}}\, ,
\label{eq_fourier}
\end{eqnarray}
where $N$ is the number of  unit cells, ${\bf r}_\ell$ and  ${\bm \rho}_\alpha$ are defined after Eq.~(\ref{rotation}), and  ${\bf k}$ sums over  the full Brillouin zone of the honeycomb lattice, shown in Fig.~\ref{fig_lattice}(d).  

After the Fourier transform, the linear spin-wave part of the Hamiltonian (\ref{HJpm_xy2}) is given by
\begin{align}
\hat{{\cal H}}^{(2)}=\sum_{\bf k}
\Big\{ &\widetilde{A} \left( a^\dagger_{\bf k} a^{\phantom{\dag}}_{\bf k} + b^\dagger_{\bf k} b^{\phantom{\dag}}_{\bf k} \right) 
\label{H2special}\\
&+\left(\widetilde{B}_{\bf k}a^\dagger_{\bf k} b^{\phantom{\dag}}_{\bf k}
+\widetilde{B}_{\bf k}a^{\dag}_{\bf k} b^{\dag}_{-{\bf k}} 
+{\rm H.c.}\right)\Big\}, \nonumber
\end{align}
with $\widetilde{A}$ and $\widetilde{B}_{\bf k}$ given by
\begin{align}
&\widetilde{A}=S\big(-J+4|J_{\pm\pm}|+3J_3\big) \, ,\nonumber\\
\label{ABzz}
&\widetilde{B}_{\bf k}=3S\big(J \gamma_\mathbf{k} -2|J_{\pm\pm}|\gamma'_\mathbf{k}
+J_3 \gamma^{(3)}_\mathbf{k}\big)/2,
\end{align}
where   the hopping amplitudes are
\begin{align}
\label{gks}
\gamma_\mathbf{k}&=\frac{1}{3}\sum_{\alpha} e^{i\mathbf{k}{\bm \delta}_\alpha},\ 
\gamma^{(3)}_{\mathbf{k}}=\frac{1}{3}\sum_{\alpha} e^{i\mathbf{k}{\bm \delta}^{(3)}_\alpha},\\
\gamma'_\mathbf{k}&=\frac{1}{3}\sum_{\alpha} \cos\tilde{\varphi}_\alpha e^{i\mathbf{k}{\bm \delta}_\alpha}.
\nonumber
\end{align}
We note that in the high-field polarized phase that we study in Sec.~\ref{sec_xfield}, the spin Hamiltonian naturally assumes the two-sublattice form, so that the linear spin-wave part of it takes the form identical to (\ref{H2special}), but with  different $\widetilde{A}$ and $\widetilde{B}_{\bf k}$ that are  given in Sec.~\ref{sec_xfield}.

A general two-sublattice bosonic model  still leads to a problem of diagonalization of the $4\!\times\!4$ matrix. While this problem is often reducible to an analytically treatable diagonalization of $(\hat{\bf g}\hat{\bf H}_{\bf k})^2$ \cite{us_anisotropic}, where $\hat{\bf g}=[1,1,-1,-1]$ is a paraunitary diagonal matrix and $\hat{\bf H}_{\bf k}$ is the LSWT Hamiltonian matrix, the resultant formalism is rather cumbersome, especially for the non-linear extension of the spin-wave theory that we  pursue, see Ref.~\cite{Smit20}. 
 
An important distinction of the model    (\ref{H2special}) in our case is that it is diagonalizable by 
much simpler means and results in a much more manageable non-linear theory. Not only this LSWT Hamiltonian has fewer elements in its matrix, but, crucially, the ``normal'' ($a^\dag b$)  and ``anomalous'' ($a^\dag b^\dag$) matrix elements  in its second line are the same. This important feature allows us to split the diagonalization problem in an intuitively clear two-step process described next.  This form of the model (\ref{H2special}) can be traced all the way back to the easy-plane nature of the spin Hamiltonian in Eq.~(\ref{HJpm_xy}), and, while it may look artificial, it should in no way be restrictive of the physical results that we obtain from it as was discussed earlier.

The first step of the diagonalization of the  Hamiltonian \eqref{H2special} is a {\it unitary} transformation from the operators $a_{1(2){\bf k}}\!\equiv\!a_{\bf k}(b_{\bf k})$  to their symmetric and antisymmetric combinations, with the phase factor  $e^{i\varphi_{\bf k}}$  of $\widetilde{B}_{\bf k}$ absorbed symmetrically in the operators of both species \cite{ushoney16}
\begin{eqnarray}
a^{\phantom{\dag}}_{\alpha{\bf k}}=
\frac{e^{i(-1)^{\alpha+1}\varphi_{\bf k}/2}}{\sqrt{2}}\sum_{\mu} V^{\alpha\mu}  c^{\phantom{\dag}}_{\mu{\bf k}}\,  ,
\label{eq_phaseshift}
\end{eqnarray}
where the $2\times 2$ matrix ${\bf \hat{V}}$ is 
\begin{eqnarray}
{\bf \hat{V}}=\left( \begin{array}{rr} 1& \ \ 1 \\ -1 & 1 \end{array}\right).
\label{V}
\end{eqnarray}
After this transformation, the LSWT Hamiltonian is block-diagonal in the new bosonic index $\mu=1,2$
\begin{eqnarray}
\label{Hkc}
\hat{\cal H}^{(2)}\!=\!\sum_{{\bf k},\mu} \Big\{
A_{\mu{{\bf k}}} c^\dagger_{\mu{\bf k}}c^{\phantom{\dag}}_{\mu{\bf k}}
\!-\!\frac{B_{\mu{\bf k}}}{2}
\left(c^{\phantom{\dag}}_{\mu{\bf k}}c^{\phantom{\dag}}_{\mu-{\bf k}}\!+\!{\rm H.c.}\right)\!\Big\}\!,\ \ \
\end{eqnarray}
with
\begin{align}
A_{\mu{{\bf k}}}=\widetilde{A}+(-1)^\mu |\widetilde{B}_{\bf k}|,~B_{\mu{\bf k}}=(-1)^{\mu+1} |\widetilde{B}_{\bf k}|.
\end{align}
The second step is the textbook {\it paraunitary} Bogolyubov transformation for the individual bosonic species
\begin{eqnarray}
c_{\mu{\bf k}}=u_{\mu{\bf k}} d^{\phantom{\dag}}_{\mu{\bf k}}+v_{\mu{\bf k}} d^{\dagger}_{\mu-{\bf k}}\, ,
\label{eq_bogolyubov}
\end{eqnarray}
with the  parameters  defined by  
$2u_{\mu{\bf k}}v_{\mu{\bf k}}\!=\!B_{\mu{\bf k}}/\varepsilon_{\mu{\bf k}}$ and 
$u^2_{\mu{\bf k}}+v^2_{\mu{\bf k}}\!=\!A_{\mu{\bf k}}/\varepsilon_{\mu{\bf k}}$,
where the linear spin-wave energies are given by
\begin{align}
\label{omegas}
\varepsilon_{\mu{\bf k}}=\sqrt{A_{\mu\mathbf{k}}^2-B_{\mu\mathbf{k}}^2}\, .
\end{align}
Thus, the spin-wave spectrum  consists of two branches, $\varepsilon_{1{\bf k}}$ and $\varepsilon_{2{\bf k}}$, which will be loosely referred to as the acoustic and the optical modes. Note that the observed spectrum still consists of four modes due to the four-sublattice structure of the zigzag state. The additional two modes are obtained from $\varepsilon_{\mu{\bf k}}$ by the shift with the ordering vector $\mathbf{Q}$, see Sec.~\ref{app_JKG} and Appendix~\ref{app_formalism}.

\begin{figure}
\includegraphics[width=0.8\linewidth]{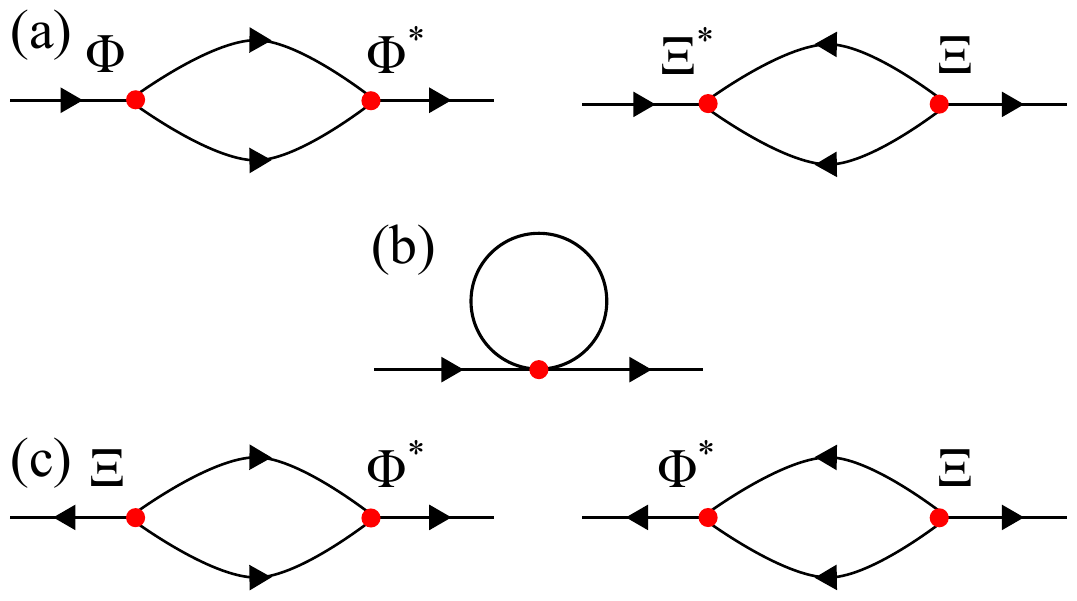}
\vskip -0.2cm
\caption{(a) Decay and source diagrams for the $1/S$ contribution to the self-energy (\ref{eq_sigma_decay}) 
and (\ref{eq_sigma_source}) from the three-magnon interactions. (b) The Hartree-Fock  diagram from the four-magnon interactions. (c) The off-diagonal diagrams contributing to the 
anomalous self-energy terms, see Sec.~\ref{sec_onshell}.}
\label{fig_diagrams}
\vskip -0.2cm
\end{figure}

\subsection{Non-linear spin-wave formalism}
\label{sec_nlswt_formalism}

To study  quantum effects  in the excitation spectra due to magnon interactions, the higher-order $1/S$ anharmonic terms in the  bosonic Hamiltonian are needed. 

The most qualitatively important effects are induced by the three-magnon terms,  which originate from  
mixing  the $S^x$ and $S^z$ spin components in the anisotropic $J_{\pm \pm}$ coupling  in (\ref{HJpm_xy2}) due to the broken SU(2) symmetry.  Skipping the technical steps of the Holstein-Primakoff, Fourier, unitary  \eqref{eq_phaseshift},  and  Bogolyubov  (\ref{eq_bogolyubov})  transformations in these terms, which are exposed in Appendix~\ref{app_formalism} in some detail,  one arrives to the general form of the  cubic Hamiltonian for the magnon normal modes 
\begin{eqnarray}
&&\hat{\cal H}^{(3)}=\frac{1}{3!\sqrt{N}}\sum_{\sum {\bf k}_i={\bf Q}}\sum_{\eta\nu\mu} \left(
\Xi^{\eta\nu\mu}_{{\bf q}{\bf k}{\bf p}} 
d^{\dagger}_{\eta{\bf q}} d^{\dagger}_{\nu{\bf k}} d^{\dagger}_{\mu{\bf p}}+{\rm H.c.}\right)\hskip 0.88cm \
\label{H3main}\\
&&\phantom{\hat{\cal H}^{(3)}}
+\frac{1}{2!\sqrt{N}}\sum_{\sum {\bf k}_i={\bf Q}}\sum_{\eta\nu\mu} \left(
\Phi^{\eta\nu\mu}_{{\bf q}{\bf k};{\bf p}} 
d^{\dagger}_{\eta{\bf q}} d^{\dagger}_{\nu{\bf k}} d^{\phantom{\dag}}_{\mu-{\bf p}}+{\rm H.c.}\right),
\nonumber
\end{eqnarray}
where the combinatorial factors are due to symmetrization in the source and decay vertices, $\sum {\bf k}_i\!=\!{\bf p}\!+\!{\bf k}\!+\!{\bf q}$,  ${\bf Q}\!=\!\left(0,2\pi/3a\right)$ is the ordering vector of the zigzag structure, and explicit expressions for the vertices $\Phi^{\eta\nu\mu}_{{\bf q}{\bf k};{\bf p}}$ and $\Xi^{\eta\nu\mu}_{{\bf q}{\bf k}{\bf p}}$ are also given in  Appendix~\ref{app_formalism}. Note that the cubic coupling in (\ref{H3main})  is umklapp-like: momentum in the decay process is conserved up to the ordering vector  ${\bf Q}$. This is reminiscent of the case of the square-lattice antiferromagnet in a field of Ref.~\cite{ZhCh99}. Similarly to the LSWT, the form of the cubic terms in (\ref{H3main}) remains the same for the field-polarized state considered in Sec.~\ref{sec_xfield}, but with the ordering vector $\mathbf{Q}=0$ and different expressions for the vertices, see Appendix~\ref{app_formalism}.

The two lowest-order diagrams from the three-magnon interactions that contribute to the spectrum, the ``decay'' and the ``source''  diagrams, are shown in Fig.~\ref{fig_diagrams}(a), with the self-energy
\begin{equation}
\Sigma_\mu^{(3)} \left({\bf k},\omega \right)=\Sigma_\mu^{d} \left( {\bf k},\omega \right)+\Sigma_\mu^{s} \left( {\bf k},\omega \right),
\label{eq_sigma3}
\end{equation}
where 
\begin{align}
\label{eq_sigma_decay}
&\Sigma_{\mu}^{d} \left({\bf k},\omega \right)=\frac{1}{2N} \sum_{\mathbf{q},\eta\nu} \frac{\big| {\Phi}^{\eta\nu\mu}_{{\bf q},{\bf k-q+Q};-{\bf k}}\big|^2}{\omega - \varepsilon_{\eta \bf q} - \varepsilon_{\nu {\bf k}- {\bf q}+{\bf Q}}+i0},\\
\label{eq_sigma_source}
&\Sigma_{\mu}^{s} \left({\bf k},\omega \right)=-\frac{1}{2N}  \sum_{\mathbf{q},\eta\nu} \frac{\big| {\Xi}^{\eta\nu\mu}_{{\bf q},{\bf -k-q+Q},{\bf k}}\big|^2}{\omega + \varepsilon_{\eta \bf q} + \varepsilon_{\nu -{\bf k}- {\bf q}+{\bf Q}}-i0}.
\end{align}

The same parts of the spin Hamiltonian (\ref{HJpm_xy2}) that give the linear spin-wave terms yield the four-boson interactions that contribute the Hartree-Fock corrections to the spectrum in the same $1/S$-order as the three-magnon interactions. Deferring technical details to Appendix~\ref{app_formalism}, the corresponding correction to the LSWT Hamiltonian $\hat{\cal H}^{(2)}$ in Eq.~(\ref{Hkc}) can be found in a standard manner and is  given by 
\begin{align}
\delta \hat{\mathcal{H}}^{(4)}\!=\!\sum_\mathbf{k,\mu} \Big\{\delta A^{(4)}_{\mu\bf{k}}c^\dagger_{\mu\bf{k}} c^{\phantom \dagger}_{\mu\bf{k}}\!-\!\frac12\Big( \delta B^{(4)}_{\mu\bf{k}} c^\dagger_{\mu\bf{k}} c^\dagger_\mathbf{\mu -k}\!+\!\text{H.c.}\Big)\Big\}.
\label{eq_ham_hf}
\end{align}
The Hartree-Fock $1/S$-correction to the spectrum, shown diagrammatically in Fig.~\ref{fig_diagrams}(b), is given by
\begin{align}
\delta\varepsilon^{(4)}_{\mu\bf{k}}=\frac{A_{\mu\bf{k}} \delta A^{(4)}_{\mu\bf{k}}-\text{Re}\big( \delta B^{(4)}_{\mu\bf{k}}\big)B_{\mu\bf{k}} }{\varepsilon_{\mu\bf{k}}},
\label{eq_hf_sigma}
\end{align}
with the explicit form of $\delta A^{(4)}_{\mu\bf{k}}$ and $\delta B^{(4)}_{\mu\bf{k}}$   given in  Appendix \ref{app_formalism}.

Thus, the  self-energy that contributes to the spectrum in the $1/S$-order is given by 
\begin{equation}
\Sigma_\mu ({\bf k},\omega)=\delta\varepsilon^{(4)}_{\mu\bf{k}}+\Sigma_\mu^{(3)} ({\bf k},\omega).
\label{eq_sigma}
\end{equation}

\begin{figure*}
\includegraphics[width=2.0\columnwidth]{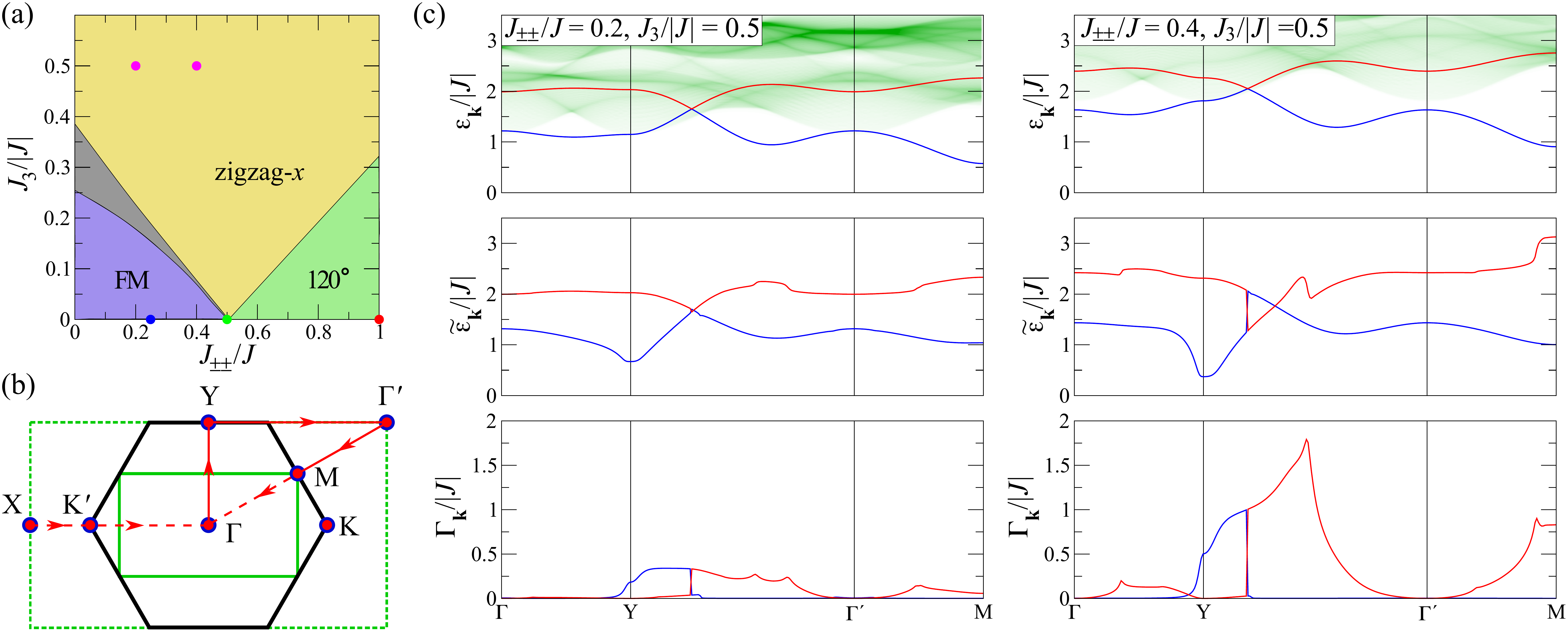}
\vskip -0.1cm
\caption{(a) Classical phase diagram of the model (\ref{HJpm_xy}) in Cartesian coordinates with parameter sets (\ref{translate}) marked by magenta dots, cf. Fig.~\ref{fig_lattice}(b). (b) BZ of the honeycomb lattice (hexagon) and the first magnetic BZ of the zigzag state (small rectangle). High-symmetry points and the representative ${\bf k}$-path are indicated. In (c), only the solid parts of this path are used due to the symmetry of the spectrum. (c) Linear spin-wave spectrum, $\varepsilon_{\mu\mathbf{k}}$, two-magnon continuum, renormalized  spectrum, $\widetilde{\varepsilon}_{\mu\mathbf{k}}$, and magnon decay rates, $\Gamma_{\mu\mathbf{k}}$, for the representative parameter sets of $J_{\pm\pm}$ and $J_3$ (\ref{translate}) and $S=1/2$ are shown.}
\label{fig_spectrum}
\vskip -0.2cm
\end{figure*}

\subsection{Quantum effects in the spectrum}

As is shown above,  calculation of the magnon spectra $\varepsilon_{\mu\mathbf{k}}$ and self-energies $\Sigma_\mu({\bf k},\omega)$ for the model \eqref{HJpm_xy} is analytically straightforward in the zigzag state. 

The results of the calculations of the spectrum within the linear and non-linear approximations are shown in Fig.~\ref{fig_spectrum} for the representative values of $J_{\pm\pm}$ and $J_3$ in that state. The selected parameter sets are shown by the two magenta points in the phase diagram in Fig.~\ref{fig_spectrum}(a). Their coordinates are $J_{\pm\pm}=0.2J$ and $0.4J$, $J_3=0.5|J|$,  all for $J<0$ (calculations for $J_3=|J|$ are shown in Appendix~\ref{app_formalism}). Rewriting them in the generalized Kitaev-Heisenberg language using Eq.~\eqref{eq_jkg_xy_transform} gives
\begin{align}
\label{translate}
J_{\pm\pm}&\!=\!0.2J\!\rightarrow\!\{J_0,K,\Gamma,\Gamma'\}\!=\!\{-1,0.5,0.75,0.25\},\\
J_{\pm\pm}&\!=\!0.4J\!\rightarrow\!\{J_0,K,\Gamma,\Gamma'\}\!=\!\{-1,0.86,0.93,0.07\},\nonumber
\end{align}
with the normalization of $J_0\!=\!-1$.  As was argued above, all  terms of the extended KH model that  correspond to our considered subspace of parameters are significant.

Fig.~\ref{fig_spectrum}(c) shows the linear spin-wave spectrum $\varepsilon_{\mu\mathbf{k}}$ from (\ref{omegas}) together with the renormalized on-shell spectrum $\widetilde{\varepsilon}_{\mu\mathbf{k}}$ and the magnon decay rate $\Gamma_{\mu\mathbf{k}}$ defined as 
\begin{equation}
\widetilde{\varepsilon}_{\mu\mathbf{k}}=\varepsilon_{\mu\mathbf{k}}+\text{Re}~\Sigma_\mu({\bf k},\varepsilon_{\mu\mathbf{k}}),  \ \ \Gamma_{\mu\mathbf{k}}=-\text{Im}\Sigma_\mu({\bf k},\varepsilon_{\mu\mathbf{k}}),
\label{eq_spectrum_onshell}
\end{equation}
with $\Sigma_\mu({\bf k},\omega)$  from (\ref{eq_sigma}), all along the contour in Fig.~\ref{fig_spectrum}(b). Note that the dashed part of the contour is omitted in (c) since the $\text{X}\rightarrow\Gamma$ part of the path is redundant to  $\text{Y}\rightarrow\Gamma'$  and the $\Gamma' \rightarrow \text{M}$  is equivalent to  $\Gamma\rightarrow \text{M}$. Again, because of the two sublattices, there are   two inequivalent magnon modes in these results that are defined in the full BZ of the honeycomb lattice. 

There are several features of the linear spin-wave spectrum that should be pointed out. Generally, the spectrum is gapped due to the presence of the symmetry-breaking bond-dependent anisotropic $J_{\pm\pm}$ term, with  the gap at the ordering vector ${\bf Q}=\text{Y}$ that increases with both $J_{\pm\pm}$ and $J_3$. However, this gap vanishes upon approaching the $\{J_{\pm\pm},J_3\}=\{0.5J,0\}$ point, which corresponds to the aforementioned  honeycomb 120${\degree}$ compass model \eqref{eq_120compass}  marked as a green dot in Figs.~\ref{fig_lattice}(a) and (b) and in Fig.~\ref{fig_spectrum}(a), see Sec.~\ref{app_JKG} for a discussion of the spectrum at this tricritical point.

Even though the spectrum is gapped at the ordering vector, there is another ``pseudo-Goldstone'' mode with a rather small gap located at the M point, which is {\it not} the ordering vector. These low-lying modes are present in the spectrum due to proximity of our model to a simpler $J_1$--$J_3$ model. This model possesses a true accidental degeneracy  because  the third-neighbor interaction $J_3$ splits the honeycomb lattice into four sublattices, while nearest-neighbor  $J_1$ can constraint only a linear combination of spins from these four sublattices, see \cite{Lhuillier2001}. The degeneracies of that nature are common in the spectra of frustrated models such as that on the triangular lattice \cite{chubukov_j1j2,us_mimicry,us_anisotropic}. As we show below, these quasi-Goldstone modes are crucial for the non-linear quantum corrections, providing the low-lying two-magnon continuum for the single-magnon modes to interact with.

The middle panels in Fig.~\ref{fig_spectrum}(c) shows $\widetilde{\varepsilon}_{\mu\mathbf{k}}$, the magnon spectrum (\ref{eq_spectrum_onshell}) with the on-shell one-loop quantum corrections. One can see  a substantial downward  renormalization of the spectrum compared to the LSWT, which is most prominently pronounced near the Y point. This is due to a direct coupling to the two-magnon continuum provided by the three-magnon terms in (\ref{H3main}). The two-magnon density of states is shown as an intensity map in the top panels of Fig.~\ref{fig_spectrum}(c). The renormalization becomes more significant with the increase of the anisotropic $J_{\pm\pm}$ term. This is obvious from the fact that the three-magnon interaction originates only from the $S^x S^z$ terms in the bond-dependent part of the Hamiltonian \eqref{HJpm_xy2}, which, in turn, exists only because of the SOC-induced anisotropic interaction $J_{\pm\pm}$. There is also an enhancement of the renormalization due to van Hove singularities in the two-magnon continuum that manifest themselves in the self-energy, which can lead to an instability in the spectrum,  $\widetilde{\varepsilon}_{\mu\mathbf{k}}<0$ at large $|J_{\pm\pm}|$. The magnitude of the third-neighbor interaction $J_3$ also affects   spectrum renormalization, but  indirectly through the larger LSWT gaps that modify the density of states in the two-magnon continuum, see Appendix~\ref{app_formalism}.

We should point out  a peculiar feature of the renormalized spectrum $\widetilde{\varepsilon}_{\mu\mathbf{k}}$.  The self-energies of the two modes are not equal at the band-crossing point $\mathbf{k}^*$ given by the LSWT, see  Fig.~\ref{fig_spectrum}(c). This effect can also be observed in the imaginary part of the self-energies in the bottom panels of  Fig.~\ref{fig_spectrum}(c). One can also see that the band crossing point of the renormalized spectra  shifts to a different ${\bf k}$-point due to that difference in renormalization. 
However,  self-energies appear to be smooth functions of the momentum at the crossings
\begin{align}
\Sigma_1(\mathbf{k}^*-\mathbf{q})\rightarrow \Sigma_2(\mathbf{k}^*+\mathbf{q}),~ \mathbf{q}\rightarrow 0\, .
\end{align}
This is an interesting feature that requires further investigation, and it is related to the phase factor $\varphi_\mathbf{k}$ in the wave function \eqref{eq_phaseshift} and the fact that the LSWT band-crossing point $\mathbf{k}^*$ for a generic set of parameters  of the model (\ref{HJpm_xy}) is not determined by the lattice symmetries and is not a high-symmetry point.

The three-magnon interaction also yields the finite lifetime of magnons, given by the imaginary part of the self-energy (\ref{eq_spectrum_onshell}), $\Gamma_{\mu\mathbf{k}}$. As is shown in Fig.~\ref{fig_spectrum}(c), the low-energy parts of the lower magnon branch are typically stable in most of the Brillouin zone because of the lack of the phase space for decays for them. An important aspect of the three-magnon coupling in (\ref{H3main}) that stems from the structure of the anisotropic coupling in (\ref{HJpm_xy2}), is that the momentum in the decay process is conserved up to the ordering vector of the zigzag structure ${\bf Q}$. This means that the single-magnon branch couples to the two-magnon continuum that is offset by that momentum. Because of that, the magnons from the lower branch in the vicinity of the Y point are significantly affected by the renormalization  and  decay processes into the pairs of the low-energy pseudo-Goldstone  magnons  modes that are in the proximity of  the M points. The high-energy modes typically acquire finite lifetime in a larger portion of the Brillouin zone with a  significant  decay rate for large $|J_{\pm\pm}|$, yielding strongly damped excitations. Qualitatively  similar features have been found by the inelastic neutron scattering  in $\alpha$-RuCl$_3$ \cite{banerjee2018,us_rucl3,us_PRR}, where stable low-lying spin-wave modes have been observed in coexistence with the higher-energy continuum.

\subsection{Quantum effects in the phase diagram, DMRG}

\begin{figure}[b]
\includegraphics[width=0.99\columnwidth]{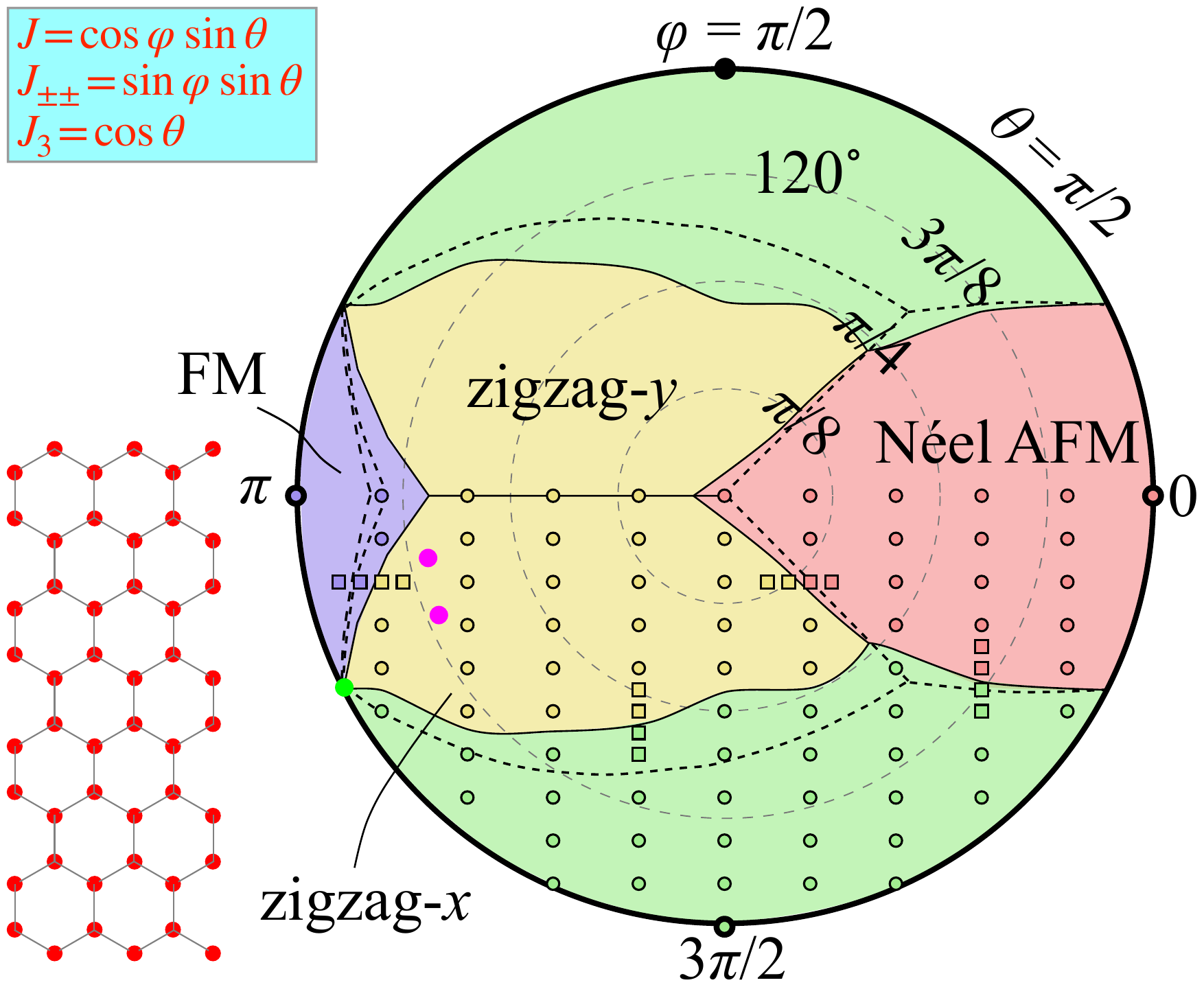}
\caption{DMRG phase diagram for $S=1/2$ (colors and solid lines) and classical Luttinger-Tisza phase diagram (dashed lines) from Fig.~\ref{fig_lattice}(b)  of the model  (\ref{HJpm_xy}). DMRG calculations were performed on a 48-site cluster for the parameters indicated with open symbols and the solid boundaries are obtained via interpolating between DMRG points, see the text. The inset shows the cluster used in DMRG.}
\label{fig_dmrg}
\end{figure}

Here we complement our discussion of the quantum effects in   excitation spectra by an analysis of the concomitant effects in the ground state phase diagram of the model (\ref{HJpm_xy}).
Specifically, we  investigate a possibility of the shifts of the phase boundaries for the considered case of the $S=1/2$ model relative to the classical phase diagram in  Fig.~\ref{fig_lattice}(b) obtained by LT calculation. For that we use density matrix renormalization group \cite{white_density_1992}. We perform calculation on the 
 $6\times8$-site cluster with open boundaries, see Fig.~\ref{fig_dmrg}, using ITensor package \cite{itensor} with 10 sweeps, and keeping up to $m=200$ states. Calculations were performed for the parameter sets shown in Fig.~\ref{fig_dmrg} by the open circles. The color coding for the states is the same as in Fig.~\ref{fig_lattice}. Note that since the model (\ref{HJpm_xy}) is symmetric under $J_{\pm\pm}\rightarrow-J_{\pm\pm}$ (with a simultaneous $\pi/2$ spin rotation), we only perform calculations for the lower half of the phase diagram, $J_{\pm\pm}<0$.
For the relatively small cluster size and for the robust ordered states, the number of  sweeps and that of the kept states appear to be sufficient. 
We identify transitions between different phases by finding the maximal value of the static spin structure factor and identifying its ordering vector. The transitions between  the phases are also verified via anomalies in the ordered on-site magnetic moment. The resultant lines of the phase boundaries are obtained by interpolations between the DMRG points and serve as guides to the eye. In order to check the validity of such interpolations, we have also performed additional calculations keeping up to $m=400$ states, which are shown n Fig.~\ref{fig_dmrg} by the open squares.

The resultant DMRG phase diagram for the $S=1/2$ case of the model \eqref{HJpm_xy} is presented in Fig.~\ref{fig_dmrg}, where the shift of the boundaries between   zigzag and $120{\degree}$ phases compared to the classical LT results is clearly demonstrated. This trend is broadly in agreement with the findings of our $1/S$ analysis of the excitation spectra provided above, which predicts strong spectrum renormalization, potentially leading to  spectrum instability of the zigzag state in the same region of the parameter space. 
The expansion of the fluctuation-prone $120{\degree}$ state at the expense of the proximate collinear state in the anisotropic-exchange models in the quantum limit is also in agreement with the similar results in the triangular-lattice case \cite{us_anisotropic}.

For the other phase boundaries in Fig.~\ref{fig_dmrg}, the classical model  works very well even in the $S=1/2$ case. One more significant change that should be pointed out is the lack of the narrow region of the incommensurate state at the FM-zigzag border, showing instead a direct transition between the two states. We note that the small cluster size used in our calculations may be insufficient to capture narrow regions of incommensurate states, but they are outside of the scope of our work.

\subsection{Special case of the $120{\degree}$ compass model}
\label{app_JKG}

Here we briefly consider    excitation spectrum and effects of magnon decay in them at a  special point of the phase diagram of the model \eqref{HJpm_xy}, the $120{\degree}$ compass model \eqref{eq_120compass}, which corresponds to $J_{\pm\pm}\!=\!J/2\!<\!0$ and $J_3\!=\!0$ in the model \eqref{HJpm_xy}, or $K\!=\!\Gamma\!=\!-J_0\!>\!0$ and $\Gamma'\!=\!0$ in the extended Kitaev-Heisenberg language \eqref{eq_H_JKGGp}. This tricritical point is marked by the green dot in the phase diagrams in Figs.~\ref{fig_lattice}(a) and (b), Fig.~\ref{fig_spectrum}(a), and Fig.~\ref{fig_dmrg}, and it was inspirational for the easy-plane model considered in this work as is discussed in Sec.~\ref{sec_model}. 

\begin{figure}
\includegraphics[width=0.99\linewidth]{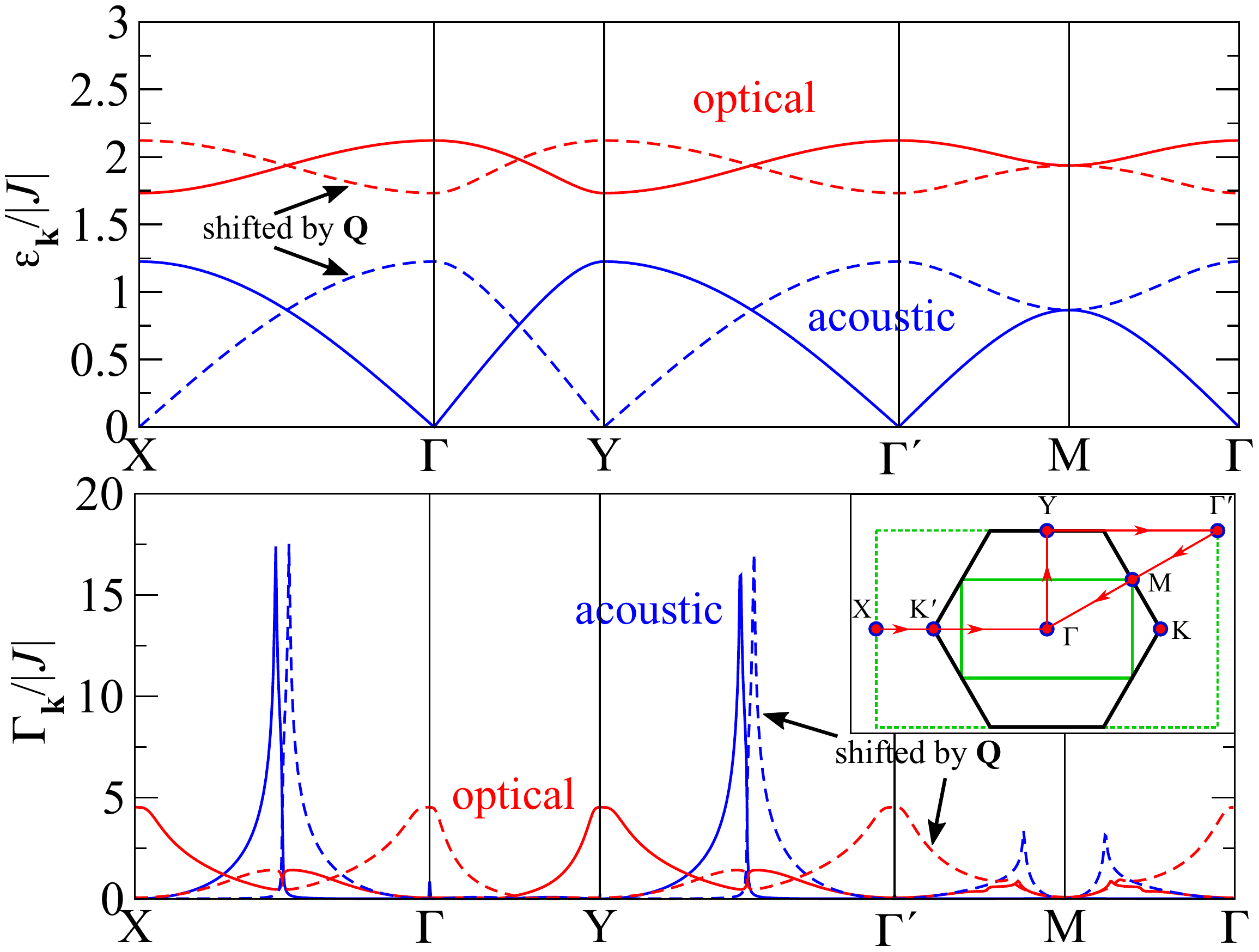}
\caption{Upper panel: LSWT magnon bands of the model \eqref{eq_120compass}, which is equivalent to  $J_{\pm\pm}=J/2<0$ and $J_3=0$ in \eqref{HJpm_xy}, or $K=\Gamma=-J_0>0$ and $\Gamma'=0$ in  \eqref{eq_H_JKGGp};
``shifted,'' acoustic, and optical modes are identified. 
Lower panel: on-shell decay rates $\Gamma_{\mu\mathbf{k}}$ (\ref{eq_spectrum_onshell}) along the full ${\bf k}$-cut of Fig.~\ref{fig_spectrum}(a). }
\label{fig_jkg_energy_gamma}
\end{figure}

This special point was discussed earlier in Ref.~\cite{Smit20}, but without presenting explicit spin-wave calculations for it.  That work  has also explored a different extension of the model around this point  and has used a more general but more cumbersome analytical procedure  that relied on the four-sublattice diagonalization \cite{Smit20}. Here,  we are using the two-sublattice approach discussed above and present the results for the on-shell magnon decay rates $\Gamma_{\mu\mathbf{k}}$ (\ref{eq_spectrum_onshell}) in Fig.~\ref{fig_jkg_energy_gamma}.

A distinctive feature of this tricritical point is that the lower  magnon mode in its zigzag spectrum is gapless, see the upper panel of Fig.~\ref{fig_jkg_energy_gamma}. The two of the four magnon branches are obtained by the shift of the two original modes of the two-sublattice approach by $\mathbf{Q}=(0,2\pi/3a)$, shown by solid and dashed lines, respectively. The lower panel of Fig.~\ref{fig_jkg_energy_gamma} shows the on-shell decay rates (\ref{eq_spectrum_onshell}) in units of $|J|$ using the same color and  line conventions. Because of the large phase space for magnon decays due to the presence of the Goldstone modes and because of the significant anisotropic term $J_{\pm\pm}$,  magnon decay rates are large in comparison with the results in Fig.~\ref{fig_spectrum}, reaching values of the magnon bandwidth if averaged over the BZ.

Somewhat unexpected is a strong divergence in the acoustic mode decay rate in the vicinity 
of the ``Dirac-like'' points of the spectra where the non-shifted and  ``shifted'' modes intersect.
This divergence can be demonstrated to be of a stronger character than that of the two-magnon density of states (not shown). Using an analytical insight into the asymptotics of the vertex, this behavior
can be identified with an emission of a Goldstone-like acoustic mode with ${\bf q}\!\rightarrow\!0$ 
and a decay into a lower Dirac branch with an umklapp of the momentum by ${\bf Q}$ due to the structure of the three-magnon coupling in (\ref{H3main}), leading to an enhancement factor $\propto \! 1/|{\bf q}|$ in the probability. 
Since this is not a decay of a Goldstone mode, such a divergent amplitude is not forbidden, and may 
have been encountered before in a different model, see Ref.~\cite{ushoney16}.

\subsection{Summary}
\label{app_bigger_picture}

Altogether, the two-sublattice approach to the spin-wave theory of the easy-plane anisotropic-exchange model \eqref{HJpm_xy} allows us to consider $1/S$ quantum effects in magnon spectrum in a rather uncomplicated manner. The three-magnon interaction leads to the renormalization of the spin-wave dispersions, decays,  and redistribution of the spectral weight. Remarkably, the strength of this interaction in our model depends only on one bond-dependent $J_{\pm\pm}$ term, which highlights the benefit of using the ``ice-like'' parametrization of the model \eqref{HJpm_xy}. A strong downward renormalization of the spectrum points to a shift of the phase boundary between the $120{\degree}$ and zigzag phases, supported by the  DMRG data.

We argue that the results presented in this section are  generic as  the studied model \eqref{HJpm_xy} transforms into the extended Kitaev-Heisenberg model with all key anisotropic  terms present and significant \eqref{eq_jkg_xy_transform}, (\ref{translate}). Similar features of a coexistence of the well-defined low-energy modes with the broadened higher-energy  continuum have also been found in the generalized KH model for very different parameter sets \cite{us_rucl3,us_PRR,Smit20} and observed in the Kitaev materials with the zigzag ground states \cite{Songvilay_Co_2020,Park_Co_2020,banerjee2018}.

\section{Polarized phase}
\label{sec_xfield}

It is  well established by now that the  magnon spectrum of the {\it collinear} ordered states in the models with anisotropic-exchange terms, which are present due to spin-orbit coupling,  may  be strongly affected by  the decay and renormalization processes \cite{Rau_Yb_NLSWT_2019,Sid_Co_2020,Smit20,us_rucl3,us_kagome}. While in the field-induced polarized phases such nonlinear effects  are often neglected under a general assumption that the high magnetic field suppresses quantum fluctuations, their significant ramifications have been recently discussed for several systems with spin-orbit coupling 
\cite{robinson_quasiparticle_2014,McClarty2018,Rau_Yb_NLSWT_2019,Sid_Co_2020,%
Vojta17_hkg_field,Vojta2020_NLSWT}. 

In this section we  study  quantum corrections to the spectrum in the high-field paramagnetic state of the easy-plane anisotropic-exchange model \eqref{HJpm_xy}. We show that the magnon interaction effects can be significant in the presence of the bond-dependent terms, especially in the vicinity of the critical field of the transition from the paramagnetic to the long-range ordered phase. Specifically, within the $1/S$ approach, such quantum corrections have been recently shown to produce unphysical divergences upon approaching this transition  \cite{Vojta2020_NLSWT}, indicating a downward renormalization of  the critical field. Here, we offer an approach to  regularize such singularities in this class of models.

\subsection{High-field spin-polarized state}
\label{sec_spin_polarized}
 
We consider the model (\ref{HJpm_xy})  in the in-plane field that induces a spin-polarized state with spins fully aligned. The classical energy of this state,
\begin{align}
\frac{E_{\rm cl}}{N}=3S^2(J+J_3)-2HS,
\label{Ecl_polarized}
\end{align}
is invariant to the field direction because  contributions of the bond-dependent terms to it from different bonds cancel out. Here $N$ is the number of atomic unit cells as before and the units of $g\mu_B$ are absorbed into the definition of the field $H$, with $g$ being the Land\'{e} $g$-factor. 

There are two principal in-plane field directions for the honeycomb lattice, along and perpendicular to the nearest-neighbor bond, see Fig.~\ref{fig_lattice}(c). Although the choice of the field direction lowers the symmetry of the model  (\ref{HJpm_xy}), the model remains invariant to the simultaneous change of the sign in the $J_{\pm \pm}$-term and switching the field direction from along to perpendicular to the bond, similarly to the zigzag-$x$ and -$y$ states, see Sec.~\ref{Sec_lswt}. 

Thus, without  loss of generality,  we consider  the in-plane magnetic field directed along the $x$-axis in Fig.~\ref{fig_lattice}(c), perpendicular to the AB bond. The critical field $H_c$ of the transition from the spin-polarized to zigzag state for $H\parallel x$, as obtained from vanishing of the magnon gap at the M point \cite{us_PRR}, is given by \footnote{Note that for smaller values of $J_3$ Eq.~\eqref{eq_hc_neg} gives negative values of $H_c$, indicating that the transition in this case is first order.}
\begin{equation}
H_c=\left\{    \begin{array}{ll}
        2S(J+3J_3-J_{\pm\pm}), & J_{\pm\pm}\!<\!0,\\
        2S(J+3J_3+2J_{\pm\pm}), & J_{\pm\pm}\!>\!0.
    \end{array}
\right.
\label{eq_hc_neg}
\end{equation}

The spin-wave formalism in the polarized phase is simplified, as the latter naturally offers the two-sublattice description of spins in the honeycomb lattice, analogous to the zero-field zigzag state in the previous section, with the spin-axes transformation from the global to local reference frame given by a  simple cyclic permutation 
\begin{align}
(S_i^x,S_i^y,S_i^z)_\text{lab}=(S_i^z,S_i^x,S_i^y)_\text{loc},
\label{lab_to_loc_FM}
\end{align}
for both A and B sublattices.

The spin  Hamiltonian \eqref{HJpm_xy}, rotated to the local axes of the polarized state  (\ref{lab_to_loc_FM}), is given by
\begin{align}
\hat{{\cal H}}_\text{loc}=&\sum_{\langle ij\rangle_1}\Big\{
\big(J+ 2 J_{\pm \pm}c_\alpha\big)S^{x}_i S^{x}_j+\big(J-2 J_{\pm \pm}c_\alpha\big)S^{z}_i S^{z}_j\nonumber\\
&\quad \quad \quad \quad \quad \quad\quad \quad +2  J_{\pm \pm}\Big( S^x_i S^z_j+S^z_i S^x_j\Big) s_\alpha \Big\}\nonumber\\
  +&J_3\sum_{\langle ij\rangle_3} \Big(S^{x}_i S^{x}_j+S^{z}_i S^{z}_j\Big)-H\sum_i S^z_i\, .
\label{HJpm_local2}
\end{align}
This local-axes form of the Hamiltonian  is the last stage of a transformation before the spin-wave expansion, cf.  Eq.~\eqref{HJpm_xy2} for the zigzag state. The technical aspects of the  $1/S$ spin-wave expansion for the spin-polarized state  replicate identically the steps that are  described in Sec.~\ref{sec_lswt_formalism} and Sec.~\ref{sec_nlswt_formalism} with the choice of $\mathbf{Q}=0$ and 
\begin{align}
&\widetilde{A}=H-3S\big(J+J_3\big),\nonumber\\
&\widetilde{B}_{\mathbf{k}}=3S\big(J \gamma_\mathbf{k}+2J_{\pm\pm}\gamma'_\mathbf{k}+J_3 \gamma^{(3)}_{\mathbf{k}}\big)/2.
\end{align}
Thus, the expressions for the LSWT spectrum in Eq.~ (\ref{omegas}) and  self-energies in Eq.~(\ref{eq_sigma}) retain the same form in the present case with the parameters given above. Below, we proceed  with the results for the dynamical structure factor in the spin-polarized state.

\subsection{Dynamical structure factor}
\label{sec_polarized_offshell}

\begin{figure*}
\centering
\includegraphics[width=1.9\columnwidth]{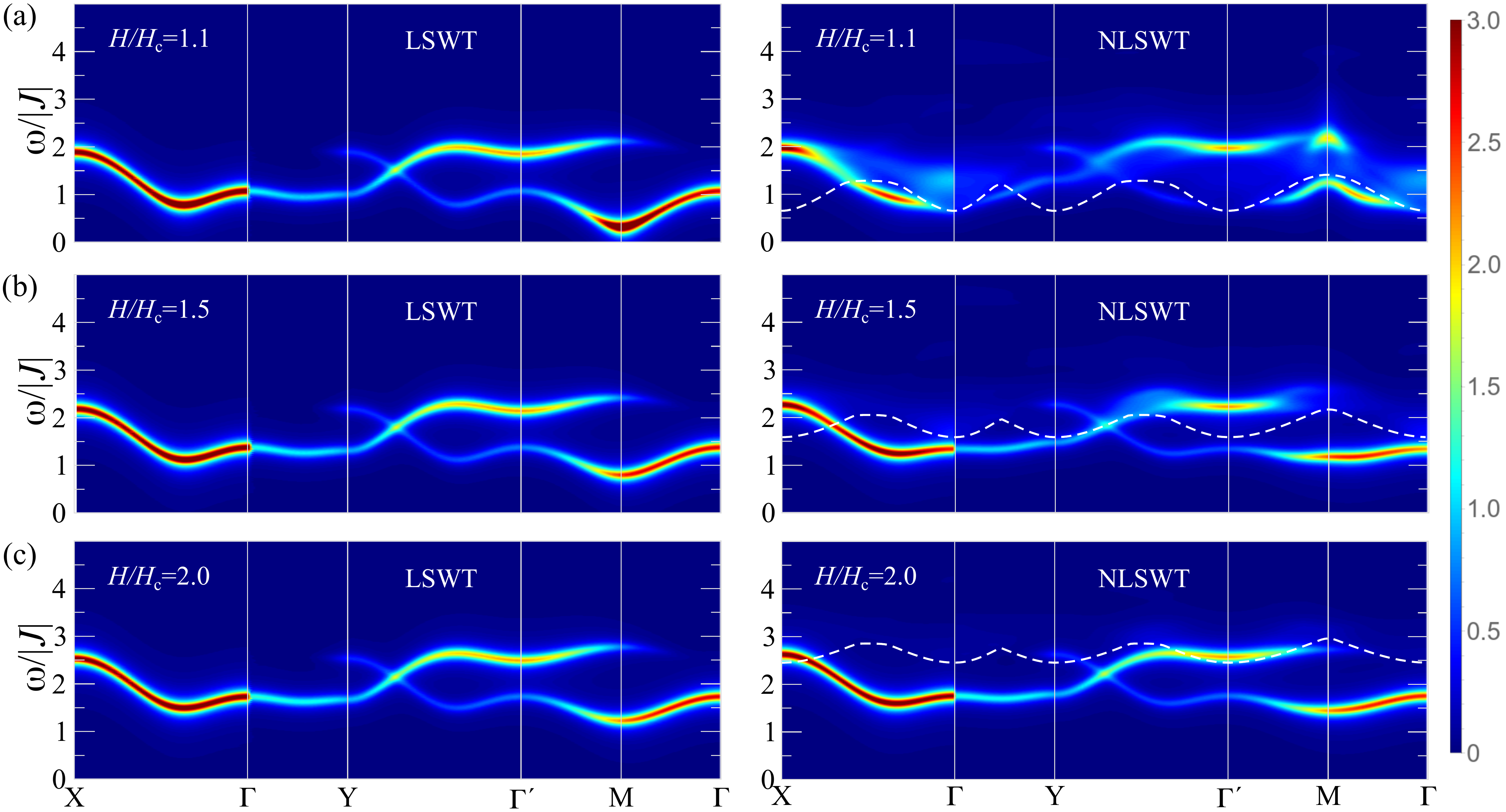}
\caption{Intensity plots of the dynamical structure factor $\cal{S}(\mathbf{k},\omega)$ (\ref{eq_Sqw}) in the polarized phase along the representative in-plane ${\bf k}$-path, see Fig.~\ref{fig_spectrum}(b),   for $H \!\parallel \!x$, $J\!<\!0$, $J_{\pm\pm}\!=\!0.2J$, $J_3\!=\!0.5|J|$, $S\!=\!1/2$, and three different $H/H_c$. Left and right panels are the LSWT and NLSWT results, see text. The  dashed lines in the right panels show the bottom of the two-magnon continuum.  Units of $J^{-1}$  and artificial broadening $\delta=0.1|J|$ are used.}
\label{fig_offshell_field}
\end{figure*}
 
An  important difference of the magnon spectrum in the polarized phase compared with the zigzag one is that the number of magnon species is equal to the number of magnetic ions in the unit cell  and, thus, there are only two branches of excitations that should be observable.
 
Here we present results for the magnon spectrum in the form of the dynamical spin structure factor 
\begin{align}
\mathcal{S}({\bf k},\omega)=\sum_{\alpha\beta} \Big( \delta_{\alpha\beta} -\frac{k_\alpha k_\beta}{k^2}\Big) \mathcal{S}^{\alpha \beta} ({\bf k},\omega),
\label{eq_Sqw}
\end{align}
where the spin-spin dynamical correlation function is
\begin{equation}
\mathcal{S}^{\alpha \beta} ({\bf k},\omega) =\frac{1}{\pi} \text{Im} \int_{-\infty}^{\infty} \, dt \, e^{i\omega t}
\, i\big\langle {\cal T} S^{\alpha}_{{\bf k}}(t) S^{\beta}_{-{\bf k}}(0) \big\rangle.
\label{eq_Sqw0}
\end{equation}
For the ordered magnetic states, the structure factor is naturally split into the transverse and longitudinal components, related to the directional and amplitude modulations of the magnetic order parameter, that have different ranking in $1/S$ sense,  see Ref.~\cite{triSqw}. In the leading  $1/S$-order, only  transverse  structure factor component is present and the dynamical correlation function is straightforwardly related to the non-interacting LSWT single-magnon spectral functions, ${\cal A}^{(0)}_{\mu}({\bf k},\omega)\!\equiv\!\delta(\omega-\varepsilon_{\mu{\bf k}})$,
\begin{equation}
\mathcal{S}^{\alpha \alpha} ({\bf k},\omega) =\sum_\mu {\cal F}_{\mu{\bf k}}^{\alpha\alpha} {\cal A}^{(0)}_{\mu{\bf k}}(\omega),
\label{eq_Skw_Akw0}
\end{equation}
dressed by the ``kinematic formfactors'' ${\cal F}_{\mu{\bf k}}^{\alpha\alpha}$ that are responsible for the intensity-modulation of the $\delta$-functional peaks of the spectral functions throughout the Brillouin zone. These formfactors depend on the order and  lattice structure, with their explicit forms for the considered case given in  Appendix~\ref{app_dsf}.

The modification of this picture within the nonlinear spin-wave theory (NLSWT) is twofold. 
First is a straightforward extension for the transverse  structure factor component that consists of taking into account  one-loop $1/S$ self-energy $\Sigma_\mu({\bf k},\omega)$ \eqref{eq_sigma}, see Sec.~\ref{sec_nlswt_formalism}, in the diagonal magnon Green's function
\begin{equation}
G_\mu ({\bf k},\omega)=-i\langle d^{\phantom \dagger}_{\mu\mathbf{k}} d^\dagger_{\mu\mathbf{k}} \rangle=\frac{1}{\omega-\varepsilon_{\mu\mathbf{k}}-\Sigma_\mu ({\bf k},\omega)},
\label{eq_Gmagnon}
\end{equation}
replacing bare spectral function ${\cal A}^{(0)}_{\mu}({\bf k},\omega)$ in (\ref{eq_Skw_Akw0}) with ${\cal A}_{\mu}({\bf k},\omega)\!\equiv\!-\frac{1}{\pi}\mbox{Im}G_\mu ({\bf k},\omega)$.

Second, we also include  the leading $1/S$ term from the longitudinal component of  $\mathcal{S}({\bf k},\omega)$,  neglected within the LSWT, which accounts for the direct contributions of the two-magnon continuum, see Appendix~\ref{app_dsf}.

Both NLSWT extensions are expected to provide insights into the effects of the spectral weight redistribution, spectrum renormalization, and  magnon spectral line broadening due to decays. 
We note that the presented NLSWT results do not constitute a strict $1/S$ correction to the LSWT. This is because some of the $1/S$ corrections, such as the off-diagonal contributions to $\mathcal{S}({\bf k},\omega)$ \cite{triSqw},  are still neglected  and the off-shell $\omega$-dependence in the self-energy  (\ref{eq_Gmagnon}) is retained. The latter allows to observe more complicated spectral features  and corresponds to the off-shell spectrum renormalization.

Figure~\ref{fig_offshell_field} demonstrates our results for $\cal{S}(\mathbf{k},\omega)$ along the high-symmetry in-plane ${\bf k}$-path in the Brillouin zone, see Fig.~\ref{fig_spectrum}(b), for various $H/H_c$. We chose a moderate value of anisotropic interaction $J_{\pm\pm}\!=\!0.2J$ and $J_3\!=\!0.5|J|$, see also Fig.~\ref{fig_spectrum} and Sec.~\ref{sec_zz}~D, and selected three values of the magnetic field:  near the transition, $H\!=\!1.1 H_c$, intermediate, $H\!=\!1.5 H_c$, and  deep in the polarized phase, $H\!=\!2.0 H_c$, Fig.~\ref{fig_offshell_field} (a), (b), and (c), respectively. The formfactor ${\cal F}_{\mu{\bf k}}^{\alpha\alpha}$ can be seen as strongly suppressing spectral intensities for some regions of BZ, and the jump in intensity at the $\Gamma$ point is due to its angular dependence. Calculations for other parameters for $H=1.1H_c$ are presented in Appendix~\ref{app_dsf}.

One can see in Fig.~\ref{fig_offshell_field}(c) that the effect of magnon interactions is negligible in strong field and LSWT provides a close description of the entire spectrum in this case. However, as  the field is lowered toward $H_c$, the effects of interaction become more pronounced. While deviations from the LSWT in the form of  energy renormalization, line broadening, intensity redistribution, and appearance of the more complicated spectral features such as spectral line splitting  can already be noticed for the fields as high as $1.5 H_c$, see Fig.~\ref{fig_offshell_field}(b), they become unmistakable in  Fig.~\ref{fig_offshell_field}(a) for the field near the critical field, $1.1H_c$. For this field and for that modest value of the anisotropic $J_{\pm\pm}$-term, both  higher- and lower-energy modes acquire finite lifetime via magnon decay processes in extended regions of the ${\bf k}$-space where they overlap  with the two-magnon continuum, the bottom of which is shown by the dashed lines in the right panels of Fig.~\ref{fig_offshell_field}. Same regions also demonstrate pronounced spectral weight redistribution  to the continuum and a strong renormalization of the spectrum. 

Two regions of the reciprocal space in Fig.~\ref{fig_offshell_field}(a) are of interest. First is the proximity of the $\Gamma$ and $\Gamma'$ points, where decays and downward renormalization are a result of the strong interaction with the two-magnon continuum, which originates from the low-lying nearly-Goldstone modes at the M point. For $J_{\pm\pm}<0$ and  field directed   perpendicular to the bond, the M point and its equivalent M$^\prime$ point become truly gapless at $H_c$. Below $H_c$, one of the zigzag domains with the ordering vector M or M$^\prime$ is  selected. The observed softening of the  magnon mode and the dominance of the two-magnon continuum at $\Gamma$ point are beyond the LSWT, but should be a generic feature of the transition from an ordered to nominally polarized phase in the presence of magnon interactions that are induced by anisotropic terms in this class of models. We discuss this effect  and its relevance to real materials in more detail in Sec.~\ref{sec_rucl3}.

The second region of interest concerns the proximity of the M point and the behavior of the lower mode near it. This mode experiences a significant  upward renormalization, which can be shown to diverge at $H\!\rightarrow\!H_c$. This divergence is unrelated to interaction with the two-magnon continuum and is unphysical. The same phenomenon has been discussed in Ref.~\cite{Vojta2020_NLSWT}  in the analysis of the $1/S$-corrections to the spectrum in a simpler $K$--$J$ model, and was interpreted as a sign of the downward renormalization of $H_c$, leaving the problem of the divergence unresolved. We discuss this problem and our approach to the  regularization of such singularities in this class of models next. 

\subsection{Divergence regularization}
\label{sec_onshell}

\begin{figure*}
\centering
\includegraphics[width=0.99\linewidth]{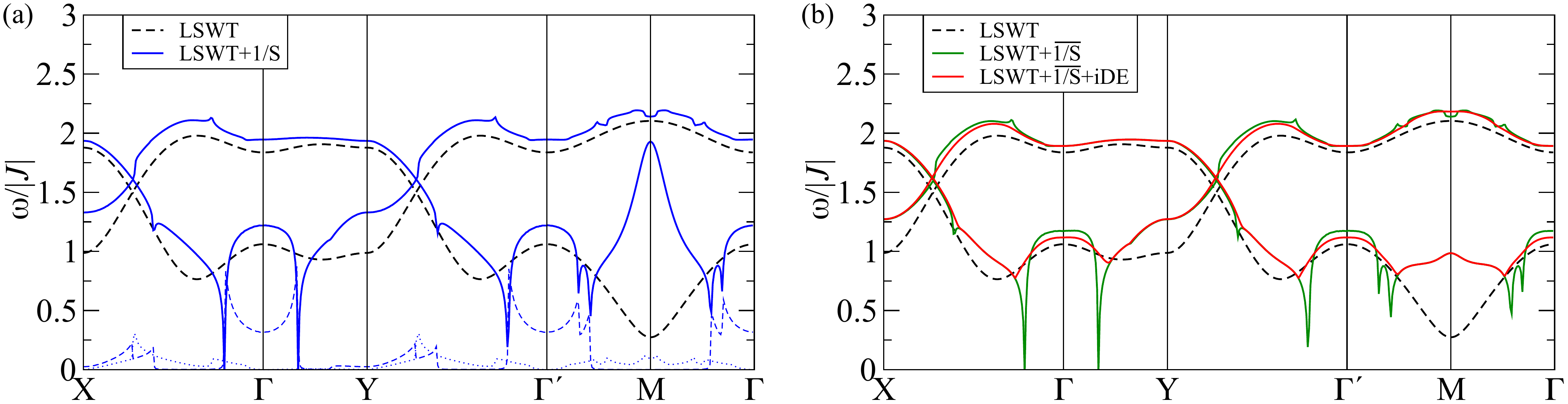}
\caption{Magnon spectrum in the polarized phase for $J_{\pm\pm}\!=\!0.2J$, $J_3\!=\!0.5|J|$, and $S\!=\!1/2$, for $H\!=\!0.75|J|$ ($H_c\!=\!0.7|J|$). (a)  LSWT results (\ref{omegas}) (dashed lines)  and  $1/S$-renormalized on-shell results  \eqref{eq_spectrum_onshell} (blue lines). Magnon decay rates for the lower and upper modes are shown with the blue dashed and dotted lines, respectively. (b) LSWT results (dashed line) and regularized spectrum as given by Eq.~\eqref{eq_hf_reg2} (green lines) and Eq.~\eqref{eq_reg3} (red lines), see the text.}
\label{fig_hf_divergence}
\end{figure*}

\begin{figure}[b]
\centering
\includegraphics[width=0.99\linewidth]{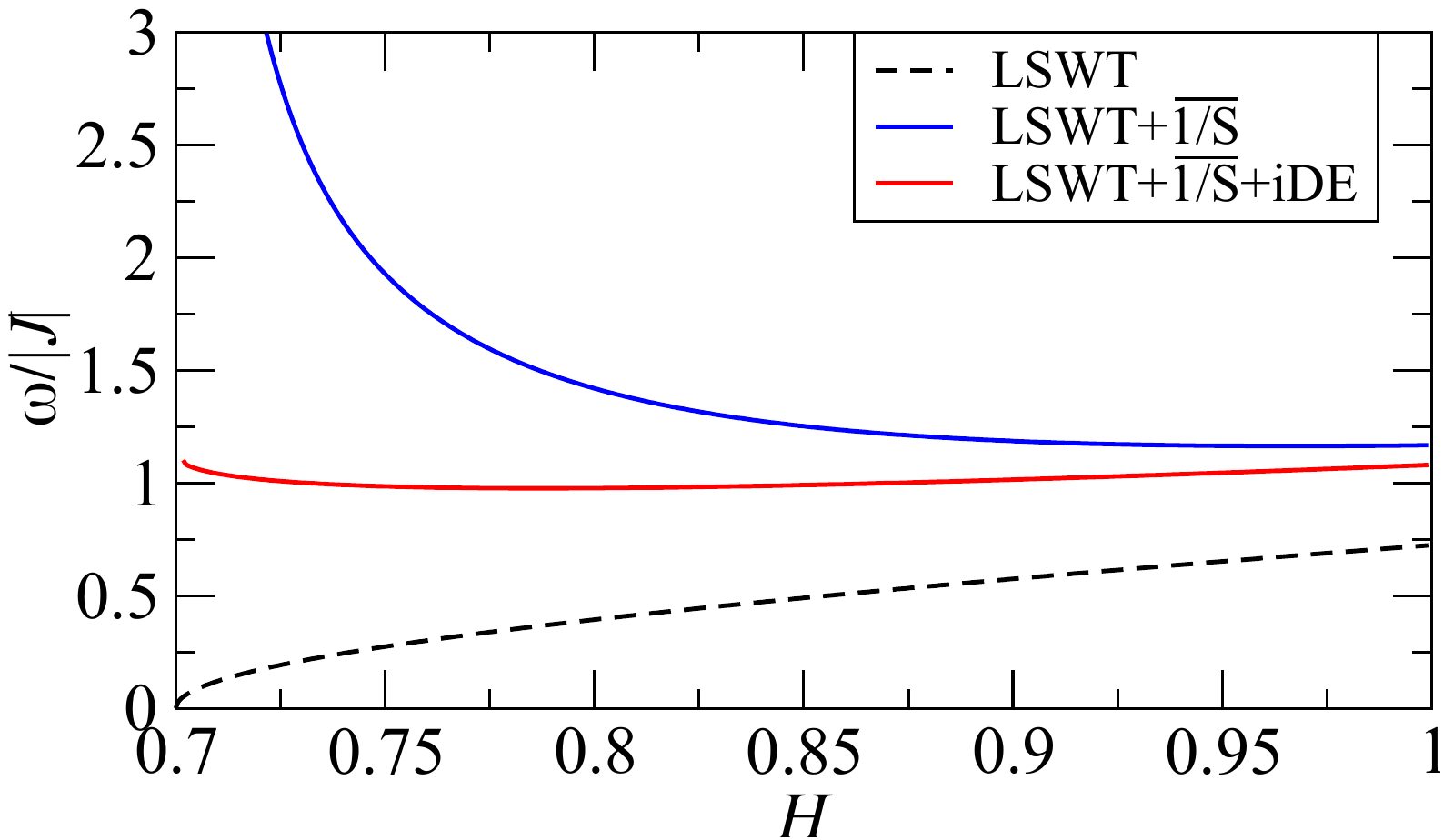}
\caption{Magnon energy at the M point vs field for $H\!\geq\!H_c$ for the same parameters as in Fig.~\ref{fig_hf_divergence}. LSWT  $\varepsilon_\mathbf{k}$  (\ref{omegas}) (dashed line), on-shell $1/S$-renormalized $\widetilde{\varepsilon}_{\mu\mathbf{k}}$ (\ref{eq_spectrum_onshell}) (blue line), and regularized energy \eqref{eq_reg3} (red line) are shown.}
\label{fig_hf_divergence1}
\end{figure}

In addition to the results in Fig.~\ref{fig_offshell_field}, we explicate the problem of the anomalous {\it hardening} of the one-magnon mode at the M point  in Fig.~\ref{fig_hf_divergence}(a)  and its divergence at $H\!\rightarrow \!H_c$ in Fig.~\ref{fig_hf_divergence1}. Figure~\ref{fig_hf_divergence}(a)  shows the LSWT magnon energies,  $\varepsilon_{\mu\mathbf{k}}$ from Eq.~(\ref{omegas}) (dashed line),  together with the renormalized on-shell spectrum, $\widetilde{\varepsilon}_{\mu\mathbf{k}}$ from Eq.~(\ref{eq_spectrum_onshell}) (solid blue line), for one of the representative parameter sets, $J_{\pm\pm}\!=\!0.2J$, $J_3\!=\!0.5|J|$, $J\!<\!0$, $S\!=\!1/2$,   field $H\!=\!0.75|J|$, and along the same ${\bf k}$-path as in  Fig.~\ref{fig_offshell_field}. According to Eq.~(\ref{eq_hc_neg}), $H_c\!=\!0.7|J|$ for this choice of parameters. Figure~\ref{fig_hf_divergence1} shows the energies of the lowest magnon mode, $\varepsilon_{1\mathbf{k}}$ and $\widetilde{\varepsilon}_{1\mathbf{k}}$,  at ${\bf k}\!=\!{\rm M}$ versus field for $H\!\geq\!H_c$ using the same line and color conventions.

Compared with Fig.~\ref{fig_offshell_field}, the effect of  one-loop  diagrams from Fig.~\ref{fig_diagrams}(a) and \ref{fig_diagrams}(b) in the renormalized on-shell spectrum $\widetilde{\varepsilon}_{\mu\mathbf{k}}$ in Fig.~\ref{fig_hf_divergence} is strictly $1/S$, see Eq.~(\ref{eq_sigma}) and Eq.~(\ref{eq_spectrum_onshell}), with two contributions, Hartree-Fock \eqref{eq_hf_sigma} and three-magnon self-energy \eqref{eq_sigma3}. The various spike-like features in the $1/S$ spectrum in Fig.~\ref{fig_hf_divergence}(a) are all clearly identifiable with the  Van Hove singularities in the two-magnon continuum, which get imprinted on the single-magnon branches by the coupling via the decay part of the three-magnon self-energy \eqref{eq_sigma_decay}. These and other non-analytic features of that nature are well-known and thoroughly documented, see Ref.~\cite{decay_review}.  It is also well-understood that their regularization requires a self-consistent approach that goes beyond the strict $1/S$-approximation, such as the self-consistent Born approximation \cite{Mourigal10}, which is useful when the products of  magnon decay are also unstable, or the technically more advantageous imaginary Dyson equation (iDE) approach \cite{umbrella,us_kagome,Smit20,us_rucl3}, which solves for the magnon energy in the complex plane and corresponds to a physical assumption of a finite lifetime in the initial state . 

While we are going to employ the iDE approach to regularize Van Hove singularities in $\widetilde{\varepsilon}_{\mu\mathbf{k}}$, the nature of the observed anomaly at the M point is unrelated to the interaction with the two-magnon continuum. One can verify that {\it all} contributions to the self-energy, Hartree-Fock, decay, and source, exhibit singular behavior at the M point as $H\!\rightarrow \!H_c$, demonstrated in Fig.~\ref{fig_hf_divergence1}. It can be shown to correspond to $\propto\!(H-H_c)^{-1/2}$, in agreement with Ref.~\cite{Vojta2020_NLSWT}. In that  work,  the choice of the simplified $K$--$J$ model and that of the high-symmetry field direction completely eliminate  three-magnon terms from the $1/S$-expansion \cite{Vojta2020_NLSWT}, leaving the Hartree-Fock terms   a  sole source of the divergent behavior. 

In this work we are dealing with a more generic case with all types of anharmonicities present, but the origin of the divergence and the type of regularization it requires can be made particularly clear from the simplified case of Ref.~\cite{Vojta2020_NLSWT}. Let us assume that only the $\omega$-independent Hartree-Fock contributions are present in the $1/S$-expansion. Because of the lower symmetry of the spin models of the studied type, the gapless  LSWT mode at the M point  at $H_c$ is necessarily relativistic. This renders all binary averages of the Holstein-Primakoff bosonic operators, which enter the Hartree-Fock decoupling of the four-boson term in the SWT Hamiltonian, finite at $H\!\rightarrow \!H_c$, see  Appendix~\ref{app_formalism} for their explicit form. Consider now the LSWT Hamiltonian \eqref{Hkc} together with the Hartree-Fock correction to  it (\ref{eq_ham_hf}) before the final Bogolyubov transformation (\ref{eq_bogolyubov}), as they take the same form 
\begin{align}
\hat{{\cal H}}^{(2)}\!+\!\delta \hat{\mathcal{H}}^{(4)}\!=\!\sum_\mathbf{k,\mu} \Big\{{\sf A}_{\mu\bf{k}}c^\dagger_{\mu\bf{k}} c^{\phantom \dagger}_{\mu\bf{k}}\!-\!\frac12\Big( {\sf B}_{\mu\bf{k}} c^\dagger_{\mu\bf{k}} c^\dagger_\mathbf{\mu -k}\!+\!\text{H.c.}\Big)\Big\},
\label{eq_H2+H4}
\end{align}
but with the functions ${\sf A}_{\mu\mathbf{k}}$ and ${\sf B}_{\mu\mathbf{k}}$ 
\begin{align}
{\sf A}_{\mu\mathbf{k}}= A_{\mu\mathbf{k}}+ \delta A^{(4)}_{\mu\bf{k}},\ \ \ {\sf B}_{\mu\mathbf{k}}= B_{\mu\mathbf{k}}+ \delta B^{(4)}_{\mu\bf{k}},
\label{eq_hf_reg1}
\end{align}
that combine terms of different orders in $1/S$ \cite{chubukov_large-s_1994,Kotov98,Kotov99}. 

It is clear that the eigenvalues of the Hamiltonian (\ref{eq_H2+H4}) 
\begin{align}
\epsilon_{\mu\mathbf{k}}= \sqrt{{\sf A}_{\mu\mathbf{k}}^2-|{\sf B}^{\phantom 2}_{\mu\mathbf{k}}|^2}\ , \
\label{eq_en_hf}
\end{align}
are perfectly regular at $H\!\rightarrow \!H_c$, and the problem of the divergence occurs solely due to its expansion in $1/S$
\begin{equation}
\!\epsilon_{\mu\mathbf{k}}\!\approx\! \varepsilon_{\mu\mathbf{k}}\!+\delta\varepsilon^{(4)}_{\mu\mathbf{k}}, \
\delta\varepsilon^{(4)}_{\mu\mathbf{k}}\!=\!\frac{A_{\mu\bf{k}} \delta A^{(4)}_{\mu\bf{k}}\!-\!\text{Re}\big( \delta B^{(4)}_{\mu\bf{k}}\big)B_{\mu\bf{k}} }{\varepsilon_{\mu\bf{k}}},
\label{eq_en_hf_exp}
\end{equation}
where we have replicated the explicit form of $\delta\varepsilon^{(4)}_{\mu\mathbf{k}}$ from (\ref{eq_hf_sigma}). 
Because of the lower symmetry of the model, the numerator in $\delta\varepsilon^{(4)}_{\mu\mathbf{k}}$ stays finite for $H\!\rightarrow \!H_c$ and ${\bf k}\!\rightarrow \!{\rm M}$, while the LSWT magnon energy in denominator vanishes, leading to a  singularity in the $1/S$ expansion.

Thus, both the nature of the divergence and the regularization procedure for it are clear. For the latter, one should retain the form of the magnon energies in (\ref{eq_en_hf}) that keeps the higher-order $1/S$ contributions, an approach also used in various contexts in the past \cite{chubukov_large-s_1994,Kotov98,Kotov99}. 

In the more general case encountered here, in the same order of the $1/S$-expansion there are also three-magnon $\omega$-dependent contributions to the self-energy that require regularization. While somewhat more involved, the approach to them will follow  similar logic. Our consideration provided above for the Hartree-Fock terms also sheds light onto the nature of the divergence in the three-magnon self-energies. It is now obvious that it comes from the LSWT magnon energy $\varepsilon_{\mu\mathbf{k}}$  in the denominators of the Bogolyubov parameters $u_{\mu\bf{k}}$ and $v_{\mu\bf{k}}$ that enter  three-magnon vertices, see Appendix~\ref{app_formalism}, having nothing to do with the two-magnon anomalies.  

Since the divergence in question does not originate from the $\omega$-dependent part of the self-energy, it suggests breaking the three-magnon self-energy in (\ref{eq_sigma3}) into the singular and regular parts, 
\begin{align}
\Sigma_\mu^{(3)}({\bf k},\varepsilon_{\mu\mathbf{k}})=\delta \Sigma_\mu({\bf k},\varepsilon_{\mu\mathbf{k}})+\Sigma^{(3)}_{\mu\mathbf{k}},
\label{eq_deltasigma}
\end{align}
where $\Sigma^{(3)}_{\mu\mathbf{k}}\!=\!\Sigma_\mu^{(3)}({\bf k},0)$ is the self-energy taken at $\omega\!=\!0$.
While $\Sigma^{(3)}_{\mu\mathbf{k}}$ is  $\omega$-independent and is free from the Van Hove singularities of the two-magnon continuum, it retains the divergence at $H\!\rightarrow \!H_c$ at the M point. On the other hand, the $\omega$-dependent part of the self-energy, $\delta \Sigma_\mu({\bf k},\omega)$, is no longer divergent at the M point at $\omega\!=\!\varepsilon_{\mu\mathbf{k}}$, because, by construction, the divergence in the  three-magnon vertices is canceled by the extra factor $\varepsilon_{\mu\mathbf{k}}$. 

The idea  is to regularize  singular $\Sigma^{(3)}_{\mu\bf{k}}$ part by converting it to the Hartree-Fock-like corrections to the LSWT Hamiltonian as in (\ref{eq_H2+H4}). However, for that one also needs to recall the off-diagonal  three-magnon self-energies, shown in Fig.~\ref{fig_diagrams}(c), that are of the same $1/S$-order, but are typically neglected as not contributing to the energy up to a higher order
\begin{align}
\Sigma_\mu^{{\rm od}}(\mathbf{k},\omega)=&-\frac{1}{2N}\sum_\mathbf{q,\eta \nu} \frac{\Xi^{\eta\nu\mu}_{{\bf q},{\bf k-q},{\bf -k}} \Phi^{\eta\nu\mu*}_{{\bf q},{\bf -k-q};{\bf k}}}{\omega+\varepsilon_\mathbf{\eta q}+\varepsilon_\mathbf{\nu k-q}-i0}\nonumber\\
&+\frac{1}{2N}\sum_\mathbf{q,\eta \nu} \frac{\Xi^{\eta\nu\mu}_{{\bf q},{\bf -k-q},{\bf k}} \Phi^{\eta\nu\mu*}_{{\bf q},{\bf k-q};{\bf -k}}}{\omega-\varepsilon_\mathbf{\eta q}-\varepsilon_\mathbf{ \nu k-q}+i0}.
\label{eq_od_selfenergy}
\end{align}
This is because we need to infer contributions to the model for $c_{\mu\bf{k}}$ ($c^\dagger_{\mu\bf{k}}$) bosons, not for the final $d_{\mu\bf{k}}$ ($d^\dagger_{\mu\bf{k}}$) quasiparticles. For the Hartree-Fock corrections, if we wanted to infer $\delta A^{(4)}_{\mu\bf{k}}$ and $\delta B^{(4)}_{\mu\bf{k}}$ from $\delta\varepsilon^{(4)}_{\mu\mathbf{k}}$, we  would also need off-diagonal $V^{{\rm od}}_{\mu\bf{k}} $ terms that are neglected as not contributing to the magnon energy in the same $1/S$-order, see Appendix~\ref{app_formalism}. The  transformation from the $c$- to $d$-language for the Hartree-Fock corrections is 
\begin{align}
\left(
\begin{array}{cc}
 \delta\varepsilon^{(4)}_{\mu\bf{k}} & V^{{\rm od}}_{\mu\bf{k}}  \\[4pt]
 V^{{\rm od}*}_{\mu\bf{k}} & \delta\varepsilon^{(4)}_{\mu\bf{k}}  \\
\end{array} 
\right)=
U \left(
\begin{array}{cc}
 \delta A^{(4)}_{\mu\bf{k}} & -\delta B^{(4)}_{\mu\bf{k}}  \\[4pt]
 -\delta B^{(4)*}_{\mu\bf{k}} & \delta A^{(4)}_\mathbf{\mu k}  \\
\end{array} 
\right)U,
\label{eq_hf_matrix}
\end{align}
with the direct and inverse Bogolyubov transformations 
\begin{align}
U=\left(
\begin{array}{cc}
 u_{\mu\bf{k}} & v_{\mu\bf{k}}  \\
 v_{\mu\bf{k}} & u_{\mu\bf{k}}  \\
\end{array} 
\right), \ \ \ 
U^{-1}=\left(
\begin{array}{cc}
 u_{\mu\bf{k}} & -v_{\mu\bf{k}}  \\
 -v_{\mu\bf{k}} & u_{\mu\bf{k}}  \\
\end{array} 
\right),
\label{UV_matrix}
\end{align}
and the $u$--$v$ parameters  defined by  
$2u_{\mu{\bf k}}v_{\mu{\bf k}}\!=\!B_{\mu{\bf k}}/\varepsilon_{\mu{\bf k}}$ and 
$u^2_{\mu{\bf k}}+v^2_{\mu{\bf k}}\!=\!A_{\mu{\bf k}}/\varepsilon_{\mu{\bf k}}$,
as before.

Since we now seek the Hartree-Fock-like corrections  from the three-magnon terms  to the Hamiltonian in the $c$-language, $\delta A^{(3)}_{\mu\bf{k}}$ and $\delta B^{(3)}_{\mu\bf{k}}$, the corresponding transformation from the $\omega\!=\!0$ self-energies is given by
\begin{align}
\left(
\begin{array}{cc}
 \delta A^{(3)}_{\mu\bf{k}} & -\delta B^{(3)}_{\mu\bf{k}}  \\[4pt]
 -\delta B^{(3)*}_{\mu\bf{k}} & \delta A^{(3)}_\mathbf{\mu k}  \\
\end{array} 
\right)= U^{-1} \left(
\begin{array}{cc}
 \Sigma^{(3)}_{\mu\mathbf{k}} & \Sigma^{{\rm od}}_{\mu\mathbf{k}}  \\[4pt]
 \Sigma^{{\rm od}*}_{\mu\mathbf{k}} & \Sigma^{(3)}_{\mu\mathbf{k}}  \\
\end{array} 
\right)U^{-1} ,
\label{eq_A3B3_matrix}
\end{align}
where $\Sigma^{{\rm od}}_{\mu\mathbf{k}}\!=\!\Sigma_\mu^{{\rm od}}(\mathbf{k},0)$ in (\ref{eq_od_selfenergy}), which can be simplified to
\begin{align}
&\delta A^{(3)}_{\mu\bf{k}}= \left( u^2_{\mu\bf{k}}+ v^2_{\mu\bf{k}}\right) \Sigma^{(3)}_{\mu\mathbf{k}}-2 u_{\mu\bf{k}} v_{\mu\bf{k}} \text{Re} \left(\Sigma^{{\rm od}}_{\mu\mathbf{k}} \right) \nonumber \\
&\delta B^{(3)}_{\mu\bf{k}}= 2 u_{\mu\bf{k}} v_{\mu\bf{k}} \Sigma^{(3)}_{\mu\mathbf{k}} -u^2_{\mu\bf{k}} \Sigma^{{\rm od}}_{\mu\mathbf{k}} -v^2_{\mu\bf{k}} \Sigma^{{\rm od}*}_{\mu\mathbf{k}}.
\label{eq_A3B3}
\end{align}
While not obviously regular, one can verify that the obtained expressions are finite at $H\!\rightarrow \!H_c$ and ${\bf k}\!\rightarrow \!{\rm M}$, as opposed to the constituent $\Sigma^{(3)}_{\mu\mathbf{k}}$ and $\Sigma^{{\rm od}}_{\mu\mathbf{k}}$. 

Altogether, the regularized spectrum is 
\begin{align}
\overline{\varepsilon}_{\mu\mathbf{k}}=\sqrt{\overline{{\sf A}}_{\mu\mathbf{k}}^2-\big|\overline{{\sf B}}^{\phantom 2}_{\mu\mathbf{k}}\big|^2}+\delta\Sigma({\bf k},\varepsilon_\mathbf{k}),
\label{eq_en_hf1}
\end{align}
with $\delta\Sigma_\mu (\mathbf{k},\omega)$ given by \eqref{eq_deltasigma} and
\begin{align}
&\overline{{\sf A}}_{\mu\mathbf{k}}= A_{\mu\mathbf{k}}+ \delta A^{(4)}_{\mu\bf{k}}+\delta A^{(3)}_{\mu\bf{k}},
\nonumber\\ 
&\overline{{\sf B}}_{\mu\mathbf{k}}= B_{\mu\mathbf{k}}+ \delta B^{(4)}_{\mu\bf{k}}+\delta B^{(3)}_{\mu\bf{k}}.
\label{eq_hf_reg2}
\end{align}

The regularized spectrum from Eq.~\eqref{eq_hf_reg2}  is shown in Fig.~\ref{fig_hf_divergence}(b) by the green lines. The anomalous hardening of the spectrum of the on-shell $1/S$-approximation at the M-point is no longer  present and  the divergence at $H\!\rightarrow \!H_c$ at the M point is removed, see Fig.~\ref{fig_hf_divergence1}. The discussed regularization can also be  successfully combined with the self-consistent iDE regularization of the Van Hove singularities due to two-magnon continuum \cite{umbrella,Smit20,us_rucl3}.  The modified spectrum is simply
\begin{align}
\overline{\varepsilon}_{\mu\mathbf{k}}=\sqrt{\overline{{\sf A}}_{\mu\mathbf{k}}^2-\big|\overline{{\sf B}}^{\phantom 2}_{\mu\mathbf{k}}\big|^2}+\delta\Sigma({\bf k},\varepsilon_\mathbf{k}+i\Gamma_{\mu\bf{k}}),
\label{eq_reg3}
\end{align}
where the magnon decay rates $\Gamma_{\mu\bf{k}}$ are determined self-consistently 
from  $\Gamma_{\mu\bf{k}}\!=\!-{\rm Im}\big(\Sigma_\mu^{(3)}({\bf k},\varepsilon_{\mu\mathbf{k}}+i\Gamma_{\mu\bf{k}})\big)$. The spectrum is shown by the red lines in Fig.~\ref{fig_hf_divergence}(b) and Fig.~\ref{fig_hf_divergence1}. One can see that the results of the combined regularization schemes   at the M point are indistinguishable from that of Eq.~\eqref{eq_hf_reg2}, while removing the unphysical Van Hove singularities elsewhere in the spectrum. 

As one can see in our Figure~\ref{fig_hf_divergence1}, the regularized gap at the ordering wavevector remains finite at the nominal critical field $H_c$ (\ref{eq_hc_neg}) obtained by LSWT, supporting the hypothesis of Ref.~\cite{Vojta2020_NLSWT}  that the divergence of the LSWT gap signals the downward renormalization of  $H_c$  due to quantum fluctuations. However, the obtained slow field-dependence of the gap for the chosen set of parameters does not yield a reliable extrapolation to determine such a renormalized critical field, which may point to a potential first-order transition between the zigzag and partially polarized state. 

\begin{figure}
\includegraphics[width=0.99\linewidth]{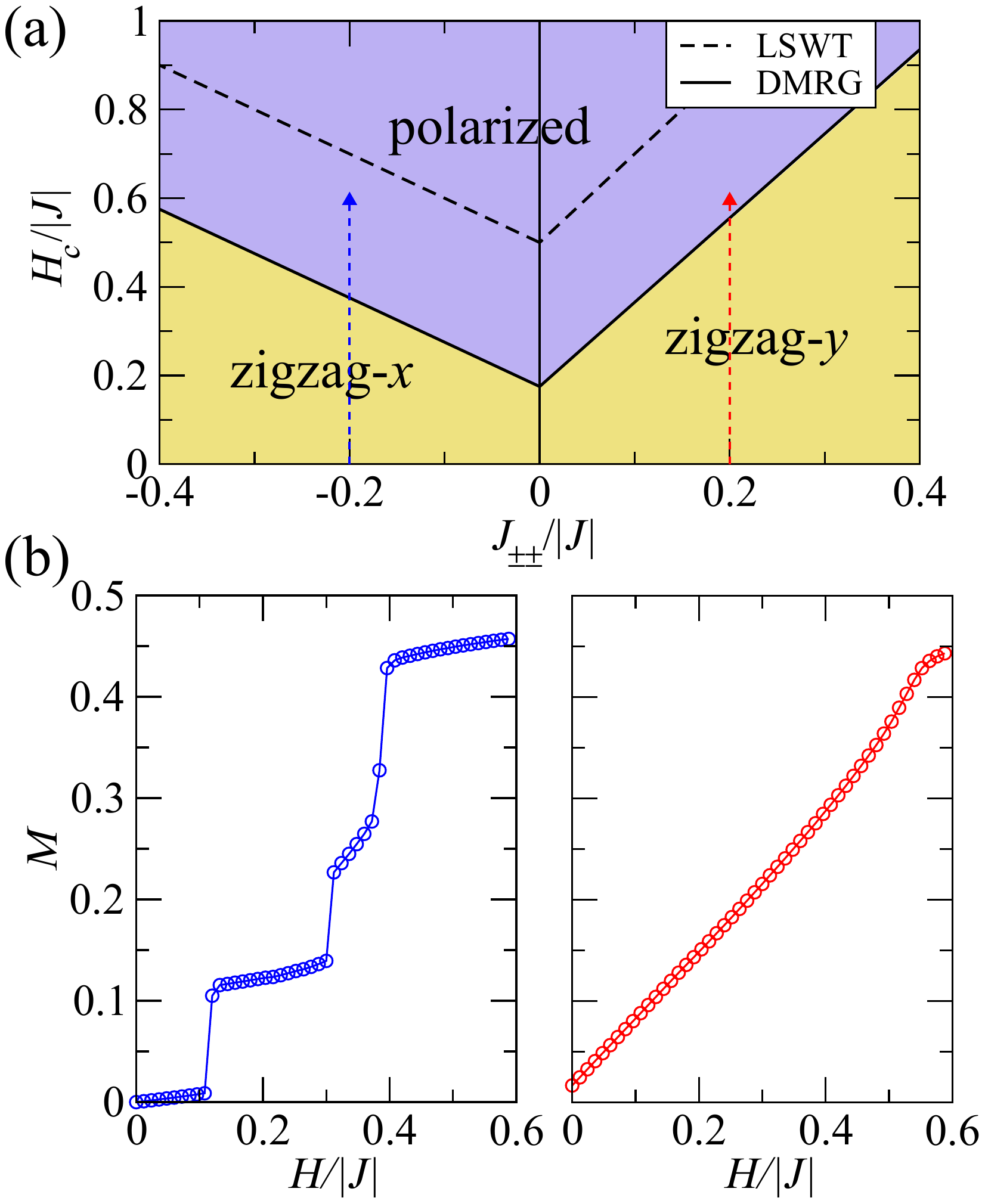}
\caption{(a) Phase diagram of the model (\ref{HJpm_xy}) in a field  as a function of $J_{\pm\pm}/|J|$ for $J_3\!=\!0.5|J|$ and $H\! \parallel \!x$. Critical field $H_c$ for the transition from  the zigzag to FM state  from LSWT \eqref{eq_hc_neg} (dashed line) and DMRG on a 128-site cluster (solid line) are shown. (b) Magnetization from DMRG as a function of magnetic field for $J_{\pm\pm}=-0.2|J|$ and $J_{\pm\pm}=0.2|J|$.}
\label{fig_dmrg_field}
\end{figure}

We complement these results by DMRG, which is used to evaluate the critical field between the zigzag and polarized  FM states for the considered case of the $S=1/2$ model (\ref{HJpm_xy}) for $H\! \parallel \!x$. The calculations were performed on an open boundary $16\times 8$-site cluster using ITensor package \cite{itensor}  with 4 sweeps, relative error $<10^{-6}$, and keeping up to $m=200$ states. For the relatively small cluster size and for the robust ordered states, the number of  sweeps and that of the kept states appear to be sufficient. The value of the critical field was extracted from the singularity in the magnetic susceptibility $\chi=dM/dH$. Our results are shown in Fig.~\ref{fig_dmrg_field}(a) for a fixed  $J_3=0.5|J|$ as a function of $J_{\pm\pm}/|J|$, scanning through 20 points along each axis, together with the LSWT results (\ref{eq_hc_neg}).  Remarkably, the transition  from the zigzag-$x$ state ($J_{\pm\pm}<0$) to the polarized state is first-order, while  transition from the zigzag-$y$ state is  second-order, which is indicated by the DMRG magnetization plots in Fig.~\ref{fig_dmrg_field}(b). We should note there are additional transitions below the critical field for $J_{\pm\pm}<0$ but they are beyond the scope of this work. Overall, the DMRG data shows that for $S=1/2$ the critical field is suppressed compared to the quasiclassical values \eqref{eq_hc_neg}, supporting the discussion provided above and also in agreement with the prior work on a related model \cite{Vojta2020_NLSWT}.

\section{Connection to Kitaev materials}
\label{sec_rucl3}

\begin{figure}
\centering
\includegraphics[width=0.99\linewidth]{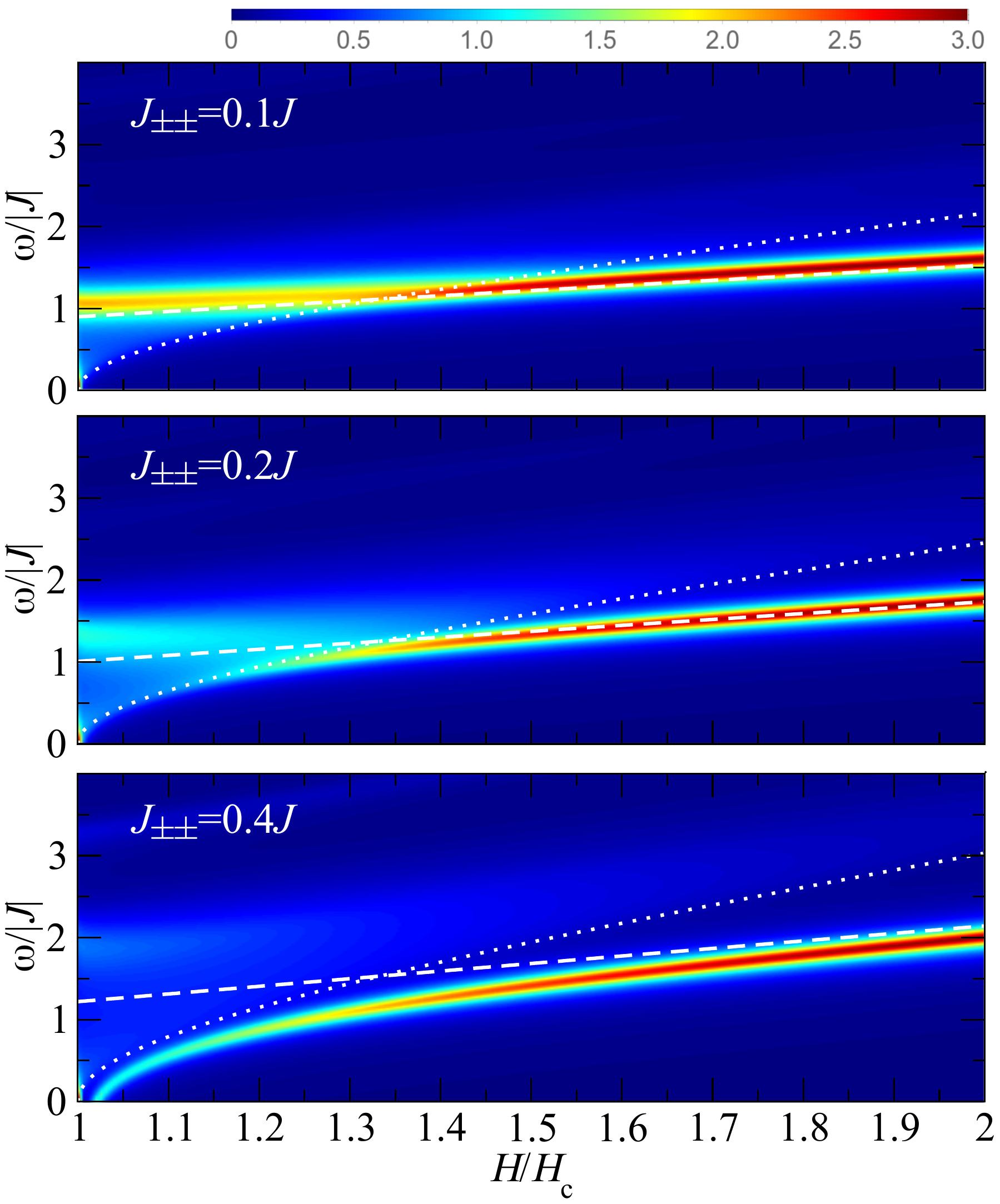}
\caption{Intensity plots of the dynamical structure factor $\cal{S}(\mathbf{k},\omega)$  at $\mathbf{k}\!=\!\Gamma$ as a function of magnetic field in the polarized phase $H\!\geq\!H_c$ for $H\!\parallel\! x$, $J\!<\!0$, $J_3\!=\!0.5|J|$, and three representative values of $J_{\pm\pm}$. The LSWT results for the lowest magnon mode $\varepsilon_{\mathbf{k}=\Gamma}$ (dashed lines) and for the bottom of the magnon continuum, $2\varepsilon_{\mathbf{k}={\rm M}}$, (dotted lines), are shown. Units of $J^{-1}$  and artificial broadening $\delta=0.1|J|$ are used.}
\label{fig_strfac_esr}
\end{figure}

In this section,  we provide a detailed look at the characteristic features of the structure factor $\mathcal{S}(\mathbf{k},\omega)$ at the ${\bf k}\!=\!\Gamma$ point in the field-polarized phase  in the proximity to a transition to the ordered zigzag phase. This analysis is important as it highlights a generic behavior of $\mathcal{S}(0,\omega)$  in a wide class of anisotropic-exchange models, which is not adequately described by the LSWT approximation  and is crucially dependent on the magnon interactions that are  induced by anisotropic terms in these models. This  ${\bf k}$-point is also special as it is often accessible by a large variety of experimental probes, such as  ESR, terahertz spectroscopy, and Raman and inelastic neutron scattering \cite{Zvyagin17,Loidl17,Balz2019,Wulferding20,Zvyagin20,Sahasrabudhe20}.

Figure~\ref{fig_strfac_esr} shows the intensity plots of $\mathcal{S}(0,\omega)$ in the polarized phase in NLSWT approximation, which includes magnon interaction effects within in the transverse component of the structure factor  and a direct contribution of the two-magnon continuum from the longitudinal part of $\mathcal{S}(\mathbf{k},\omega)$, see Sec.~\ref{sec_polarized_offshell}. Results for $J\!<\!0$, $J_3\!=\!0.5|J|$, and three different $J_{\pm\pm}/J$ are presented as a function of magnetic field for $H\!\geq\!H_c$ and $H \!\parallel \!x$ together with the LSWT results for the lowest magnon energy (dashed lines) and the bottom of the two-magnon continuum (dotted lines)  at the $\Gamma$ point (calculations for $J_3=|J|$ are presented in Appendix~\ref{app_dsf}). The LSWT energy of the lowest magnon mode at ${\bf k}\!=\!\Gamma$  for the easy-plane exchanges $J$ and $J_3$ obeys  \cite{Akhiezer_1961,us_PRR}      
\begin{align}
\varepsilon_{1,\mathbf{k}=0}=\sqrt{ H \big(  H-3S(J+J_3)\big)},
\label{eq_e1_lswt}
\end{align}
which remains gapped at the critical field $H_c$, while the bottom of the two-magnon continuum is determined by the vanishing gap at the M point and is given by $2\varepsilon_{\mathbf{k}={\rm M}}$. 

At high enough field,  quantum effects in the spectrum are negligible and LSWT gives a close description of the structure factor, see also Fig.~\ref{fig_offshell_field} and Sec.~\ref{sec_polarized_offshell}. Although for the smaller values of anisotropic-exchange term most of the weight remains in the single-magnon branch and its deviations from the LSWT energy are minimal, one can already see a substantial contribution of the two-magnon continuum at low energies in the vicinity  of the critical field and a noticeable broadening of the single-magnon branch upon entering this continuum, see the upper panel of Fig.~\ref{fig_strfac_esr}. For the progressively larger values of $J_{\pm\pm}$, the effects of broadening and spectral weight redistribution become substantially more pronounced. As one can see in the middle panel of Fig.~\ref{fig_strfac_esr}, it becomes difficult to make out the broadened single-magnon branch within the dominant two-magnon continuum intensity. 

Upon the further increase of anisotropic term, the one-magnon mode is strongly renormalized  and is  repelled from the continuum, shifting lower in energy and losing a substantial spectral weight near $H_c$, see Fig.~\ref{fig_offshell_field} and the bottom panel of Fig.~\ref{fig_strfac_esr}. Moreover, there is a significant transfer of the spectral weight to the continuum for the fields approaching $H_c$, with some remnants of the broad single-magnon mode still vaguely detectable.

Thus, we demonstrate that in anisotropic-exchange models,  the behavior of $\mathcal{S}(0,\omega)$  near $H_c$ is  dominated by the effects of strong coupling of the single-magnon mode to the two-magnon continuum,  induced by anisotropic terms. We believe that our observations capture  essential spectral behavior relevant to such models, and while the model \eqref{HJpm_xy} is not  applicable to $\alpha$-RuCl$_3$ \cite{us_PRR}, the provided analysis can be  relevant to the features that were observed in it experimentally. 

For the latter,  ESR, terahertz, Raman, and neutron-scattering experiments in $\alpha$-RuCl$_3$  have discussed their observations of the strongly field-dependent $\mathbf{k}\!=\!0$ mode in the polarized phase near $H_c$ using variable gyromagnetic ratio and have also reported  the incoherent higher-energy signal in the vicinity of the critical field \cite{Zvyagin17,Loidl17,Balz2019,Wulferding20,Zvyagin20,Sahasrabudhe20}. In agreement with the qualitative analysis of the prior work \cite{us_rucl3}, we propose that the redistribution of the spectral weight  and the  curvature of the $\mathbf{k}\!=\!0$ mode can be understood as the result of  interaction between the single-magnon mode  and the continuum that is made of quasi-Goldstone modes at the M point, whose gap closure marks a transition to the ordered zigzag state. 

Our results for the model \eqref{HJpm_xy} are also potentially relevant to the recently proposed new class of the $3d$ Co-based honeycomb compounds with Kitaev interactions and strong easy-plane anisotropy  \cite{Motome_Co_2018,Khaliullin_3d_2020,Liu_Co_review,Winter_Co_2022,BaCoAsO_22}  and  also to the experimental studies of a variety of real materials with zigzag ground states, such as Na$_2$Co$_2$TeO$_6$, Na$_3$Co$_2$SbO$_6$ \cite{Cava_Co_2007,NCSO_zigzag16,NCSO_McGuire_2019,Park_Co_2020,Ma_Co_2020,Songvilay_Co_2020,Hess_Co_2021,NCTO_Li_2021,NCTO_Garlea_2021,
Rachel_Co_2021,Li_Co_2022,Li_NCSO_2022}, CoPS$_3$ \cite{Wildes_Co_2017}, and Ag$_3$Co$_2$SbO$_6$ \cite{Zvereva_Co_2016}. Our study can also be relevant to the other zigzag antiferromagnets with strong Kitaev interactions, such as sodium iridate Na$_2$IrO$_3$ \cite{Suchitra_Ir_field_2019} and $S\!=\!1$ Kitaev magnet Li$_3$NiSbO$_6$ \cite{Zvereva_Ni_2017}.

\section{Conclusions}
\label{sec_discussion}

In the present study, we have provided a series of analytical and numerical insights into the phase diagram and spectral properties of the extended Kitaev-Heisenberg model on the honeycomb lattice in the parameter subspace that corresponds to the easy-plane limit, in which interactions are restricted to the spin projections  onto the crystallographic plane of magnetic ions. As we have emphasized, the studied easy-plane anisotropic-exchange model can be also be seen as an extension of the highly-degenerate honeycomb 120$^\circ$ compass model, with the original compass point being a tricritical point in its phase diagram. If translated to the standard parametrization in the cubic axes, the explored parameter subspace also corresponds to a general choice of variables, with all symmetry-allowed $K$, $J$, $\Gamma$, and $\Gamma'$ terms present and significant, suggesting that the offered considerations are relevant to a much wider parameter space.   

The key purpose of the present work is to offer an efficient analytical path for a consistent account of the nonlinear effects of magnon interactions in the anisotropic-exchange models, which may allow to draw convincing quantitative conclusions on the generic features of spin excitations in their ordered phases. As we have demonstrated,  one can significantly simplify the diagonalization of the harmonic spin-wave Hamiltonian by a judicious choice of the parameter subspace leading to the studied model, which allows to convert the calculation of the nonlinear terms in both zero-field zigzag and field-polarized phases into a fairly systematic procedure without losing generality of the consideration.

We have employed this approach to calculate the quantum self-energy corrections to the spin-wave spectrum in the zigzag state and demonstrated that they are strongly enhanced due to the three-magnon terms, induced  by the anisotropic interaction in the model. We have found them leading to decays and renormalization in the magnon spectrum at higher energies, extending results of the prior works. Strong renormalization of the spectrum  for larger  anisotropic term has been taken as indicative of the instability of the zigzag state, supported by our exploratory DMRG study of the phase diagram of the $S\!=\!1/2$ model, which has suggested shifting of the phase boundaries and shrinking of the zigzag phase. 

Due to  anisotropic interactions, field-polarized phase is not free from quantum fluctuations, especially in the proximity to the critical point separating ordered and nominally polarized states. Unfortunately,  strong unphysical divergences in the $1/S$ spectrum at the critical field, previously interpreted as a sign of the downward renormalization of the transition, have been observed. As a significant technical development, our study has offered  a regularization scheme of such divergences based on the renormalization of the Bogolyubov transformation, which should be applicable to a large class of anisotropic-exchange models with complex ground states. The downward shift of the critical field due to quantum effects has been also  supported by our investigative DMRG calculation of the model.

Lastly, we have provided a consideration of the characteristic features of the structure factor $\mathcal{S}(\mathbf{k},\omega)$ at the ${\bf k}\!=\!\Gamma$ point in the field-polarized phase  in the proximity to a transition to the ordered zigzag phase. This analysis has highlighted a generic behavior of $\mathcal{S}(0,\omega)$ near the critical field  in a wide class of anisotropic-exchange models, which is shown to be dominated by the effects of strong coupling of the single-magnon mode to the two-magnon continuum that are not adequately described by the LSWT approximation. We believe that our analysis captures  essential spectral behavior in many materials including $\alpha$-RuCl$_3$ and should also be relevant to the other Kitaev honeycomb magnets with strong easy-plane anisotropy.

\begin{acknowledgments}

We would like to thank Peter Kopietz for an important insight and previous collaboration that has precipitated this study.  A.~L.~C. is thankful to Matthias Vojta for a discussion and a comment. 

This work was supported by the U.S. Department of Energy,
Office of Science, Basic Energy Sciences under Award No. DE-SC0021221 (A.~L.~C.).

\end{acknowledgments}

\appendix
\section{The model and its classical states}
\label{app_model}

The model \eqref{HJpm} is written in the crystallographic axes that are related to the honeycomb plane of magnetic ions with $x$ and $y$ axes perpendicular and parallel to the AB bond, respectively, and $z$ axis normal to the plane. However, the nearest-neighbor Hamiltonian can also be written in the quantization axes that are related to the ligand octahedra \cite{rau_jkg,rau2014trigonal}, which were originally used to introduce Kitaev model in the transition metal oxides \cite{Jackeli}.

The general nearest-neighbor anisotropic-exchange Hamiltonian in the crystallographic axes is given by
\begin{align}
{\cal H}=&\sum_{\langle ij\rangle_1}\Big\{
J \Big(S^{x}_i S^{x}_j+S^{y}_i S^{y}_j+\Delta S^{z}_i S^{z}_j\Big)\nonumber\\-&2 J_{\pm \pm} \Big( \Big( S^x_i S^x_j - S^y_i S^y_j \Big) c_\alpha 
-\Big( S^x_i S^y_j+S^y_i S^x_j\Big)s_\alpha \Big)\nonumber\\
 -&J_{z\pm}\Big( \Big( S^x_i S^z_j +S^z_i S^x_j \Big) c_\alpha 
 +\Big( S^y_i S^z_j+S^z_i S^y_j\Big)s_\alpha \Big)\Big\},
\label{HJpm}
\end{align}
where we included all the nearest-neighbor exchange interactions allowed by the symmetry of the edge-sharing ligand octahedra \cite{rau2014trigonal,chaloupka2015}. The transformation from the cubic to crystallographic reference frame, ${\bf S}_{\rm cryst}\!=\!
\hat{\mathbf{R}}_c{\bf S}_{\rm cubic}$, is given by 
\begin{align}
\hat{\mathbf{R}}_c=\left(
\begin{array}{ccc}
 \frac{1}{\sqrt{6}} & \frac{1}{\sqrt{6}} & -\frac{2}{\sqrt{6}} \\
 -\frac{1}{\sqrt{2}} & \frac{1}{\sqrt{2}} & 0 \\
 \frac{1}{\sqrt{3}} & \frac{1}{\sqrt{3}} & \frac{1}{\sqrt{3}} \\
\end{array} 
\right).
\label{eq_cubic_transform}
\end{align}
The exchange parameters of models \eqref{eq_H_JKGGp} and \eqref{HJpm} are related through the linear transformation
\begin{align}
J_0&=\frac{1}{3}\left( 2J+\Delta J+2J_{\pm\pm}-\sqrt{2} J_{z\pm}\right),\nonumber\\
\label{eq_jkg_transform}
K&=-2J_{\pm\pm}+\sqrt{2}J_{z\pm},\\
\Gamma&=\frac{1}{3} \left( -J+\Delta J-4J_{\pm\pm}-\sqrt{2} J_{z\pm}\right),\nonumber\\
\Gamma'&=\frac{1}{6} \left( -2J+2\Delta J+4J_{\pm\pm}+\sqrt{2} J_{z\pm}\right),\nonumber
\end{align}
and the inverse transformation is given by
\begin{align}
J&=J_0+\frac{1}{3} \left( K-\Gamma-2\Gamma'\right),\nonumber\\
\Delta J&=J_0+\frac{1}{3} \left( K+2\Gamma+4\Gamma'\right),\nonumber\\
2J_{\pm\pm}&=\frac{1}{3} \left( -K-2\Gamma+2\Gamma'\right),\nonumber\\
\sqrt{2}J_{z\pm}&=\frac{1}{3} \left( 2K-2\Gamma+2\Gamma'\right).
\end{align}
\begin{figure}
\centering
\includegraphics[width=0.99\linewidth]{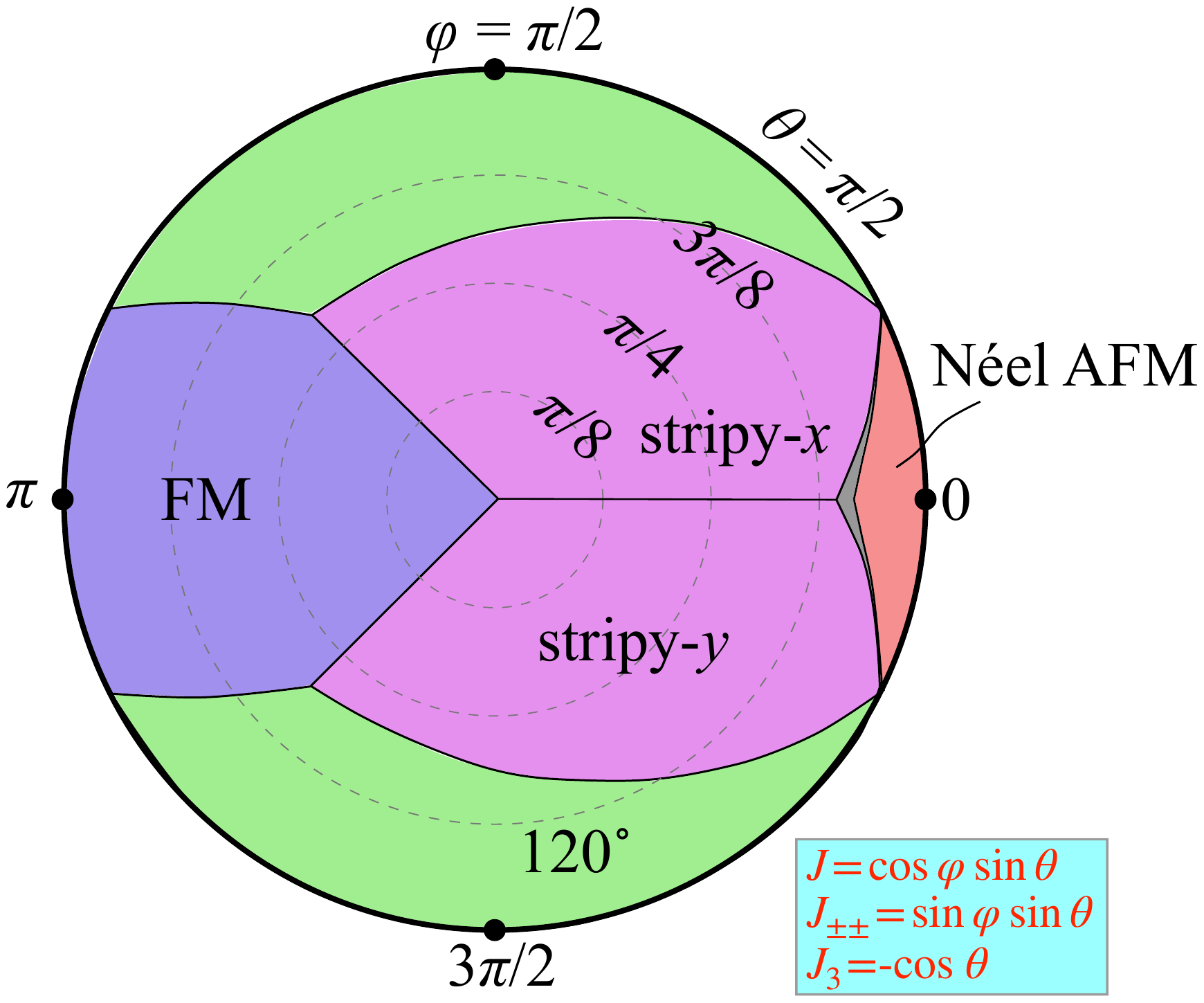}
\caption{The classical phase diagram of the $J$-$J_{\pm\pm}$-$J_3$ model \eqref{HJpm_xy} for $J_3<0$.}
\label{fig_app_pd}
\end{figure}
The phase diagrams in Fig.~\ref{fig_lattice} and Fig.~\ref{fig_app_pd} are obtained using the Luttiger-Tisza method \cite{lt_original}. Here we briefly outline its basics. Generally, the energy of the interacting classical spins $\mathbf{S}_i$ on a lattice is given by
\begin{align}
E_\text{cl}=\sum_{\langle ij\rangle_n}\sum_{\alpha\beta} \left(\mathbf{S}^\alpha\right)^{\rm T}_i \hat{\bm J}^{\alpha\beta}_{ij} \mathbf{S}^\beta_j,
\end{align}
where $\mathbf{S}^\alpha$ is a 3-dimensional vector and $\alpha$ and $\beta$ are sublattice indices. The energy minimization is also a subject of the constraint on the spin length 
\begin{align}
\left| \mathbf{S}^\alpha_i \right|^2=S^2.
\label{eq_stronglt}
\end{align}
The most common version of the LT method uses an approximation of the ``strong'' spin-length constraint by  an average, or ``weak'' constraint
\begin{align}
\sum_{i=1}^N \sum_\alpha \left| \mathbf{S}^\alpha_i \right|^2=N N_s S^2,
\end{align}
where $N$ is the number of unit cells and $N_s$ number of sublattices. The Fourier transform
\begin{align}
\mathbf{S}^\alpha_i=\frac{1}{\sqrt{N}}\sum_\mathbf{k} \mathbf{S}^\alpha_\mathbf{k} e^{i\mathbf{k}\mathbf{r}_i},
\end{align}
yields the classical energy 
\begin{align}
E_\text{cl}=\sum_\mathbf{k}\sum_{\alpha\beta} \mathbf{S}^\alpha_\mathbf{k} \hat{\bm J}^{\alpha\beta}(\mathbf{k}) \mathbf{S}^\beta_\mathbf{-k},
\end{align}
with $\hat{\bm J}^{\alpha\beta}(\mathbf{k})$ being the Fourier transform of the spin interaction matrix
\begin{align}
\hat{\bm J}^{\alpha\beta}(\mathbf{k})=\frac{1}{2}\sum_{{\bm \delta}=\mathbf{r}_j-\mathbf{r}_i} \hat{\bm J}^{\alpha\beta}_{ij} e^{i\mathbf{k} \bm \delta},
\end{align}
and the weak constraint is given by
\begin{align}
\sum_\mathbf{k} \mathbf{S}^\alpha_\mathbf{k} \mathbf{S}^\alpha_\mathbf{-k}=N S^2.
\label{eq_weak_lt_constraint}
\end{align}
The minimization of the classical energy with the weak constraint \eqref{eq_weak_lt_constraint} yields the eigenvalue equation
\begin{align}
\hat{\bm J}^{\alpha\beta}(\mathbf{k}) \mathbf{S}^\beta_\mathbf{-k}=\lambda \mathbf{S}^\beta_\mathbf{-k}.
\end{align}
The energy of the classical state corresponds to  the minimal eigenvalue $\lambda_\text{min}$ achieved at the ordering vector  $\mathbf{k}_\text{min}$, with the latter determining the type of the spin arrangement. If the corresponding state also satisfies the strong constraint \eqref{eq_stronglt}, then LT method gives the correct classical ground state. In our case, the strong constraint is  satisfied for the FM, zigzag, N\'{e}el, and stripy phases in Figs.~\ref{fig_lattice} and \ref{fig_app_pd}. Otherwise, it breaks down, suggesting a more complicated incommensurate or multi-${\bf Q}$ phases, which is the case of the grey regions in Fig.~\ref{fig_lattice}.

Here we also present the classical energies of the single-$\mathbf{Q}$ ordered states, which are shown in the phase diagrams in Fig.~\ref{fig_lattice} and Fig.~\ref{fig_app_pd}:
\begin{align}
E_\text{FM}&=3J+3J_3,\\
E_\text{AFM}&=-3J-3J_3,\nonumber\\
E_\text{stripy}&=-J+3J_3-4|J_{\pm\pm}|,\nonumber\\
E_\text{zigzag}&=J-4|J_{\pm\pm}|-3J_3,\nonumber\\
E_\text{120${\degree}$}&=-6|J_{\pm\pm}|.\nonumber
\end{align}
Similarly to the zigzag-$x$ and zigzag-$y$, there are two types of stripy states, as shown in the classical phase diagram in Fig.~\ref{fig_app_pd}: stripy-$x$ with magnetic moments in the $x-y$ plane oriented along the $x$ axis for $J_{\pm\pm}>0$, and stripy-$y$ with magnetic moments along the $y$-axis for $J_{\pm\pm}<0$.

\section{Non-linear spin-wave theory formalism}
\label{app_formalism}
In this section we present details of calculations of three-magnon and four-magnon self-energies in Fig.~\ref{fig_diagrams} and Eqs.~\eqref{eq_sigma3} and \eqref{eq_hf_sigma}. We should note that the steps presented here are applicable for both zigzag and polarized states.
\subsection{Three-magnon interaction}

The anisotropic terms of the Hamiltonian \eqref{HJpm_xy} induce three-magnon interaction due to the broken SU(2) symmetry. For the collinear zigzag and polarized states, shown in Fig.~\ref{fig_lattice}, the formalism of the three-magnon interaction can be structured in a general manner presented below. The second member of the $J_{\pm\pm}$ term in  (\ref{HJpm_xy}) is solely responsible for the anharmonic (cubic) coupling of magnons. 
After rotation to the local reference frames (\ref{rotation}) and (\ref{lab_to_loc_FM}), as can be seen from Eqs.~\eqref{HJpm_xy2} and \eqref{HJpm_local2},
it is given by
\begin{eqnarray}
\label{H3_local}
&&{\cal H}^{(3)}\Rightarrow 2 J_{\pm \pm}\sum_{i\in A}\Big[ e^{i{\bf Q}{\bf r}_{i}}S_{i}^{z}
\sum_{{\bm \delta}_\alpha} S^{x}_j \sin \tilde{\varphi}_\alpha\\
&&\phantom{{\cal H}^{(3)}\Rightarrow +2 J_{\pm \pm}}
+S^{x}_i \sum_{{\bm \delta}_\alpha}  e^{i{\bf Q}({\bf r}_{i}+{\bm \delta}_\alpha-{\bm \delta}_1)}S_{j}^{z} 
\sin \tilde{\varphi}_\alpha\Big],\nonumber \ \ \ \ \ \ \ 
\end{eqnarray}
where, $i\!\in\! A$, $j\!\in\! B$, 
${\bf r}_{j}\!=\!{\bf r}_{i}+{\bm \delta}_\alpha$, see Fig.~\ref{fig_lattice}.
Note that since $\tilde{\varphi}_1\!=\!0$,  cubic terms are generated only due to couplings 
along the ${\bm \delta}_2$ and ${\bm \delta}_3$ bonds. With a little bit of algebra, the phase factor in the second term
of (\ref{H3_local}) simplifies to the same $e^{i{\bf Q}{\bf r}_{i}}$ as in the first term. The ordering vector $\mathbf{Q}$ characterizes the ground state, thus, $\mathbf{Q}=(0,2\pi/3)$ for the zigzag state and $\mathbf{Q}=0$ in the uniformly polarized state. This expression is valid for zigzag-$x$ state and polarized state for $H \parallel x$, while the results for zigzag-$y$ and $H \parallel y$ polarized state are obtained with $J_{\pm\pm}\rightarrow -J_{\pm\pm}$ due to the symmetry of the Hamiltonian \eqref{HJpm_xy}.

After  Holstein-Primakoff  and Fourier transforms (\ref{eq_fourier})  and using  
$\sin \tilde{\varphi}_{2(3)}\!=\!\pm\sqrt{3}/2$ the three-magnon terms are given by
\begin{align}
{\cal H}^{(3)}=\frac{\tilde{J}^{(3)}}{\sqrt{N}}\sum_{\sum {\bf k}_i={\bf Q}} \Big( &\gamma^{\prime\prime}_{{\bf q}-{\bf Q}} a^{\dag}_{\bf q} b^{\dag}_{\bf k} b_{-{\bf p}}  \nonumber\\
 + &\gamma^{\prime\prime *}_{\bf q} b^{\dag}_{\bf q} a^{\dag}_{\bf k} a_{-{\bf p}}+{\rm H.c.}\Big) ,
\label{H3k0}
\end{align}
where $\tilde{J}^{(3)}\!=\!-3\sqrt{2S} J_{\pm\pm}$,
$\sum {\bf k}_i\!=\!{\bf p}\!+\!{\bf k}\!+\!{\bf q}$, and $\sum {\bf k}_i\!=\!-{\bf Q}$ in the H.c.-terms. The amplitude $\gamma^{\prime\prime}_{{\bf q}}$ is
\begin{align}
\gamma^{\prime\prime}_{{\bf q}}=\frac{1}{3}\sum_{\alpha} s_\alpha e^{i\mathbf{q}{\bm \delta}_\alpha}=\frac1{2\sqrt3}\Big(e^{i{\bf q} \bm{\delta}_2}- e^{i{\bf q} \bm{\delta}_3}\Big)=
|\gamma^{\prime\prime}_{{\bf q}}|e^{i\psi_{\bf q}},
\label{gk11}
\end{align}

Note that the three-magnon vertex strength $\tilde{J}^{(3)}$ is only dependent on the bond-anisotropic interaction $J_{\pm\pm}$. The unitary transformation \eqref{eq_phaseshift}  transforms (\ref{H3k0}) to
\begin{eqnarray}
{\cal H}^{(3)}\!=\!\frac{\tilde{J}^{(3)}}{\sqrt{N}}\!\!\!\sum_{\sum {\bf k}_i={\bf Q}} \sum_{\eta\nu\mu} \left(
F^{\eta\nu\mu}_{{\bf q},{\bf k}{\bf p}} c^{\dagger}_{\eta{\bf q}} c^{\dagger}_{\nu{\bf k}} 
c^{\phantom{\dag}}_{\mu-{\bf p}}\!+\! {\rm H.c.}_{-{\bf Q}}\right)\!, \ \ \ \
\label{H3ktr}
\end{eqnarray}
where $F^{\eta\nu\mu,i}_{{\bf q},{\bf k}{\bf p}}$ is the dimensionless vertex
\begin{eqnarray}
F^{\eta\nu\mu}_{{\bf q},{\bf k}{\bf p}}=\frac{|\gamma^{\prime\prime}_{\bf q}|e^{\frac{i\psi_{{\bf Q}}}{2}}}
{2\sqrt{2}}\sum_{\alpha\neq\beta}
 V^{\alpha\eta}V^{\beta\nu}V^{\beta\mu}
 \, e^{i(-1)^\beta \widetilde{\varphi}_{{\bf q}, {\bf k p}}} \ \ , \ \ \ \ 
\label{eq_Fphase}
\end{eqnarray}
where the ``total'' phase factor 
\begin{eqnarray}
\widetilde{\varphi}_{{\bf q}, {\bf k p}}=\frac{\psi_{{\bf Q}}}{2}+
\psi_{\bf q}+\frac{\varphi_{{\bf k}}+ \varphi_{{\bf p}} - \varphi_{{\bf q}}}{2},
\label{psikqp}
\end{eqnarray}
is introduced.

Using the symmetry of the vertex in (\ref{eq_Fphase})  to permutations of ${\bf k}$  and ${\bf p}$ 
momenta (together with the $\nu$ and $\mu$ boson indices) 
${F}^{\eta\nu\mu}_{{\bf q},{\bf k}{\bf p}}\!=\!{F}^{\eta\mu\nu}_{{\bf q},{\bf p}{\bf k}}$, 
antisymmetry of the phase    
$\widetilde{\varphi}_{-{\bf q}, -{\bf k}\, -{\bf p}}\!=\!-\widetilde{\varphi}_{{\bf q}, {\bf k p}}$
that gives $\left({F}^{\eta\nu\mu}_{-{\bf q}, -{\bf k}\, -{\bf p}}\right)^*\!=\!{F}^{\eta\mu\nu}_{{\bf q},{\bf k}{\bf p}}$,
and  explicit expression for ${F}^{\eta\nu\mu}_{{\bf q},{\bf k}{\bf p}}$ in (\ref{eq_Fphase}), one can considerably simplify 
individual terms of the tensor to
\begin{eqnarray}
&&{F}^{111}_{{\bf q},{\bf k}{\bf p}}\!=\!{F}^{122}_{{\bf q},{\bf k}{\bf p}}\!=\!
- {F}^{221}_{{\bf q},{\bf k}{\bf p}}\!=\!- {F}^{212}_{{\bf q},{\bf k}{\bf p}}\!=\!
\frac{i|\gamma^{\prime\prime}_{\bf q}|e^{\frac{i\psi_{{\bf Q}}}{2}}\!\!
\sin\widetilde{\varphi}_{{\bf q}, {\bf k p}}}{\sqrt{2}}, \ \  \ \ \
\label{Fcompact}
\\
&&{F}^{222}_{{\bf q},{\bf k}{\bf p}}\!=\!{F}^{211}_{{\bf q},{\bf k}{\bf p}}\!=\!
-{F}^{112}_{{\bf q},{\bf k}{\bf p}}\!=\!-{F}^{121}_{{\bf q},{\bf k}{\bf p}}\!=\! 
\frac{|\gamma^{\prime\prime}_{\bf q}|e^{\frac{i\psi_{{\bf Q}}}{2}}\!
\cos\widetilde{\varphi}_{{\bf q}, {\bf k p}}}{\sqrt{2}}. \ \ \  \nonumber
\end{eqnarray}
In anticipation of the decay rates, it is already clear that different decay channels, say $2\!\rightarrow\!\{1,1\}$, 
do not mix  terms of the tensor of different symmetries ($\cos\widetilde{\varphi}$ with 
$\sin\widetilde{\varphi}$).

Finally, the Bogolyubov transformation (\ref{eq_bogolyubov}) yields the cubic Hamiltonian for the 
magnon normal modes in the following form
\begin{eqnarray}
&&\hat{\cal H}^{(3)}=\frac{1}{3!\sqrt{N}}\sum_{\sum {\bf k}_i={\bf Q}}\sum_{\eta\nu\mu} \left(
\Xi^{\eta\nu\mu}_{{\bf q}{\bf k}{\bf p}} 
d^{\dagger}_{\eta{\bf q}} d^{\dagger}_{\nu{\bf k}} d^{\dagger}_{\mu{\bf p}}+{\rm H.c.}\right)\hskip 0.88cm \
\label{Hsource2}
\\
&&\phantom{\hat{\cal H}^{(3)}}
+\frac{1}{2!\sqrt{N}}\sum_{\sum {\bf k}_i={\bf Q}}\sum_{\eta\nu\mu} \left(
\Phi^{\eta\nu\mu}_{{\bf q}{\bf k};{\bf p}} 
d^{\dagger}_{\eta{\bf q}} d^{\dagger}_{\nu{\bf k}} d^{\phantom{\dag}}_{\mu-{\bf p}}+{\rm H.c.}\right),
\label{Hdecay2}
\end{eqnarray}
where the combinatorial factors are due to symmetrization in the source (\ref{Hsource2}) and decay (\ref{Hdecay2}) vertices
\begin{align}
\Xi^{\eta\nu\mu}_{{\bf q}{\bf k}{\bf p}}=\tilde{J}^{(3)}\, \widetilde{\Xi}^{\eta\nu\mu}_{{\bf q}{\bf k}{\bf p}},~
\Phi^{\eta\nu\mu}_{{\bf q}{\bf k};{\bf p}}=\tilde{J}^{(3)}\, \widetilde{\Phi}^{\eta\nu\mu}_{{\bf q}{\bf k};{\bf p}}
\end{align}
with the corresponding dimensionless vertices given by 
\begin{eqnarray}
&&\widetilde{\Xi}^{\eta\nu\mu}_{{\bf q}{\bf k}{\bf p}} = 
F^{\eta\nu\mu}_{{\bf q},{\bf k}{\bf p}} 
(u_{\eta{\bf q}}+v_{\eta{\bf q}})
(u_{\nu{\bf k}}v_{\mu{\bf p}}+v_{\nu{\bf k}}u_{\mu{\bf p}})
\label{eq_sourcevertex}
\\
&&\phantom{\widetilde{\Xi}^{\eta\nu\mu}_{{\bf q}{\bf k}{\bf p}}}
+ F^{\nu\eta\mu}_{{\bf k},{\bf q}{\bf p}} 
(u_{\nu{\bf k}}+v_{\nu{\bf k}})
(u_{\eta{\bf q}}v_{\mu{\bf p}}+v_{\eta{\bf q}}u_{\mu{\bf p}})\nonumber\\
&&\phantom{\widetilde{\Xi}^{\eta\nu\mu}_{{\bf q}{\bf k}{\bf p}}}
+ F^{\mu\eta\nu}_{{\bf p},{\bf q}{\bf k}} 
(u_{\mu{\bf p}}+v_{\mu{\bf p}})
(u_{\eta{\bf q}}v_{\nu{\bf k}}+v_{\eta{\bf q}}u_{\nu{\bf k}})\, ,\nonumber\\
&&\widetilde{\Phi}^{\eta\nu\mu}_{{\bf q}{\bf k};{\bf p}} =  
F^{\eta\nu\mu}_{{\bf q},{\bf k}{\bf p}} 
(u_{\eta{\bf q}}+v_{\eta{\bf q}})
(u_{\nu{\bf k}}u_{\mu{\bf p}}+v_{\nu{\bf k}}v_{\mu{\bf p}})
\label{eq_decayvertex2}
\\
&&\phantom{\widetilde{\Phi}^{\eta\nu\mu}_{{\bf q}{\bf k};{\bf p}}}
+ F^{\nu\mu\eta}_{{\bf k},{\bf p}{\bf q}} 
(u_{\nu{\bf k}}+v_{\nu{\bf k}})
(u_{\eta{\bf q}}u_{\mu{\bf p}}+v_{\eta{\bf q}}v_{\mu{\bf p}})\nonumber\\
&&\phantom{\widetilde{\Phi}^{\eta\nu\mu}_{{\bf q}{\bf k};{\bf p}}}
+ F^{\mu\eta\nu}_{{\bf p},{\bf q}{\bf k}} 
(u_{\mu{\bf p}}+v_{\mu{\bf p}})
(u_{\eta{\bf q}}v_{\nu{\bf k}}+v_{\eta{\bf q}}u_{\nu{\bf k}}). \nonumber\quad\quad
\end{eqnarray}

The decay rate in the lowest Born approximation due to the cubic term (\ref{eq_decayvertex2}) of the ${\bf k}$ magnon from a branch $\mu$ in the $\mu\! \rightarrow\! \{ \eta, \nu \}$ 
channel is given by
\begin{align}
\label{eq_gammak}
&\Gamma^{\mu \rightarrow \{ \eta, \nu \}}_{{\bf k}} = \frac{\pi}{2N} \sum_{{\bf q}} \big| \Phi^{\eta\nu\mu}_{{\bf q},{{\bf k}-{\bf q}+{\bf Q}};-{\bf k}}\big|^2 \\
 &\quad\quad\quad\quad\quad\quad\quad\quad\quad\quad
\times\delta\left(\varepsilon_{\mu{\bf k}}-\varepsilon_{\eta{\bf q}}-\varepsilon_{\nu{\bf k-q+Q}}\right),\nonumber
\end{align}
where integration is over the Brillouin zone of the honeycomb lattice, see Fig.~\ref{fig_lattice}(d).
\subsection{Hartree-Fock corrections}
In this section we show the details of calculations of the four-magnon corrections from Fig.~\ref{fig_diagrams}(b). The four-magnon correction terms originate from ``even'' parts of the spin-wave Hamiltonian, which in the case of zigzag state is given by
\begin{align}
\mathcal{H}^\text{even}_\text{zigzag}=&\sum_{i,{\bm \delta}_1}\Big\{
\big(J- 2 \left| J_{\pm \pm}\right|\big)S^{x}_i S^{x}_j-\big(J+2 \left| J_{\pm \pm}\right|\big)S^{z}_i S^{z}_j\Big\}\nonumber\\
+&\sum_{i,{\bm \delta}_{2,3}}\Big\{
\big(J+  \left| J_{\pm \pm}\right|\big)S^{x}_i S^{x}_j+\big(J- \left| J_{\pm \pm}\right|\big)S^{z}_i S^{z}_j\Big\} 
\nonumber\\
  +&J_3\sum_{i,{\bm \delta}_\alpha^{(3)}} \Big(S^{x}_i S^{x}_j-S^{z}_i S^{z}_j\Big),
\end{align}
and in the case of the polarized phase, respectively,
\begin{align}
\mathcal{H}^\text{even}_\text{pol}=&\sum_{\langle ij\rangle_1}\Big\{
\big(J+ 2 J_{\pm \pm}c_\alpha\big)S^{x}_i S^{x}_j+\big(J-2 J_{\pm \pm}c_\alpha\big)S^{z}_i S^{z}_j \Big\}\nonumber\\
  +&J_3\sum_{\langle ij\rangle_3} \Big(S^{x}_i S^{x}_j+S^{z}_i S^{z}_j\Big)\, 
\end{align}
as can be inferred from Eqs.~\eqref{HJpm_xy2} and \eqref{HJpm_local2}. Keeping higher-order terms in the Holstein-Primakoff expansion \eqref{eq_HP}, such as
\begin{align}
S^x\approx \sqrt{\frac{S}{2}}\left( a+a^\dagger-\frac{a^\dagger aa+a^\dagger a^\dagger a}{4S}\right),
\end{align}
yields the four-magnon terms in the spin-wave Hamiltonian:
\begin{align}
&S^x_i S^x_j \rightarrow -\frac{1}{8} \left( a^{\phantom{\dagger}}_i b^\dagger_j b^{\phantom{\dagger}}_j b^{\phantom{\dagger}}_j+a^{\phantom{\dagger}}_i b^\dagger_j b^\dagger_j b^{\phantom{\dagger}}_j+\left(a\rightarrow b\right)+\text{H.c.}\right)\nonumber\\
&S^z_i S^z_j \rightarrow a^\dagger_i a^{\phantom{\dagger}}_i b^\dagger_j b^{\phantom{\dagger}}_j
\end{align}
Decoupling of the four-magnon terms is given by
\begin{align}
&S^{x}_i S^{x}_j \rightarrow - \frac{1}{2} \left[ \left( n_i +n_j \right) \left( \bar{\Delta}^{(n)}_\alpha+m^{(n)}_\alpha \right)  \right. \nonumber \\
&+ \left( n+ \frac{\delta}{2} \right) \left( a^\dagger_i b^{\phantom{\dagger}}_j+b^\dagger_j a^{\phantom{\dagger}}_i + a^{\phantom{\dagger}}_i b^{\phantom{\dagger}}_j+a^{{\dagger}}_i b^{{\dagger}}_j \right)  \nonumber\\
 &+\left.  \frac{1}{4} \left( \bar{\Delta}^{(n)}_\alpha +m^{(n)}_\alpha \right) \left(  b^{\phantom{\dagger}}_j b^{\phantom{\dagger}}_j+a^{{\dagger}}_i a^{{\dagger}}_i + \text{H.c.} \right) \right],
\end{align}
\begin{align}
S^{z}_i S^{z}_j \rightarrow &n \left( n_i +n_j \right) + m^{(n)}_\alpha \left( a^\dagger_i b^{\phantom{\dagger}}_j+b^\dagger_j a^{\phantom{\dagger}}_i \right)  \nonumber\\+ &\bar{\Delta}^{(n)}_\alpha \left( a^{\phantom{\dagger}}_i b^{\phantom{\dagger}}_j+a^{{\dagger}}_i b^{{\dagger}}_j\right) ,
\end{align}
where the Hartree-Fock averages are defined as
\begin{eqnarray}
n&=&\langle a^\dagger_i a^{\phantom{\dagger}}_i \rangle =\langle b^\dagger_j b^{\phantom{\dagger}}_j \rangle=\frac{1}{2N}\sum_{\bf k} \left(v_{1\bf k}^2+v_{2\bf k}^2\right),\\
m^{(1)}_\alpha&=& \langle a^\dagger_i b^{\phantom{\dagger}}_j \rangle =\frac{1}{2N}\sum_{\bf k} \cos \phi_\mathbf{k,\alpha} \left(v_{2\bf k}^2- v_{1\bf k}^2\right),\nonumber\\
m^{(3)}_\alpha&=& \langle a^\dagger_i b^{\phantom{\dagger}}_j \rangle =\frac{1}{2N}\sum_{\bf k} \cos \phi_\mathbf{k,\alpha}^{(3)} \left(v_{2\bf k}^2- v_{1\bf k}^2\right),\nonumber\\
\delta &=&\langle a^{\phantom{\dagger}}_i a^{\phantom{\dagger}}_i \rangle= \langle b^{\phantom{\dagger}}_j b^{\phantom{\dagger}}_j \rangle=\frac{1}{2N}\sum_{\bf k} \left(u_{1\bf k} v_{1\bf k} +u_{2\bf k} v_{2\bf k}\right),\nonumber\\
\bar{\Delta}^{(1)}_\alpha &=&\langle a^{\phantom{\dagger}}_i b^{\phantom{\dagger}}_j \rangle =\frac{1}{2N}\sum_{\bf k}\cos \phi_\mathbf{k,\alpha} \left( u_{2\bf k} v_{2\bf k}-u_{1\bf k} v_{1\bf k}\right),\nonumber\\
\bar{\Delta}^{(3)}_\alpha &=&\langle a^{\phantom{\dagger}}_i b^{\phantom{\dagger}}_j \rangle =\frac{1}{2N}\sum_{\bf k}\cos \phi_\mathbf{k,\alpha}^{(3)}  \left( u_{2\bf k} v_{2\bf k}-u_{1\bf k} v_{1\bf k}\right),\nonumber
\end{eqnarray}
where 
\begin{align}
\phi_\mathbf{k,\alpha}&=\varphi_\mathbf{k}-\mathbf{k} \bm{\delta}_\alpha,\\
\phi_\mathbf{k,\alpha}^{(3)}&= \varphi_\mathbf{k}-\mathbf{k} \bm{\delta}_\alpha^{(3)},\nonumber
\end{align}
and $i\!\in$~A, ${\bf r}_j\!=\!{\bf r}_i+{\bm \delta}({\bm \delta}^{(3)})_\alpha$. We should note that all of these averages are purely real.

\begin{figure*}
\includegraphics[width=2.0\columnwidth]{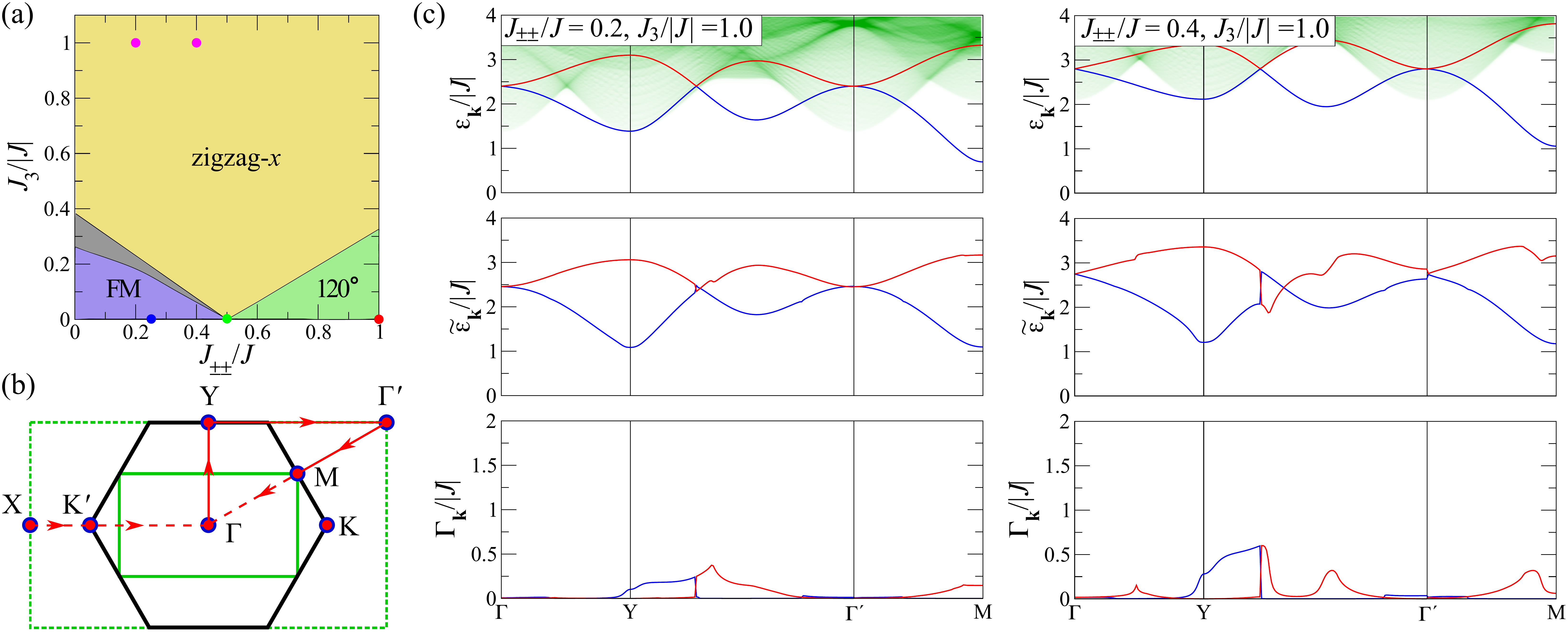}
\vskip -0.1cm
\caption{Same as in Fig.~\ref{fig_spectrum} for $J_3/|J|=1.0$.}
\label{fig_spectrum_app}
\vskip -0.2cm
\end{figure*}

After the Fourier transform \eqref{eq_fourier} four-magnon corrections are given by
\begin{align}
&\delta \mathcal{H}^{(4)}=\sum_\mathbf{k} \Big\{\delta A_\mathbf{k} \left( a^\dagger_\mathbf{k} a^{\phantom{\dagger}}_\mathbf{k} +b^\dagger_\mathbf{k} b^{\phantom{\dagger}}_\mathbf{k}\right)+\left(\delta B_\mathbf{k} a^\dagger_\mathbf{k} b^{\phantom{\dagger}}_\mathbf{k} +\text{H.c.}\right)\nonumber\\
&+\left(\delta C_\mathbf{k} a^\dagger_\mathbf{k} b^\dagger_\mathbf{-k} +\text{H.c.}\right)+\delta D_\mathbf{k} \left( a^{\phantom{\dagger}}_\mathbf{k} a^{\phantom{\dagger}}_\mathbf{-k} +b^{\phantom{\dagger}}_\mathbf{k} b^{\phantom{\dagger}}_\mathbf{-k} +\text{H.c.}\right)\Big\},
\label{eq_app_hf_fourier}
\end{align}
The unitary transform \eqref{eq_phaseshift} yields Eq.~\eqref{eq_ham_hf} where
\begin{align}
\delta A^{(4)}_\mathbf{\mu k}&=\delta A_\mathbf{k} +(-1)^\mu \text{Re} \left( \delta B_\mathbf{k} e^{-i\varphi_\mathbf{k}}\right),\nonumber\\
\delta B^{(4)}_\mathbf{\mu k}&=(-1)^{\mu+1}\text{Re}\left(\delta C_\mathbf{k} e^{-i\varphi_\mathbf{k}}\right)-2\delta D_\mathbf{k}.
\end{align}
Finally, Bogolyubov transformation \eqref{eq_bogolyubov} gives the four-magnon corrections to the  magnons as
\begin{align}
&\delta \mathcal{H}^{(4)}=\sum_{{\bf k},\mu} \Big\{
\delta\varepsilon^{(4)}_{\mu{{\bf k}}} d^\dagger_{\mu{\bf k}}d^{\phantom{\dag}}_{\mu{\bf k}}
\!+\!\frac{1}{2}\left(V^\text{od}_{\mathbf{\mu k}}
d^\dag_{\mu{\bf k}}d^\dag_{\mu-{\bf k}}\!+\!{\rm H.c.}\right)\!\Big\}\!,\ \ \
\end{align}
where $\delta\varepsilon^{(4)}_{\mu{{\bf k}}}$ is defined in Eq.~\eqref{eq_hf_sigma}, and $V^\text{od}_{\mathbf{\mu k}}$ is given by
\begin{align}
V^\text{od}_{\mathbf{\mu k}}=\frac{B_\mathbf{\mu k} \delta A^{(4)}_\mathbf{\mu k}-A_\mathbf{\mu k} \delta B^{(4)}_\mathbf{\mu k} }{\varepsilon_\mathbf{\mu k}}.
\end{align}
\begin{figure*}
\vskip -0.3cm
\includegraphics[width=2.0\columnwidth]{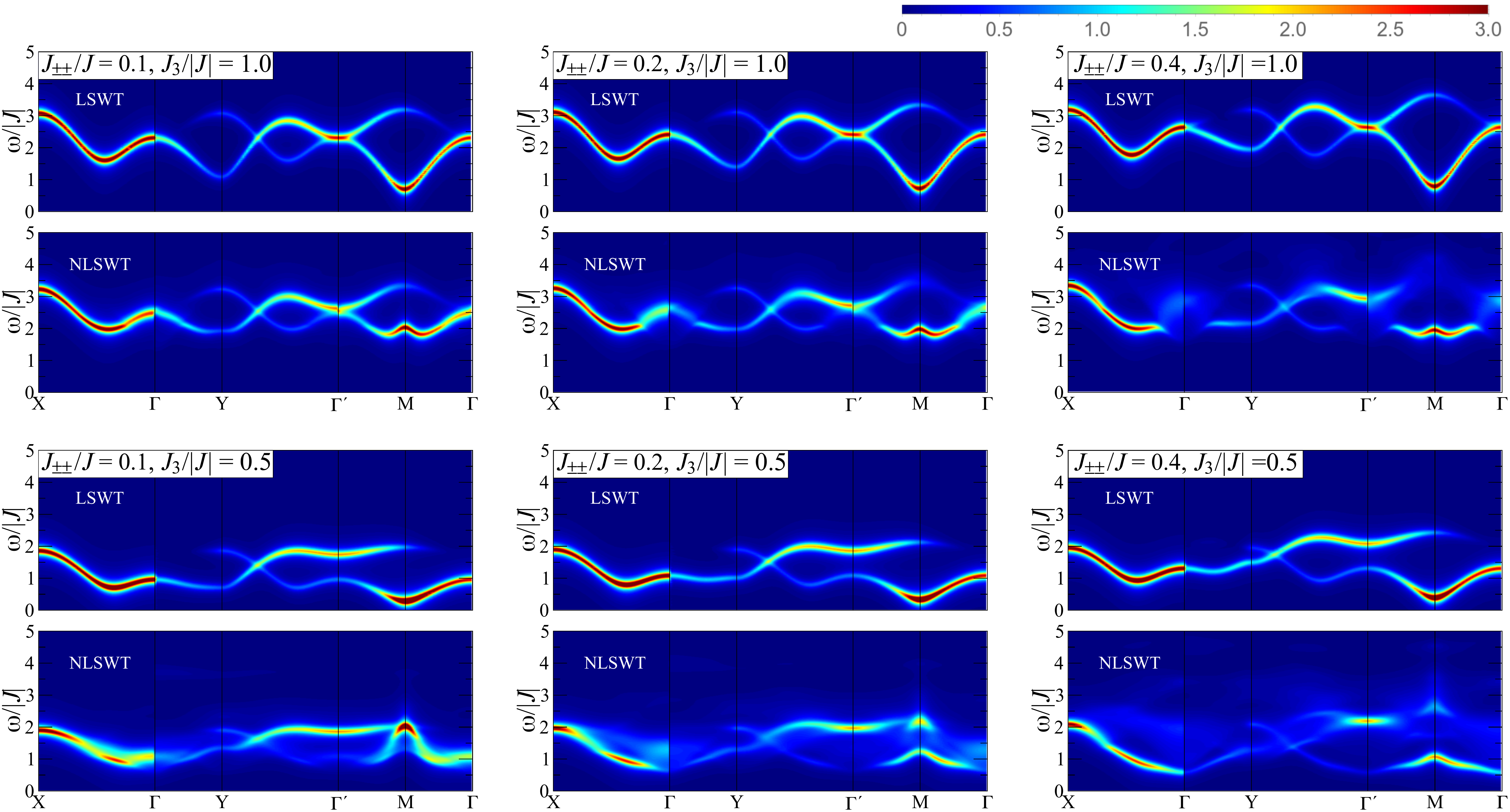}
\vskip -0.2cm
\caption{Same as in Fig.~\ref{fig_offshell_field} for various $J_{\pm\pm}$, $J_3$ and $H=1.1 H_c$.}
\label{fig_spectrum_polarized}
\end{figure*}
The procedure above is applicable for both zigzag and polarized states. The only difference is in the $\delta A_\mathbf{k}$, $\delta B_\mathbf{k}$, $\delta C_\mathbf{k}$, $\delta D_\mathbf{k}$ terms in Eq.~\eqref{eq_app_hf_fourier}. In the case of the zigzag phase, they are given by
\begin{widetext}
\begin{align}
\delta A_\mathbf{k}=\frac{J}{2}\left[2n-\sum_\alpha \left(\bar{\Delta}_\alpha+m_\alpha\right)\right]-\frac{J_3}{2}\left[6n+\sum_\alpha \left(\bar{\Delta}_\alpha^{(3)}+m_\alpha^{(3)}\right)\right]-\left| J_{\pm\pm}\right|\left[4n -\sum_\alpha c_\alpha\left(\bar{\Delta}_\alpha+m_\alpha\right)\right],
\end{align}
\begin{align}
\delta B_\mathbf{k}=&J\left[m_2 e^{i\mathbf{k} \bm{\delta}_2}+m_3 e^{i\mathbf{k} \bm{\delta}_3}-m_1 e^{i\mathbf{k} \bm{\delta}_1} -\frac{3}{4}(2n+\delta)\gamma_\mathbf{k}\right]-J_3\left[\sum_\alpha m_\alpha^{(3)} e^{i\mathbf{k} \bm{\delta}_\alpha^{(3)}} +\frac{3}{4}(2n+\delta)\gamma_\mathbf{k}^{(3)}\right]\nonumber\\
-&\left| J_{\pm\pm}\right|\left[ 2m_1 e^{i\mathbf{k} \bm{\delta}_1}+m_2 e^{i\mathbf{k} \bm{\delta}_2}+m_3 e^{i\mathbf{k} \bm{\delta}_3} -\frac{3}{4}(2n+\delta)\gamma'_\mathbf{k}\right],
\end{align}
\begin{align}
\delta C_\mathbf{k}=&J\left[\Delta_2 e^{i\mathbf{k} \bm{\delta}_2}+\Delta_3 e^{i\mathbf{k} \bm{\delta}_3}-\Delta_1 e^{i\mathbf{k} \bm{\delta}_1} -\frac{3}{4}(2n+\delta)\gamma_\mathbf{k}\right]-J_3\left[\sum_\alpha \bar{\Delta}_\alpha^{(3)} e^{i\mathbf{k} \bm{\delta}_\alpha^{(3)}} +\frac{3}{4}(2n+\delta)\gamma_\mathbf{k}^{(3)}\right]\nonumber\\
-&\left| J_{\pm\pm}\right|\left[2\Delta_1 e^{i\mathbf{k} \bm{\delta}_1}+\Delta_2 e^{i\mathbf{k} \bm{\delta}_2}+\Delta_3 e^{i\mathbf{k} \bm{\delta}_3} -\frac{3}{4}(2n+\delta)\gamma'_\mathbf{k}\right],
\end{align}

\begin{align}
\delta D_\mathbf{k}=-\frac{1}{8}\left[J\sum_\alpha \left(\bar{\Delta}_\alpha+m_\alpha\right)+J_3\sum_\alpha \left(\bar{\Delta}^{(3)}_\alpha+m^{(3)}_\alpha\right)-2\left| J_{\pm\pm}\right|\sum_\alpha c_\alpha \left(\bar{\Delta}_\alpha+m_\alpha\right)\right].
\end{align}
In the polarized phase the terms in Eq.~\eqref{eq_app_hf_fourier} are given by
\begin{align}
\delta A_\mathbf{k}=\frac{J}{2}\left[6n-\sum_\alpha \left(\bar{\Delta}_\alpha+m_\alpha\right)\right]+\frac{J_3}{2}\left[6n-\sum_\alpha \left(\bar{\Delta}_\alpha^{(3)}+m_\alpha^{(3)}\right)\right]-J_{\pm\pm}\sum_\alpha c_\alpha\left(\bar{\Delta}_\alpha+m_\alpha\right),
\end{align}
\begin{align}
\delta B_\mathbf{k}=&J\left[\sum_\alpha m_\alpha e^{i\mathbf{k} \bm{\delta}_\alpha} -\frac{3}{4}(2n+\delta)\gamma_\mathbf{k}\right]+J_3\left[\sum_\alpha m_\alpha^{(3)} e^{i\mathbf{k} \bm{\delta}_\alpha^{(3)}} -\frac{3}{4}(2n+\delta)3\gamma_\mathbf{k}^{(3)}\right]\nonumber\\
-&2J_{\pm\pm}\left[\sum_\alpha \cos \tilde{\varphi}_\alpha m_\alpha e^{i\mathbf{k} \bm{\delta}_\alpha} +\frac{3}{4}(2n+\delta)3\gamma'_\mathbf{k}\right],
\end{align}
\begin{align}
\delta C_\mathbf{k}=&J\left[\sum_\alpha \bar{\Delta}_\alpha e^{i\mathbf{k} \bm{\delta}_\alpha} -\frac{3}{4}(2n+\delta)\gamma_\mathbf{k}\right]+J_3\left[\sum_\alpha \bar{\Delta}_\alpha^{(3)} e^{i\mathbf{k} \bm{\delta}_\alpha^{(3)}} -\frac{3}{4}(2n+\delta)\gamma_\mathbf{k}^{(3)}\right]\nonumber\\
-&2J_{\pm\pm}\left[\sum_\alpha \cos \tilde{\varphi}_\alpha \bar{\Delta}_\alpha e^{i\mathbf{k} \bm{\delta}_\alpha} +\frac{3}{4}(2n+\delta)\gamma'_\mathbf{k}\right],
\end{align}
\begin{align}
\delta D_\mathbf{k}=-\frac{1}{8}\left[J\sum_\alpha \left(\bar{\Delta}_\alpha+m_\alpha\right)+J_3\sum_\alpha \left(\bar{\Delta}^{(3)}_\alpha+m^{(3)}_\alpha\right)+2J_{\pm\pm}\sum_\alpha c_\alpha \left(\bar{\Delta}_\alpha+m_\alpha\right)\right].
\end{align}
\end{widetext}

Figure \ref{fig_spectrum_app} shows the results of the calculations of the spectrum within the linear and non-linear approximations for the representative values of $J_{\pm\pm}$ and $J_3=|J|$ in the zero-field zigzag state. The selected parameter sets are shown by the two magenta points in the phase diagram in Fig.~\ref{fig_spectrum_app}(a). Their coordinates are $J_{\pm\pm}=0.2J$ and $0.4J$, $J_3=|J|$,  all for $J<0$.

\section{Dynamical structure factor}
\label{app_dsf}
Dynamical structure factor in the local magnetization axes for zigzag-$x$ and $H\parallel x$ polarized state is given by 
\begin{align}
\mathcal{S}\left(\mathbf{q},\omega\right) =\frac{q^2_y}{q^2} \mathcal{S}^{zz} \left(\mathbf{q},\omega\right)+\frac{q^2_x}{q^2} \mathcal{S}^{x x} \left(\mathbf{q},\omega\right)+ \mathcal{S}^{yy} \left(\mathbf{q},\omega\right)\nonumber,
\label{eq_app_dsf}
\end{align}
where $\mathbf{q}$ is defined in the crystallographic $\{x,y,z\}$ axes, while components of the structure factor $\mathcal{S}^{ab}$ in the local spin axes, see Fig.~\ref{fig_lattice}. Here we omit higher order off-diagonal terms and set $q_z=0$.

\subsection{Transverse fluctuations}
The one-magnon terms of the dynamical structure factor \eqref{eq_app_dsf} are given by Eq.~\eqref{eq_Skw_Akw0}, which intensity factors are given by
\begin{align}
{\cal F}_{1{\bf k}}^{xx}&=\sin^2 \frac{\varphi_\mathbf{k}}{2}\left(u_{1\mathbf{k}}+v_{1\mathbf{k}} \right)^2\nonumber\\
{\cal F}_{2{\bf k}}^{xx}&=\cos^2 \frac{\varphi_\mathbf{k}}{2}\left(u_{2\mathbf{k}}+v_{2\mathbf{k}} \right)^2
\end{align}
\begin{align}
{\cal F}_{1{\bf k}}^{yy}&=\cos^2 \frac{\phi_\mathbf{k,Q}}{2}\left(u_{1\mathbf{k+Q}}-v_{1\mathbf{k+Q}} \right)^2\nonumber\\
{\cal F}_{2{\bf k}}^{yy}&=\sin^2 \frac{\phi_\mathbf{k,Q}}{2}\left(u_{2\mathbf{k+Q}}-v_{2\mathbf{k+Q}} \right)^2,
\end{align}
where 
\begin{align}
\phi_\mathbf{k,Q}=\varphi_\mathbf{k+Q}-\mathbf{Q}{\bm \delta}_1,
\end{align}
and the ordering vector $\mathbf{Q}$ is chosen appropriately for the zigzag and polarized states.

\subsection{Longitudinal fluctuations}

The two-magnon (longitudinal) component is given by
\begin{align}
\mathcal{S}^{zz}\left( \mathbf{k},\omega \right)=\sum_{\mathbf{q},\mu\nu} \mathcal{F}^{\mu \nu}_{\mathbf{k}\mathbf{q}} \delta\left(\omega - \varepsilon_{\mu \bf q} - \varepsilon_{\nu \bf k-q} \right).
\label{eq_twomag}
\end{align}
The sum is over two magnon bands $\mu,\nu=1,2$.

The expression for the intensity of two-magnon contribution to the dynamical structure factor \eqref{eq_twomag} can be seen as a sum over two-magnon density of states with each component having different intensities. The expression for these intensities is generally given by
\begin{align}
\mathcal{F}^{\mu \nu}_{\mathbf{k}\mathbf{q}}=\sum_{\alpha \alpha'}\left( u^\mu_{\mathbf{q}\alpha} u^{\mu *}_{\mathbf{q}\alpha'} v^\nu_{\mathbf{q-k}\alpha'} v^{\nu *}_{\mathbf{q-k}\alpha}+u^\mu_{\mathbf{q}\alpha} v^{\mu *}_{\mathbf{-q}\alpha'} v^{\nu *}_{\mathbf{q-k}\alpha} u^{\nu *}_{\mathbf{-q+k}\alpha'}\right)
\end{align}
The sum is over $\alpha,\alpha'=A,B$. The elements of transformation matrix are given by
\begin{align}
u^\mu_{\mathbf{k}\alpha}= \begin{pmatrix} u^1_{\mathbf{k}A} & u^1_{\mathbf{k}B} \\ u^2_{\mathbf{k}A} & u^2_{\mathbf{k}B} \end{pmatrix}=\begin{pmatrix} \frac{u_{1\mathbf{k}}}{\sqrt{2}}e^{i\varphi_\mathbf{k}/2-i\mathbf{Q}\bm{\delta}_1} & \pm\frac{u_{1\mathbf{k}}}{\sqrt{2}}e^{-i\varphi_\mathbf{k}/2} \\ \frac{u_{2\mathbf{k}}}{\sqrt{2}}e^{i\varphi_\mathbf{k}/2-i\mathbf{Q}\bm{\delta}_1} & \frac{u_{2\mathbf{k}}}{\sqrt{2}}e^{-i\varphi_\mathbf{k}/2} \end{pmatrix},
\end{align}
same for $v^\mu_{\mathbf{k}\alpha}$. The plus sign and $\mathbf{Q}=0$ are for the polarized state, minus sign and ${\bf Q}\!=\!\left(0,\frac{2\pi}{3a}\right)$ are for the zigzag state.
\begin{figure}
\centering
\includegraphics[width=0.99\linewidth]{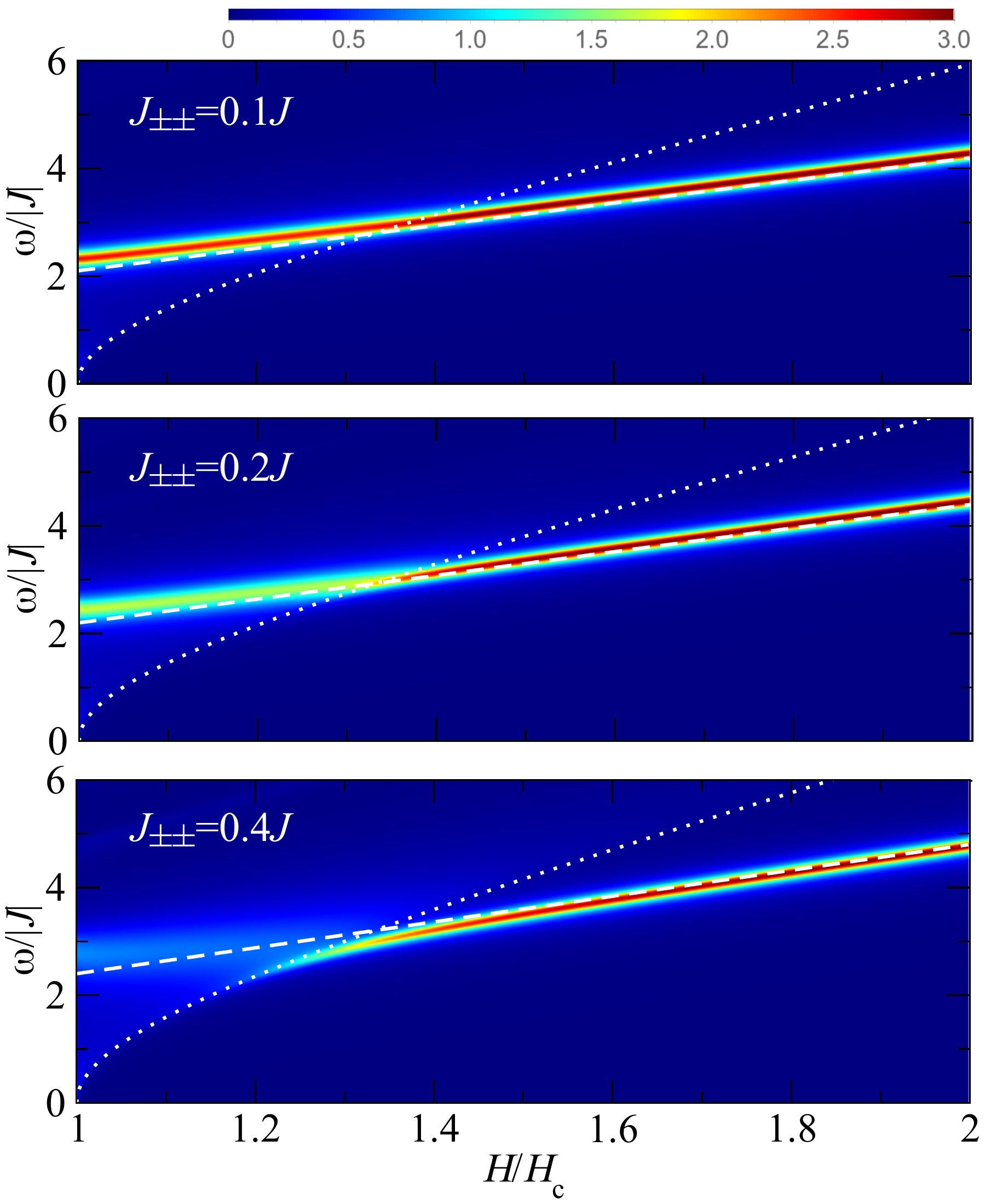}
\caption{Same as in Fig.~\ref{fig_strfac_esr} for $J_3=|J|$.}
\label{fig_spectrfunc_esr_1}
\end{figure}
Using the explicit form of the transformation one obtains
\begin{align}
\mathcal{F}^{11}_{\mathbf{k}\mathbf{q}}=\sin^2  \frac{\widetilde{\phi}_\mathbf{qk,Q}}{2}\left[ \left(u_{1\mathbf{q}} v_{1\mathbf{q-k}} \right)^2 +u_{1\mathbf{q}} v_{1\mathbf{q}} u_{1\mathbf{q}-\mathbf{k}} v_{1\mathbf{q-k}}\right],
\end{align}
\begin{align}
\mathcal{F}^{12}_{\mathbf{k}\mathbf{q}}=\cos^2 \frac{\widetilde{\phi}_\mathbf{qk,Q}}{2} \left[ \left(u_{1\mathbf{q}} v_{2\mathbf{q-k}} \right)^2 +u_{1\mathbf{q}} v_{1\mathbf{q}} u_{2\mathbf{q}-\mathbf{k}} v_{2\mathbf{q-k}}\right],
\end{align}
\begin{align}
\mathcal{F}^{21}_{\mathbf{k}\mathbf{q}}=\cos^2  \frac{\widetilde{\phi}_\mathbf{qk,Q}}{2} \left[ \left(u_{2\mathbf{q}} v_{1\mathbf{q-k}} \right)^2 +u_{2\mathbf{q}} v_{2\mathbf{q}} u_{1\mathbf{q}-\mathbf{k}} v_{1\mathbf{q-k}}\right],
\end{align}
\begin{align}
\mathcal{F}^{22}_{\mathbf{k}\mathbf{q}}=\sin^2  \frac{\widetilde{\phi}_\mathbf{qk,Q}}{2} \left[ \left(u_{2\mathbf{q}} v_{2\mathbf{q-k}} \right)^2 +u_{2\mathbf{q}} v_{2\mathbf{q}} u_{2\mathbf{q}-\mathbf{k}} v_{2\mathbf{q-k}}\right],
\end{align}
where 
\begin{align}
\widetilde{\phi}_\mathbf{qk,Q}=\varphi_\mathbf{q} +\varphi_\mathbf{k-q}-\mathbf{Q}{\bm \delta}_1.
\end{align}

Figure~\ref{fig_spectrum_polarized} shows our results for $\cal{S}(\mathbf{k},\omega)$ along the high-symmetry in-plane ${\bf k}$-path in the Brillouin zone in the polarized state for $H=1.1 H_c$ and various $J_{\pm\pm}$ and $J_3$ in comparison to the linear spin-wave theory. Magnon interactions effects are stronger for larger $J_{\pm\pm}$ due to stronger three-magnon interactions, see Sec.~\ref{sec_nlswt_formalism} and \ref{app_formalism}. Third-neighbor exchange, while not directly related to the three-magnon interactions, affects the magnon spectrum and four-magnon interactions. One can see that quantum effects are generally stronger for smaller $J_3$ due to smaller gap in the spin-wave spectrum.

Our results for the dynamical structure factor $\mathcal{S}(\mathbf{k}=0,\omega)$, which includes both transversal and longitudinal components of the spin-spin correlator, as a function of magnetic field in the polarized phase are shown in Fig.~\ref{fig_spectrfunc_esr_1} for $J_3=|J|$. Similarly to the results in Fig.~\ref{fig_strfac_esr}, larger values $J_{\pm\pm}$ yield stronger renormalization of the spectrum near the critical point of transition to the long-range-ordered state.

\bibliographystyle{apsrev4-2}
\bibliography{jpp_bib}
\end{document}